\documentclass[12pt]{article}
\usepackage[english]{babel}
\usepackage[pdftex]{color}
\usepackage{amsmath}
\usepackage{graphicx}
\usepackage{hyperref}
\usepackage{amsmath}
\usepackage{amsfonts}
\usepackage{amssymb}

\renewcommand{\theequation}{\thesection.\arabic{equation}}

\begin{document}

\title{ \textbf{{}BRST-BFV and BRST-BV Descriptions for Bosonic   Fields with   Continuous Spin on  $R^{1,d-1}$}}

\author{\sc \v{C}. Burd\'{\i}k${}^{a}$\thanks{burdik@kmlinux.fjfi.cvut.cz \hspace{0.5cm} ${}^{\dagger}$vipulvaranasi@gmail.com
\hspace{0.5cm} ${}^{\ddagger}$reshet@ispms.tsc.ru}, V.K. Pandey$^{b\dagger}$,
 A. Reshetnyak$^{c,d,e\ddagger}$ \\[0.5cm]
\it
  \it ${}^a$Department of Mathematics, Czech Technical University,\\
  \it  Prague  12000,  Czech
Republic,\\[0.2cm]
\it ${}^{b}$Department of Physics, Banaras Hindu University, 221005\\
  \it  Varanasi, India,\\[0.2cm]
\it ${}^{c}$Laboratory of Computer-Aided Design of Materials, Institute of
\\ \it Strength Physics and Materials Science SB RAS, 634055 Tomsk, Russia\\[0.2cm]
\it ${}^{d}$Tomsk State Pedagogical University, 634041, Tomsk,  Russia\\[0.2cm]
\it ${}^{e}$National Research Tomsk State University,  634050 Tomsk, Russia}
\date{}
\maketitle
\begin{abstract}
Gauge-invariant  descriptions for a free bosonic scalar field of
continuous spin in a $d$-dimensional Minkowski space-time using a
metric-like formulation are constructed on the basis of a
constrained BRST--BFV approach we propose. The resulting BRST--BFV
equations of motion for a scalar field augmented by ghost
operators contains different sets of auxiliary fields, depending
on the manner of a partial gauge-fixing and a resolution of some
of the equations of motion for a BRST-unfolded first-stage
reducible gauge theory. To achieve an equivalence of the resulting
BRST-unfolded constrained equations of motion with the initial
irreducible Poincare group conditions of a Bargmann--Wigner type,
it is demonstrated that one should replace the field in these
conditions by a class of gauge-equivalent configurations.
Triplet-like, doublet-like constrained descriptions,
as well as an unconstrained quartet-like non-Lagrangian and
Lagrangian formulations, are derived  using both Fronsdal-like and
new tensor fields. In particular, the BRST--BV equations of motion
and Lagrangian using an appropriate set of Lagrangian multipliers
in the minimal sector of the respective field and antifield
configurations are constructed in a manifest way.
\end{abstract}

\emph{The paper is dedicated to the loving memory of the father of A.A.R., Alexander Stepanovich Reshetnyak.}

\section{Introduction}

The Poincare group is a cornerstone of relativistic quantum field theories. For the first time, its representations in $\mathbb{R}^{1,3}$  were studied by E. Wigner  \cite{Wigner0}.The number of group representations describes the quantum states found in a local field theory, being some massless particles of fixed helicity (photon) and massive particles of integer (for vector and Higgs bosons) and half-integer (for quarks and leptons) spin. In higher space-time dimensions, the Poincare group $ISO(1,d-1)$ is shown to be useful in (super)string theories \cite{superstring}, \cite{BrinkHenneaux}, \cite{superstring1}.
Until now, no examples have been found to realize any other representations that exist in the Nature. So, a tachyon representation of imaginary mass, which appears to be an excitation of the lowest energy in the spectrum of bosonic string theories, is used as an indicator of instabilities, for instance, in spontaneous symmetry breaking.
The other representations are known as \emph{continuous spin representations} (CSR) which describe a massless object with an infinite number of helicities for which eigenstates of various helicities are mixed under the Lorentz transformations, in a way similar to the set
of massive particles, leading to an infinite heat capacity of the vacuum, due to Wigner's argumentation \cite{Wigner01}.

Numerous attempts have been undertaken to associate CSR with physical systems. It appears that the actual discovery of this procedure is yet to come. At the same time, the  single-valued (bosonic) and double-valued (fermionic) CSR with an infinite number of degrees of freedom (see, e.g., \cite{Iverson}) have not
yet been observed with confidence \cite{1302.4771} in the respective spectra of second-quantized bosonic strings and superstrings, as compared to the massless higher-spin (HS) fields of all the integer (0, 1, 2,...) and half-integer (1/2, 3/2, 5/2, ... ) helicities (each having a finite number of degrees of freedom), so as to be extracted using the (super)string tensionless limit  \cite{tensionlessl0}, \cite{tensionlessl}.  However, there are ways to construct a special tensionless string
limit \cite{Savvidy}, \cite{Mourad} in which CSR in the truncated
string field may be found.

The above property of CS particles is quite attractive nowadays due to an intense development of higher-spin theory
\cite{Vasiliev0}, \cite{Vasiliev1}, \cite{Vasiliev2}, \cite{Vasiliev_inter}; see the reviews \cite{Vasiliev4}, \cite{review1bek}, the discussion in the string-theory context \cite{Vasiliev5} and references therein.

Unitary irreducible  representations (UIR) using CS for the Poincare and super-Poincare groups in a $d$-dimensional Minkowski space-time with $d>4$ were first studied by the team of L. Brink and P. Ramond \cite{BrinkRamondKhan}, and, in further detail, by X. Bekaert and N. Boulanger \cite{BekaertBoulanger}.  It was shown by A.M. Khan and P. Ramond \cite{RamondKhan} that it is possible to consider CSR with CS $\Xi$ as a special limit for an HS particle of mass $m$ and spin $s$, when $\lim_{m \to 0; s\to \infty} ms = \Xi$, used to derive the Fronsdal- and Fronsdal--Fang-like equations \cite{Fronsdalfint}, \cite{Fronsdalhalfint}, albeit having CSR in the limit corresponding to massive HS particles \cite{Bekaert}, and shown to be equivalent to the Wigner and Wigner--Bargmann equations \cite{Wigner1} (for a review, see, e.g., \cite{Sorokin}).

 In turn,  a search for Lagrangian formulations (LF) and forms of relativistic field equations, not necessarily Lagrangian ones, which are to equivalently reproduce the conditions selecting massless UIR with CS, has been variously developed for $\mathbb{R}^{1,d-1}$, in both $d=4$ and higher dimensions. So, a local covariant action for bosonic CS particle  formulated using an auxiliary Lorentz vector $\eta_m$
and localized to the unit hyperboloid $\eta^2 = -1$ has been presented by an integral over $d^4x d^4\eta$ in \cite{ShusterToro} (see also \cite{Rivelles}). An LF for a scalar bosonic CSR field in terms of an infinite set of (double-)traceless totally-symmetric tensor fields of any rank in con\-stant-curvature $d$-dimensional spaces has been realized using an oscillator formalism  (in accordance with tensor representation \cite{ShusterToro})  by R.~Metsaev \cite{Metsaevcontbos}, which was used in \cite{Metsaev3} to construct  a quantum action for CSR field in $\mathbb{R}^{1,d-1}$,  whereas a twistor description for massless particles with CS has been suggested in \cite{FedorukBuch} (for relationship between the Fronsdal-like and Fang-Fronsdal-like equations \cite{Najafizadeh1} and ones obtained in \cite{Metsaevcontbos} and for  interactions, see as well, \cite{Metsaevint1}, \cite{Najafizadeh2},  \cite{Rivelles1}, \cite{Metsaevgenint1}).

Some of the most efficient tools to reconstruct a local gauge-invariant LF from the initial UIR of the Poincare or anti-de-Sitter groups  previously used merely for particles of discrete spin on a basis of the BRST--BFV approach originating from the BFV method \cite{BFV}, \cite{BFV1}, invented to quantize dynamical constrained systems, and applied, nevertheless, to a solution of the inverse problem, in fact,
to formulate an LF in terms of Hamiltonian-like objects using an auxiliary Hilbert  space whose vectors  consists of HS (spin)-tensor fields. It is not surprising that a first application in this way of the BRST--BFV method to CS fields in $R^{1,3}$ has been recently proposed by A.~Bengtsson \cite{Bengtssoncont}, one of the inventors of the constrained BRST--BFV approach to lower-spin fields \cite{BRST-BFV1}, \cite{firstBRST}.
An inclusion of holonomic (traceless and mixed-symmetry) constraints, together with differential ones, into a total system of constraints which is to be closed with respect to Hermitian conjugation with an appropriate conversion procedure for a subsystem with second-class constraints, has resulted in augmenting the original method by an unconstrained BRST--BFV method, with no restrictions imposed on the entire set of initial and auxiliary HS fields.
The application of this method  have been initiated by  A.~Pashnev, M.~Tsulaia \cite{Pashnev1}, followed by \v{C}.~B., I.~Buchbinder, V.~Krykhtin and A~R. \cite{Pashnev2}, \cite{BuchPashnev}, \cite{totfermiMin}, \cite{0505092}, \cite{totfermiAdS}, for totally-symmetric HS fields and mixed-(anti)symmetric HS fields in $R^{1,d-1}$ and AdS${}_d$ \cite{Pashnev3}, \cite{ReshMosh}, \cite{BurdikResh}, \cite{BuchbResh}, \cite{Reshetnyk2}, \cite{Reshetnyak_mas} (for a review and the interaction problem, see \cite{reviews3}).    A detailed correspondence between constrained and unconstrainedp BRST--BFV methods for arbitrary massless and massive IR of the $ISO(1,d-1)$ group with a generalized discrete spin has been recently studied in \cite{Reshetnyak_con}, where a constrained BRST--BFV LF for fermionic HS fields subject to an arbitrary Young tableaux $Y(s_1,...,s_k)$,  $k\leq [d/2]$  was first suggested and an equivalence between the underlying constrained and unconstrained LF was established. A development of this topic has resulted in an (un)constrained BRST--BV method of finding minimal BV actions necessary to construct a quantum action within the BV quantization \cite{BV} presented in \cite{ReshBRSTBV} (for bosonic HS fields, also see \cite{Barnich}, \cite{BRST-BV2}, \cite{BRST-BV3}).  An application of the BRST--BFV method to a scalar bosonic CS field in $\mathbb{R}^{1,3}$ on the basis of a so-called \emph{four-constraint formalism} \cite{Bengtssoncont} was recently proposed using the Weyl spinor notation in \cite{BuchKrychcont} (for recent developments, see also \cite{Metsaev2},  \cite{bekaertSkvortsov}, \cite{Zinoviev}, \cite{Alkalaev}, \cite{FedorukBuch2}). A prescription for a four-constraint formalism to derive an unconstrained BRST--BFV LF for a CS field \cite{Bengtssoncont} is different from the one applied to HS fields of any discrete helicity, because the set of conditions extracting a massless bosonic UIR of any integer spin and the one having CS \cite{Wigner1} contains the respective $2$ and $4$ equations, so that the ``naive'' numbers of the respective constraints being linear in the ghost approximations of Hermitian BRST operators should be $3$ and $7$.

Having in mind the equivalence between unconstrained and constrained BRST--BFV LF for one and the same HS field of a generalized discrete spin in $\mathbb{R}^{1,d-1}$  \cite{Reshetnyak_con}, we shall assume that the same property is to be valid for BRST--BFV unconstrained and constrained descriptions for the equations of motion (EoM), we intend to construct an BRST--BFV descriptions for free massless CSR particles propagating in $\mathbb{R}^{1,d-1}$. The article is devoted to the following problems:
\begin{enumerate}
\item Derivation of a constrained BRST--BFV approach to constrained gauge-invariant both non-Lagrangian description for EoM  (and Lagrangians) for a scalar CS field
in $\mathbb{R}^{1,d-1}$ in the Bargmann--Wigner form, with a compatible set of off-shell BRST-extended constraints in the metric formulation;

\item Study of an equivalence between the resulting BRST-EoM for a scalar CS field in
$\mathbb{R}^{1,d-1}$ with initial conditions extracting UIR of the $ISO(1,d-1)$ group with CS and making a comparison
with a Fronsdal-like representation;
з
\item Construction of constrained BRST--BV descriptions for EoM (and action) in the minimal sector of the field-antifield formalism on a basis of the suggested  gauge-invariant constrained EoM (and action) for a scalar CS field  in $\mathbb{R}^{1,d-1}$ in the Bargmann--Wigner form;

\item Construction of an unconstrained gauge-invariant EoM (and action) from a constrained BRST--BFV description for EoM (and action) on a basis of additional compensating field.
\end{enumerate}

The paper is organized as follows. In Section~\ref{HSsymm}, we find an higher continuous spin (HCS) symmetry algebra for a massless bosonic field with a given CS  in $\mathbb{R}^{1,d-1}$  in Bargmann--Wigner form and suggest (in Section~\ref{BRSTBFV}) a constrained BRST--BFV formulation for EoM. In the latter point, we construct a constrained BRST operator with an off-shell holonomic constraint, obtain a properly gauge-invariant EoM and action with help of the Lagrangian multipliers, find its representations in terms of Fronsdal-like fields and resolve the problem~2 concerning an equivalence with the initial set of UIR CS conditions. BRST--BV minimal formulations for EoM (and action) are derived in Section~\ref{mincBRSTBV} and an unconstrained quartet-like EoM and action  are presented in Section~\ref{tripBRSTBFV}. The short-list of the results is presented in the Conclusion.  Finally, in
Appendix~\ref{addalgebra} we construct a representation
for higher continuous spin symmetry   algebra with two pairs of  oscillators, then demonstrate in the Appendix~\ref{addalgebra2} the problem of Fock space realization and, thus, LF for the single pair of oscillators within Wigner--Bargmann form of CSR equations as well as suggest a new way to find CSR in the special tensionless limit for open bosonic string in the Appendix~\ref{addalgebra3}.

The convention $\eta_{mn} = diag (+,
-,...,-)$ for the metric tensor, with the Lorentz indices $m, n = 0,1,...,d-1$,  and the notation $\epsilon(A)$, $[gh_{H},gh_{L}, gh_{\mathrm{tot}}](A)$ for the respective values of Grassmann parity, BFV, $gh_{H}$, BV, $gh_{L}$ and total, $gh_{\mathrm{tot}}=gh_{H}+gh_{L}$, ghost numbers of a quantity $A$ are used. The totally symmetric in indices $m_1,...,m_k$ quantities $\Phi^{m_1...m_k}$ and $A^{m_1}...A^{m_k}$ are denoted respectively as $\Phi^{(m)_k}$ and $(\prod A)^{(m)_k}$.  The supercommutator $[A,\,B\}$ of quantities $A, B$
with definite values of Grassmann parity is given by $[A\,,B\}$ = $AB -(-1)^{\epsilon(A)\epsilon(B)}BA$. The Heaviside $\theta$-symbol determined as $\theta_{k,l}= 1(0)$ for $k>l (k\leq l)$.

%%%%%%%%%%%%%%%%%%%%%%%%%%%%%%%%%%%%%%%%%%%%%%%%%%%%%%%%%%%%%%%%%
\section{Higher Continuous Spin symmetry algebra $\mathcal{A}(\Xi;
\mathbb{R}^{1,d-1})$}
\label{HSsymm} %%%%%%%%%%%%%%%%%%%%%%%%%%%%%%%%%%%%%%%%%%%%%%%%
\setcounter{equation}{0}

The irreducible Poincare group massless  bosonic representation with CS in $\mathbb{R}^{1,d-1}$ is described by the $\mathbb{R}$-valued function $\Phi(x,\omega)$
of two independent variables $x^m, \omega^m$ (being by  scalar CS field \cite{BrinkRamondKhan}, \cite{ShusterToro}) on which the quadratic $C_2 = P^mP_m$ and quartic, $ C_4=W_{m_1...m_{d-3}}W^{m_1...m_{d-3}} $  Casimir operators take the values
\begin{equation}\label{Casoperator}
  C_2 \Phi(x,\omega) = 0, \quad C_4  \Phi(x,\omega) = \nu \Xi^2 \Phi(x,\omega)\ \mathrm{with} \  W^{m_1...m_{d-3}}=\epsilon^{m_1...m_{d}}P_{m_{d-2}}M_{m_{d-1}m_{d}}.
\end{equation}
$W^{m_1...m_{d-3}}$ is the generalized Pauli-Lubanski $(d-3)$-rank tensor\footnote{For $d>4$ there exist additional  Pauli-Lubanski  tensors  $W^{m_1...m_e}$= $\epsilon^{m_1...m_{d}}P_{m_{e+1}} M_{m_{e+2}m_{e+3}}\times...$ $\times M_{m_{d-1}m_{d}}$, such that $[P_{m},W^{m_1...m_e}]=0 $, thus providing for the operators $C_{2e}$ = $W_{m_1...m_e}W^{m_1...m_e} $, $e=1,3,...,d-3$ for $d=2N$, ($e=0,2,...,d-3$ for $d=2N-1$)  to be by Casimir operators \cite{BrinkRamondKhan} which are characterized by the parameters $\nu, \Xi$ and integer spin-like parameter $s_1,...,s_{k}$ for $k=[(d-4)/2]$.} with Levi-Civita tensor $\epsilon^{m_1...m_{d}}$, momentum $ P_{m}=- \imath \frac{\partial}{\partial x^m} $,  angular momentum  $M_{mn}= \widehat{M}_{mn}+S_{mn}$, for  orbital and spin parts:
 \begin{equation}\label{angspinmom}
   \widehat{M}_{mn} =  \imath x_m\frac{\partial}{\partial x^{n}}-\imath x_n\frac{\partial}{\partial x^{m}}, \ \  S_{mn}=   \imath \omega_m\frac{\partial}{\partial \omega^{n}}-\imath \omega_n\frac{\partial}{\partial \omega^{m}}.
 \end{equation}
  and with the real positive constant $\Xi$, enumerating the value of CS in $\mathbb{R}^{1,d-1}$ when $\nu=1$.  Explicitly,  the field $\Phi(x,\omega)$ should satisfy  to the 4  relations [as it was suggested for $d=4$ case by Wigner  and Bargmann \cite{Wigner1} when $\nu=1$ for the field $\widetilde{\Phi}(p,\xi)$ in momentum representation, being Fourier transform  of  $\Phi(x,\omega)$:  $\widetilde{\Phi}(p,\xi) = (2\pi)^{-d/2}\int d^dx%%
   d^d\omega \exp\{i p_{m} x^{m} +i \xi_{m} \omega^{m}\}\Phi(x,\omega)$],  In terms  of $\widetilde{\Phi}(p,\xi)$ and ${\Phi}(x,\omega)$ the equations read:
 \begin{eqnarray}\label{Eqb01w}
% \nonumber to remove numbering (before each equation)p
 &&  \Big(\eta^{mn} p_mp_n,\  \eta^{mn}{ \xi_m}p_n ,\  \imath\eta^{mn}\frac{\partial}{\partial \xi^m}p_n - \Xi,\  \eta^{mn}\xi_m \xi_n +\nu  \Big)\widetilde{\Phi}(p,\xi) = (0,0,0,0) ,  \\
 \label{Eqb01}
 && \label{Eqb23m} \eta^{mn} \frac{\partial}{\partial x^m}\frac{\partial}{\partial x^n}\Phi(x,\omega) = 0 ,  \qquad  \eta^{mn}\frac{\partial}{\partial \omega^n}\frac{\partial}{\partial x^m}\Phi(x,\omega) = 0 ,\\
  && \label{Eqb23}-\imath\omega^m\frac{\partial}{\partial x^m}\Phi(x,\omega) = \Xi \Phi(x,\omega), \qquad  \eta^{mn}\frac{\partial}{\partial \omega^m}\frac{\partial}{\partial \omega^n}\Phi(x,\omega) =   \nu \Phi(x,\omega),
\end{eqnarray}
 with some  dimensionless   parameter $\nu \in \mathbb{R}$ (being the squared  length for the  space-like  internal  vector $\xi^m$, $\xi^2 = - \nu $),  expressing the fact of ambiguity in definition of internal variables $\omega^m$ and determining  the value of the quartic Casimir operator $C_4$ on the elements of IR space of the Poincare algebra $iso(1,d-1)$ as $\nu\Xi^2$.\footnote{For $\Xi=1 $, $\nu = \mu^2$ from above equations the relations given by (1.1)--(1.4) in \cite{FedorukBuch} are obtained, whereas for $\nu=1 $ the Wigner  and Bargmann equations \cite{Wigner1} hold.}  The equations (\ref{Eqb01}), (\ref{Eqb23}) are non-Lagrangian.

Because of absence of the non-trivial solutions, with except for, $\Phi(x,\omega)=0$, due to the first equation in (\ref{Eqb23}), when expanding   $\Phi(x,\omega)$ in powers  of  only non-negative degrees in $\omega^m$, we consider the representation of $\Phi(x,\omega)$ in the form of series both in powers of   $\omega^m$ and in powers of its inverse degrees, $(\omega^m/ \omega^2)$\, in terms of independent usual in HS field theory tensor fields $\Phi^{{0}}_{(m)_k,(n)_0}(x) \equiv \Phi_{(m)_k}(x)$ and new ones for $l>0$: ${\Phi}^{{l}}_{(m)_k,(n)_l}(x)$:
\begin{equation}\label{presentaPhi}
\Phi(x,\omega) = \sum_{l\geq 0}\sum_{k\geq 0}\frac{1}{k!l!}\Phi^{{l}}_{(m)_k,(n)_l}(x)\prod_{i=1}^k\omega^{m_i}\prod_{j=1}^l\frac{\omega^{n_j}}{\omega^2}\equiv  \big(\Phi^{0} +\sum_{l\geq 1}\Phi^{l}\big)(x,\omega) ,
\end{equation}
(for $\omega^2 = \omega^m\omega_m$) to  provide  the completeness property when resolving the respective BRST complex  in the corresponding  vector (or Hilbert with endowing by an appropriate finite scalar product) space. Because of the monomial $\omega^{(m)_k}\frac{\omega^{(n)_l}}{\omega^{2l}}$ does not uniquely determine the positive and negative  degrees of $\omega^m$  its dual element $\Phi^{{l}}_{(m)_k,(n)_l}$  should have the respective dual property for $\forall l \geq 0$:
   \begin{eqnarray}
&&     \Big(\omega^{(m)_k}\frac{\omega^{(n)_l}}{\omega^{2l}} = \omega^{(m)_k}\Big\{\prod_{j=1}^i\omega^{m_{k+j}}\Big\}\frac{\omega^{(n)_{l}}\Big\{\prod_{j=1}^i\omega^{n_{l+j}}\Big\}}{\omega^{2(l+i)}}\prod_{j=1}^i\eta_{m_{k+j}n_{l+j}}\Big)\Rightarrow \nonumber \\
&& \Rightarrow \Big(\Phi^{{l}}_{(m)_k,(n)_l} \to \widetilde{\Phi}{}^{{l}}_{(m)_k,(n)_l} = \sum_{i\geq 0}d^{-i}\Phi^{{l+i}}_{(m)_{k+i},(n)_{l+i}}\prod_{j=1}^i\eta^{m_{k+j}n_{l+j}}\Big),\label{unambiguity}
   \end{eqnarray}
   where for the latter row a decomposition  for the monomial $\omega^{(m)_k}\frac{\omega^{(n)_l}}{\omega^{2l}}$ on trace and traceless parts was used (with account of the  notation for totally-symmetric sets of indices $(m)_k\equiv m_1...m_k$ and $(n)_l\equiv n_1...n_l$ and $\frac{\partial}{\partial x^n}\equiv \partial_n$)
   \begin{eqnarray}
   \hspace{-0.5em}&\hspace{-0.5em}&\hspace{-0.5em}
     \Big(\omega^{m}\frac{\omega^{n}}{\omega^{2}}=d^{-1}\eta^{mn}+ \big(\delta_{\rho}^{m}\delta_{\sigma}^{n}-d^{-1}\eta^{mn}\eta_{\rho\sigma} \big)\omega^{\rho}\frac{\omega^{\sigma}}{\omega^{2}}  \Big)\  \Rightarrow \  \Big(\omega^{(m)_{k}}\frac{\omega^{(n)_l}}{\omega^{2l}} = \nonumber\\
\hspace{-0.5em}&\hspace{-0.5em}&\hspace{-0.5em}  = \hspace{-0.5em}\prod_{i=1}^{\min(k,l)}\hspace{-0.5em}\Big\{d^{-1}\eta^{m_in_i}+ \big(\delta_{\rho_i}^{m_i}\delta_{\sigma_i}^{n_i}-d^{-1}\eta^{m_in_i}\eta_{\rho_i\sigma_i} \big)\omega^{\rho_i}\frac{\omega^{\sigma_i}}{\omega^{2}}\Big\}\Big\{\hspace{-0.1em}\theta_{k,l}\omega^{(m)_{k-l}}+\theta_{l,k-1}\frac{\omega^{(n)_{l-k}}}{\omega^{2(l-k)}}\hspace{-0.1em}\Big\}\hspace{-0.1em}\Big)\label{decompmon}
   \end{eqnarray}
  (for $\omega^{(n)_{0}}\equiv 1$).  So, the trace part of  $\omega^{m}\frac{\omega^{n}}{\omega^{2}}$ corresponds to scalar $d^{-1}\eta^{mn}\Phi^{{1}}_{m,n}$ which can be added to pure scalar $\Phi^0$.  The component fields $\Phi^{{l}}_{(m)_k,(n)_l}$ are  totally symmetric not only with respect to separate group of indices $(m)_k$ and $(n)_l$, but with respect to whole set of indices $(m)_k,(n)_l$ due to the relation: $\omega^{m_1}\frac{\omega^{n_1}}{\omega^{2}}= \omega^{n_1}\frac{\omega^{m_1}}{\omega^{2}}$.  In particular, the  property holds
 \begin{equation}\label{unambiguity1}
      \Phi^{{l}}_{(m)_k,(n)_l} = \left\{\begin{array}{c}
                                          \Phi^{{l}}_{(n)_lm_{l+1}...m_k,(m)_l}, \ \texttt{ for }\ k\geq l, \\
                                          \Phi^{{l}}_{(n)_k,(m)_k n_{k+1}...n_l} , \ \texttt{ for }\  k< l
                                        \end{array}
       \right.
   \end{equation}
   true. The symmetry properties (\ref{unambiguity1}) mean that the different traces of the functions $\Phi^{{l}}_{(m)_k,(n)_l}$ are equal:
   \begin{equation}\label{difftrace}
     \Phi^{{l}}_{(m)_k,(n)_l} \eta^{m_{k}n_{l}} \ = \ \Phi^{{l}}_{(m)_{k-2}m_{k}m_{k+1},m_{k-1}(n)_{l-1}} \eta^{m_{k}m_{k+1}}\ = \ \Phi^{{l}}_{(m)_{k-1}n_{l-1},(n)_{l-2}n_{l}n_{l+1}} \eta^{n_{l}n_{l+1}}.
   \end{equation}
   Indeed, we have the sequence of relations, e.g. for the first equality, $\Phi^{{l}}_{(m)_k,(n)_l} \eta^{m_{k}n_{l}}$  =  $\Phi^{{l}}_{(m)_{k-2}m_{k}n_{l},m_{k-1}(n)_{l-1}} \eta^{m_{k}n_{l}}$ = $\Phi^{{l}}_{(m)_{k-2}m_{k}m_{k+1},m_{k-1}(n)_{l-1}} \eta^{m_{k}m_{k+1}}$. In particular, for $k-1=l=1$ ($l-1=k=1$) we have,
   \begin{equation}\label{exdiff}
   \Phi^{{1}}_{(m)_2,n} \eta^{m_{2}n} = \Phi^{{1}}_{nm_2, m_1} \eta^{m_{2}n},\   \big(\Phi^{{2}}_{m,(n)_2} \eta^{n_{1}n_2} = \Phi^{{2}}_{n_1, m n_2} \eta^{n_{1}n_2}\big)
.  \end{equation}
  The relations (\ref{Eqb01}),  (\ref{Eqb23}) take equivalent, [but ambiguous(!) due to freedom in the definition of the monomials  and therefore component functions (\ref{unambiguity}), (\ref{decompmon})] representation   in powers of $\omega^m$ (\ref{Eqb01q}) and of  $\prod_{i=1}^k\omega^{m_i}\times $ $\prod_{j=1}^l{\omega^{n_j}}/{\omega^2}$ (\ref{Eqb2q}):
\begin{eqnarray}\label{Eqb01q}
 \hspace{-0.5em}&\hspace{-0.5em}&\hspace{-0.5em}
 \omega^{(m)_k}:\left\{\hspace{-0.5em}\begin{array}{ll}\displaystyle
 \eta^{mn}{\partial_m}\partial_n\Phi^0_{(m)_k}(x)=0,  &  \displaystyle {\partial^{m_{k+1}}}\Phi^0_{(m)_{k+1}}(x)=0, \\
\displaystyle    -\imath{\partial_{\{m_{k}}}\Phi^0_{(m)_{k-1}\}}(x) =  \Xi \Phi^0_{(m)_{k}}(x), & \displaystyle\eta^{m_{k+1}m_{k+2}}\Phi^0_{(m)_{k+2}}(x) =  \nu\Phi^0_{(m)_k}(x);
\end{array}\right.\\
 \label{Eqb2q}
 \hspace{-0.5em}&\hspace{-0.5em}&\hspace{-0.5em}
 \omega^{(m)_k}\frac{\omega^{(n)_l}}{\omega^{2l}}:\hspace{-0.2em}\left\{\hspace{-0.5em}\begin{array}{l}\displaystyle
 \eta^{mn} {\partial_m}{\partial_n}\Phi^{{l}}_{(m)_k,(n)_l}=0, \\
\displaystyle {\partial^{m_{k+1}}}\Phi^{{l}}_{(m)_{k+1},(n)_{l}} + \eta_{\{n_{l-1}n_{l}}{\partial^{n}}{\Phi}^{{l-1}}_{(m)_{k},(n)_{l-2}\}n} - 2(l-1) {\partial_{\{n_{l}}}{\Phi}^{{l-1}}_{(m)_{k},(n)_{l-1}\}} =0, \\
%%%%%%%%%%%%%%%%%%%%%
\displaystyle
   \imath{\partial_{\{m_{k}}}{\Phi}^{{l}}_{(m)_{k-1}\},(n)_l} + \Xi \eta_{\{m_{k}\{n_{l}} {\Phi}^{{l-1}}_{(m)_{k-1}\},(n)_{l-1}\}}=0  , \\
   %%%%%%%%%%%%%%%%%%%%%%%%%%%
\Phi^{{l}}_{(m)_{k}m}{}^m{}_{,(n)_l} \hspace{-0.1em}-  \hspace{-0.1em} \nu{\Phi}^{{l}}_{(m)_{k},(n)_l} \hspace{-0.1em} + \hspace{-0.1em}2  {\Phi}^{{l-1}}_{(m)_{k}n,\{(n)_{l-2}}{}^n \eta_{n_{l-1}n_{l}\}} \hspace{-0.1em}- \hspace{-0.1em}4(l-1) {\Phi}^{{l-1}}_{(m)_{k}\{n_l, (n)_{l-1}\}}   \\
   + {\Phi}^{{l-2}}_{(m)_{k},\{\{(n)_{l-4}n}{}^n \eta_{n_{l-3}n_{l-2}\}}\eta_{n_{l-1}n_{l
   }\}} + 2(l-2)(2-d){\Phi}^{{l-2}}_{(m)_{k},\{(n)_{l-2}}\eta_{n_{l-1}n_{l}\}} =0
\end{array}\right.
\end{eqnarray}
(for $\Phi^{{l}} \equiv 0 $ when $l<0$;  $k\in \mathbb{N}_0$) being respectively for each systems by D'Alambert,  divergentless,   gradient  and generalized traceless  equations.
Note, first, that all the component tensor  functions $\Phi^0_{(m)_k}$ in (\ref{Eqb01q}) and $\Phi^{{l}}_{(m)_k,(n)_l}$ in (\ref{Eqb2q}) are determined with accuracy up to the transformations (\ref{unambiguity}), i.e. may be changed on the respective  functions $\widetilde{\Phi}{}^0_{(m)_k}$ and $\widetilde{\Phi}{}^{{l}}_{(m)_k,(n)_l}$, Second,  we have used the symmetrization in indices $m_{k}, (m)_{k-1}$: $\{m_{k}, (m)_{k-1}\}$; in $n_{l-1}n_{l},  (n)_{l-2}$: $\{n_{l-1}n_{l},  (n)_{l-2}\}$ and
in 4 indices $n_{l-3}n_{l-2},...,n_{l}$ with   $(n)_{l-4}$
in (\ref{Eqb2q}) without numerical factor. Third, the left-hand side of the last equation in  (\ref{Eqb01q})
may be equivalently written as, $\eta^{m_{k+1}m_{k+2}}\Phi^0_{(m)_{k+2}}(x)= \Phi^0_{(m)_{k}m}{}^m(x)$ as it was done in the similar traceless  equations in (\ref{Eqb2q}).
The representation (\ref{presentaPhi}) leads to non-empty set of non-trivial solutions for the systems (\ref{Eqb01q}), (\ref{Eqb2q}) with account for the first note.
Fourth, the set of the equations (\ref{Eqb01q})  appears by the subsystem of the set  (\ref{Eqb2q})  for $l=0$ with allowance  made for  ambiguity  (\ref{unambiguity}), (\ref{decompmon}).
For instance, from the third equations in (\ref{Eqb2q}) for $l=1, k=1$   (with $\eta_{\{m_{1}\{n_{1}} {\Phi}^0_{(m)_{0}\},(n)_{0}\}} \equiv 2\eta_{m_{1}n_{1}} {\Phi}^{0} $) it follows:
\begin{eqnarray}\label{Eqb01q0}
\imath{\partial_{\{m_1}}{\widetilde{\Phi}}^1_{n_1\}}(x) +  2\eta_{m_1n_1} \Xi \widetilde{\Phi}{}^{0}(x)= 0 \Longleftrightarrow \left\{\begin{array}{c}
\imath d^{-1}{\partial^{n}}{\widetilde{\Phi}}^1_{n}(x) +   \Xi \widetilde{\Phi}{}^{0}(x) =0 , \\
\imath\big( \delta^{\rho}_{\{m}\delta^{\sigma}_{n\}}-d^{-1}\eta_{mn}\eta^{\rho\sigma} \big) {\partial_{\rho}}{\widetilde{\Phi}}^1_{\sigma}(x) =0.                                                                                                                            \end{array}
\right.
    \end{eqnarray}
    Note, that the similar equivalent equations take place for the third equations in (\ref{Eqb2q}) for $l>1$.\footnote{Another variant of solutions for (\ref{Eqb01}), (\ref{Eqb23}) due to the first equation in (\ref{Eqb23}) can be chosen without
poles in $\omega^m$ as its explicit solution:
$$
\Phi(x,\omega) = \delta(\omega p - \Xi) \varphi(x,\omega) ; \quad  \varphi(x,\omega)= \sum_{k\geq 0}\frac{1}{k!}\varphi_{(m)_k}(x)\omega^{m_1}...\omega^{m_k}, \ p_m = -\imath\frac{\partial}{\partial x^m}.
$$
To get the UIR with CS for the field $\varphi(x,\omega)$ one should to modify the rest equations in order to include the value of CS $\Xi$ in it that should provide the fulfillment of the equations
on the Casimir operators (\ref{Casoperator}) that was done, e.g. in \cite{Metsaevcontbos}, \cite{BuchKrychcont}. We develop the above procedure in \cite{PR2}.}

To describe the  dynamics of the fields $\Phi^{l}$ jointly, we may follow by two ways both being based on BRST-BFV approach. First variant consists in   applying the algorithm  \cite{Pashnev2}--\cite{BuchbResh}, \cite{Reshetnyk2} for construction of the Lagrangian formulation, second variant is concentrated within scope of non-Lagrangian equations of motion, as it was realized for UIR with discrete spin, see e.g. \cite{BRST-BV2}  and for CSR \cite{Alkalaev}. Whereas, the former case requires an introduction of more complicated set of oscillators to provide string-inspired Fock space structure presented in the Appendix~\ref{addalgebra}, the latter one may be realized without using of finite scalar product.

In spite of these problems the equations (\ref{Eqb01q}), (\ref{Eqb2q}) can be derived from the Lagrangian actions both in  unconstrained  with help of 4  sets of Lagrangian multipliers $\lambda^{{l}|(m)_k,(n)_l}_i$, $i=1,2,3,4$, $k,l\in \mathbb{N}_0$  and in constrained form with  3 sets of Lagrangian multipliers without $\lambda^{{l}|(m)_k,(n)_l}_4$ as follows,
\begin{eqnarray}
 S_{C|\Xi} & = &\int d^dx \sum_{k,l\geq 0}\bigg\{ \lambda^{{l}|(m)_k,(n)_l}_1 {\partial^2}\Phi^{{l}}_{(m)_k,(n)_l} +  \lambda^{{l}|(m)_k,(n)_l}_2 \Big({\partial^{m_{k+1}}}\Phi^{{l}}_{(m)_{k+1},(n)_{l}}  \nonumber\\
   && + \eta_{\{n_{l-1}n_{l}}{\partial^{n}}{\Phi}^{{l-1}}_{(m)_{k},(n)_{l-2}\}n}  - 2(l-1) {\partial_{\{n_{l}}}{\Phi}^{{l-1}}_{(m)_{k},(n)_{l-1}\}} \Big) \nonumber\\
   &&+ \lambda^{{l}|(m)_k,(n)_l}_3 \Big(   \imath{\partial_{\{m_{k}}}{\Phi}^{{l}}_{(m)_{k-1}\},(n)_l} + \Xi \eta_{\{m_{k}\{n_{l}} {\Phi}^{{l-1}}_{(m)_{k-1}\},(n)_{l-1}\}} \Big)\bigg\}, \label{costrlagr}\\
   S_{\Xi} &=&  S_{C|\Xi} + \int d^dx \sum_{k,l\geq 0} \lambda^{{l}|(m)_k,(n)_l}_4\Big( \Phi^{{l}}_{(m)_{k}m}{}^m{}_{,(n)_l} -   \nu{\Phi}^{{l}}_{(m)_{k},(n)_l} + 2  {\Phi}^{{l-1}}_{(m)_{k}n,\{(n)_{l-2}}{}^n \eta_{n_{l-1}n_{l}\}} \nonumber\\
   && - 4(l-1)   {\Phi}^{{l-1}}_{(m)_{k}\{n_l, (n)_{l-1}\}}+ {\Phi}^{{l-2}}_{(m)_{k},\{\{(n)_{l-4}n}{}^n \eta_{n_{l-3}n_{l-2}\}}\eta_{n_{l-1}n_{l
   }\}} \nonumber\\
   &&+ 2(l-2)(2-d){\Phi}^{{l-2}}_{(m)_{k},\{(n)_{l-2}}\eta_{n_{l-1}n_{l}\}}\Big). \label{uncostrlagr}
\end{eqnarray}
In this respect, note that since the Bargmann-Wigner equations is
linear in $\Phi$ the respective actions (\ref{costrlagr}),
(\ref{uncostrlagr}) have a quadratic form which leads, of course,
to a non-diagonal form of the respective propagator and to a wider
set of the degrees of freedom due to Lagrangian multipliers.
However, the EoM for $\Phi(x,\omega)$ and those for
$\lambda^{{l}|(m)_k,(n)_l}_i(x)$ are completely decoupled from
each other, and thereby form independent systems. By selecting the
appropriate initial and boundary conditions for
$\lambda^{{l}|(m)_k,(n)_l}_i(x)$ one can always fix the unwanted
degrees of freedom completely. We will use  this form of free
actions as  auxiliary ones in order to deduce the EoM. \footnote{The actions (\ref{costrlagr}), (\ref{uncostrlagr})
reflect the points of a general procedure known as the
augmentation method for the classical and quantum descriptions of
(non)-Lagrangian systems \cite{Sharapovaug}, which, owing to an
idea also suggested in \cite{Olver}, may be used to obtain the
respective actions entirely in terms of the field
$\Phi(x,\omega)$. To do so, one should consider the actions
$S^A_{C|\Xi}, S^A_{\Xi}$ augmented by the terms quadratic in
$\lambda^{{l}|(m)_k,(n)_l}_i(x) \equiv \lambda^{M_i}_i(x)$:
$$ S^A_{(C|)\Xi}= S_{(C|)\Xi}+
\int d^dx
\frac{1}{2}\lambda^{M_i}_i(x)G_{M_iN_i}\lambda^{N_i}_i(x) \ :
\frac{\delta S^A_{(C|)\Xi}}{\delta \lambda^{N_i}_i(x)} =
\lambda^{M_i}_i(x)G_{M_iN_i} + F(\Phi(x))_{N_i{}i} = 0,$$ for
$F(\Phi)_{N_i{}i} \equiv  F_{N_i{}i}^{M_i}\Phi_{M_i{}i}$ being EoM
(\ref{Eqb2q}) for constrained at $i=1,2,3$ and unconstrained
$i=1,2,3,4$ cases. For a nondegenerate matrix  $\|G_{M_iN_i}\|$,
we have actions being quadratic at the extremals $F(\Phi)_{N_i{}i}
$:
$$\mathcal{S}^A_{(C|)\Xi} = S^A_{(C|)\Xi}\vert_{\big(\lambda_{M_i{}i}
= - F(\Phi)^{N_i}_{i}G^{-1}_{N_iM_i}\big)}
=  -  \frac{1}{2}\int d^dx F(\Phi)^{M_i}_iG^{-1}_{M_iN_i} F(\Phi)^{N_i}_i $$
$$\phantom{\mathcal{S}^A_{(C|)\Xi}}\equiv
-  \frac{1}{2}\int d^dx \sum_{k,l,k',l'\geq 0}
\Phi^{{l}|(m)_k,(n)_l} \mathcal{G}^{l l'}_{(m)_k,(n)_l ;
(m_1)_{k'},(n_1)_{l'} }\Phi^{{l'}|(m_1)_{k'}, (n_1)_{l'}}.$$ In
the case of a gauge presence for $\Phi(x,\omega)$: $\delta
\Phi(x,\omega) =
\mathcal{R}(x,\partial_x;\omega,\partial_\omega)\varsigma(x,\omega)$,
we have the determinant $\det \|G_{M_iN_i}\| =0$, so that there
exist (local) generators $R^{N_i}_\alpha$ of gauge transformations
(dual to the (local) generators
$\mathcal{R}(x,\partial_x;\omega,\partial_\omega)$): $\delta
\lambda^{M_i}_{i} = R^{N_i}_\alpha \sigma^\alpha$ with their own
gauge parameters $\sigma^\alpha$. The quantities  $R^{N_i}_\alpha$
are proper eigenvectors of $\|G_{M_iN_i}\|$, so that we should
find an invertible supermatrix
$\|G_{\widetilde{M}_i\widetilde{N}_i}\|$ in a larger configuration
space $M_{\lambda}=\big\{\lambda^{M_i}_{i},
C_{\lambda}{}^{\alpha_i}_{i}, \bar{C}_{\lambda}{}^{\alpha_i}_{i},
b_{\lambda}{}^{\alpha_i}_{i}\big\}$ having its own ghost
$C_{\lambda}$, antighost $\bar{C}_{\lambda}$ and
Nakanishi--Lautrup $b_{\lambda}$ fields in addition to the ones
for the EoM $F(\Phi)_{N_i{}i}$. The only problem here which may be
controlled by an appropriate choice of the initial conditions is
that the maximal order of the EoM following from the
$\mathcal{S}^A_{(C|)\Xi}$ is greater than the one for EoM implied
by the respective actions ${S}_{(C|)\Xi}$.}.

 The  Poincare group IR relations (\ref{Eqb01}),  (\ref{Eqb23})  take the equivalent form in terms of the operators,
\begin{eqnarray}
  \hspace{-0.55em}&\hspace{-0.55em}&\hspace{-0.55em} \big(l_0,\, l_1,\,m^+_{1},\, m_{11} \big)\Phi(x,\omega)\footnotemark \ = \  0
  ; \label{irrepconts}\\
  \hspace{-0.55em}&\hspace{-0.55em}&\hspace{-0.55em} \big(l_0,\, l_1,\,m^+_{1},\, m_{11} \big)\ =\ \left(\eta^{mn} {\partial_m}{\partial_n},\, -\frac{\partial}{\partial \omega^m}{\partial^m},\,-\omega^m{\partial_m}+\imath\Xi,\, - \eta^{mn}\frac{\partial}{\partial \omega^m}\frac{\partial}{\partial \omega^n} + \nu \right)\hspace{-0.25em} \label{in-opersf}.
 \end{eqnarray}\footnotetext{For quartic Casimir operator $C_4 = (M_{mn}P^n)^2 $ evaluated for massless case on  $\Phi(x,\omega)$ we have after explicit calculation with allowance made for the equations (\ref{Eqb01}), (\ref{Eqb23}) that: $C_4\Phi(x,\omega)= (l_1^+)^2 (m_{11}-\nu)\Phi(x,\omega) $ = $\Xi^2\nu\Phi(x,\omega) $, so that the relations (\ref{Casoperator}) hold.}

It is impossible as it is shown in the appendix~\ref{addalgebra2} to realize a Fock space structure on a set of the generating functions $\Phi(x,\omega)$ given by (\ref{presentaPhi}) and its dual  with finite scalar product with standard Hermitian conjugation property, when  using  the one set of oscillators: $(a_m, a^{+n})\equiv  -\imath({\partial}/{\partial \omega^m},\omega^n)$, $[a^m, a^{+n}]= - \eta^{mn}$ for $a^m|0\rangle$ when acting on a vacuum vector $|0\rangle$. However, we enlarge the set of the operators (\ref{in-opersf}) by its (formal) duals $\big(l^+_1,\,m_{1},\, m^+_{11} \big)$:
\begin{equation}
  \big( l^+_1,\,m_{1},\, m^+_{11}\big)\ = \ \left(- \omega^{m}{\partial_m},\, -\frac{\partial}{\partial \omega^m}{\partial^m}-\imath\Xi,\, - \omega^{m}\omega_{m} +\nu\right), \label{add-opers}
\end{equation}
and by the "number particle" operator,
\begin{equation}\label{g01}
g_0 = \frac{1}{4}\big[m_{11},\,m^+_{11}\big\}:\  g_0= \omega_m \frac{\partial}{\partial \omega_m}  +\frac{d}{2}\equiv \frac{1}{2}\big\{\omega_m ,\,\frac{\partial}{\partial \omega_m}\big\},
\end{equation}
being characterized by its action on $\Phi(x,\omega) = \sum_s \big\{\Phi^{s,0}(x,\omega) + \sum_{l>0}\Phi^{s,l}(x,\omega)\big\}$, for the values of the map degree $\big(\mathrm{deg}_{\omega}$, $\mathrm{deg}_{\omega/\omega^2}\big)\Phi^{s, l}$ = $(s, l)$:
\begin{equation}\label{npvec1}
(g_0-d/2)\Phi(x,\omega) = \sum_{s}\Big\{s\Phi^{s,0}(x,\omega) + \sum_{l>0}(s-l) \Phi^{s,l}(x,\omega)\Big\}.
\end{equation}
so that  the component tensors $\Phi^{s,l}(x,\omega)$ from $\Phi(x,\omega)$, for $s=l$, belong to the kernel of the operator $(g_0-d/2)$.
From the commutators:
\begin{equation}\label{nucentral}
\big[g_0,\,m^+_{11}\big\}=2\big(m^+_{11}-\nu \big),\qquad \big[g_0,\,m_{11}\big\}=-2\big(m_{11}-\nu \big)
\end{equation}
it follows, that the non-zero number $\nu$ should be considered as a non-central charge.

Because of any linear combination of  the constraints $o_I=(o_\alpha,\,o_\alpha^+)$
should be constraint, we have, that
 \begin{eqnarray}\label{charge1}
   && m_1^+ - l_1^+ = \imath\Xi,\ \ \ m_1 - l_1= -\imath\Xi,
 \end{eqnarray}
and  $\Xi$ should be considered as the non-central charge too, because of  not  extending the zero-mode constraint, $l_0$. Note, because of the operators $l_1^+, m_1$
 can not be imposed as the constraints on $\Phi(x,\omega)$ we could  ignore the reducibility above.

  Being combined, the total set of bosonic operators $o_I$ =  $\{o_A,o_a, o_a^+; g_0, \Xi, \nu \}$, for $\{o_A\}= \{{l}_0, l_1, l^+_1,
m_1,\ m^+_1\}$,  $\{o^{(+)}_a\} = \{ m_{11}^{(+)}\}$  can be interpreted within the Hamiltonian analysis of the
dynamical systems as the respective operator-valued $5$ first-class    and
$2$  second-class  constraints subsystems among $\{o_I\}$ for a
topological gauge system, with  additional operators $g_0, \Xi, \nu $, which are not the constraints due to (\ref{g01}),  and  because of  from the
commutation relations for the operators $o_I$ (forming a Lie algebra)
\begin{equation}\label{geninalg}
    [o_I,\ o_J\}= f^K_{IJ}o_K, \  f^K_{IJ}= - f^K_{JI},
\end{equation}
the following subsets can be extracted:
\begin{eqnarray}
[o_a,\; o_b^+\} = f^c_{ab} o_c +\Delta_{ab}(g_0),\ \ [o_A,\;o_B\}
= f^C_{AB}o_C, \  [o_a,\; o_B\} = f^C_{aB}o_C .
\label{inconstraintsd}
\end{eqnarray}
Here, the constants $f^c_{ab}, f^C_{AB}, f^C_{aB}$ are
determined by the Multiplication Table~\ref{table in} and possess the
 antisymmetry property with respect to permutations of
lower indices, whereas the quantities $\Delta_{ab}(g_0)$ form a
non-degenerate $(2\times 2)$ matrix:
$\|\Delta\|$=$\mathrm{antidiag}(-4g_0, 4g_0)$, in the space $\mathcal{V}$ of vectors $\{\Phi(x,\omega)\}$ (\ref{presentaPhi}) on the surface
$\Sigma \subset \mathcal{V}$: $\|\Delta\|_{|\Sigma} \ne 0 $,
which is determined by the equations: $\big(o_A,\,o_a\big)\Phi(x,\omega) $ = $(0,0)$.
\hspace{-1ex}{\begin{table}[t] {{\small
\begin{center}
\begin{tabular}{||c||c|c|c|c|c|c|c||c||}\hline\hline
$\hspace{-0.2em}[\; \downarrow, \rightarrow
\}\hspace{-0.5em}$\hspace{-0.7em}&
 $l_0$ & $m_1$ &
$m_1^+$ & $l_1$ &
$l_1^+$& $m_{11}$ &$m_{11}^+$ &
$g_0$ \\ \hline
\hline $l_0$
    & $0$ & $0$
& $0$   &
    $0$\hspace{-0.5em} & \hspace{-0.3em}
    $0$\hspace{-0.3em}
    & $0$ & $0$ & $0$ \\
\hline $m_1$
   & \hspace{-0.5em}$0$ \hspace{-0.5em} &
   \hspace{-0.5em}$0$ \hspace{-0.9em}  & \hspace{-0.3em}$l_0$ \hspace{-0.3em} & $0$&
   \hspace{-0.3em}
   $l_0$\hspace{-0.3em}
    & $0$ & \hspace{-0.5em}$-{2}l_{1}^+$
    \hspace{-0.9em}&$l_1$  \\
\hline $m^{+}_1$ & \hspace{-0.5em}$0$\hspace{-0.7em} & \hspace{-0.7em}
   $-l_0$ \hspace{-1.0em} &
   $0$&\hspace{-0.3em}
      \hspace{-0.3em}
   $-l_0$\hspace{-0.3em}
    \hspace{-0.3em}
   &\hspace{-0.5em} $0$\hspace{-0.5em}
    &\hspace{-0.7em} $ 2l_{1}
    $\hspace{-0.7em} & $0$ &$-l^{+}_1$  \\
\hline $l_1$
   & \hspace{-0.5em}$0$ \hspace{-0.5em} &
   \hspace{-0.5em}$
   0$ \hspace{-0.9em}  & \hspace{-0.3em}$l_0$ \hspace{-0.3em} & $0$&
   \hspace{-0.3em}
   $l_0$\hspace{-0.3em}
    & $0$ & \hspace{-0.5em}$-2l_1^+$
    \hspace{-0.9em}&$l_1$  \\
\hline $l^{+}_1$ & \hspace{-0.5em}$0$\hspace{-0.7em} & \hspace{-0.7em}
   $-l_0$ \hspace{-1.0em} &
   $0$&\hspace{-0.3em}
      \hspace{-0.3em}
   $-l_0$\hspace{-0.3em}
    \hspace{-0.3em}
   &\hspace{-0.5em} $0$\hspace{-0.5em}
    &\hspace{-0.7em} $ 2l_1
    $\hspace{-0.7em} & $0$ &$-l^{+}_1$  \\
\hline $m_{11}$
    & \hspace{-0.3em}$0$
    \hspace{-0.5em} &\hspace{-0.5em} $\hspace{-0.4em}
    0\hspace{-0.3em}$\hspace{-0.3em}
   & $-2l_1$&\hspace{-0.3em}
    $0$\hspace{-0.5em} & \hspace{-0.3em}
    $ \hspace{-0.7em}-2l_1
    \hspace{-0.5em}$\hspace{-0.3em}
    & $0$ & \hspace{-0.7em}$\hspace{-0.3em}
4g_0\hspace{-0.3em}$\hspace{-0.7em}& $\hspace{-0.7em}  2\big(m_{11}-\nu \big)\hspace{-0.7em}$\hspace{-0.7em} \\
\hline $m_{11}^+$
    & $0$ & $ 2l_1^+$
   & $0$&\hspace{-0.3em}
    $\hspace{-0.2em} 2l_1^+$\hspace{-0.5em} & \hspace{-0.3em}
    $0$\hspace{-0.3em}
    & $-4g_0$ & $0$ &$\hspace{-0.5em}  -2\big(m^+_{11}-\nu\big)\hspace{-0.3em}$\hspace{-0.2em} \\
\hline\hline $g_0$
    & $0$ & $-l_1$
   &$l_1^+$& \hspace{-0.3em}
    $\hspace{-0.2em}-l_1$\hspace{-0.5em} & \hspace{-0.3em}
    $l_1^+$\hspace{-0.3em}
    &$  -2\big(m_{11}- \nu \big)$ & $ 2\big(m^+_{11}-\nu \big)$&$0$ \\
   \hline\hline
\end{tabular}
\end{center}}} \vspace{-2ex}\caption{Higher continuous spin symmetry  algebra  $\mathcal{A}(\Xi;
\mathbb{R}^{1,d-1})$.\label{table in} }\end{table}

We call the algebra of the  operators $o_I$ as the \emph{higher continuous spin symmetry algebra in Minkowski space} with notation $\mathcal{A}(\Xi;
\mathbb{R}^{1,d-1})$ (shortly HCS symmetry algebra)\footnote{one should not identify the
term "\emph{higher continuous spin symmetry algebra}" using here for free HS
formulation starting from the paper \cite{0505092} with the
algebraic structure known as "\emph{higher-spin algebra}" (see,
for instance Ref.\cite{Vasiliev_inter}) arising to describe the (half-)integer HS
interactions.}. Note, first, that we omitted in the Table~\ref{table in}  the center of $\mathcal{A}(\Xi;
\mathbb{R}^{1,d-1})$ consisting from $\Xi, \nu$. Second, the linear dependence of    $o_k=(m_1 , l_1 , \Xi)$ and $o^+_k=(m_1^+ , l_1^+ , \Xi)$ for  $ k=1,2,3$ means
the existence of independent bosonic proper zero eigen-vectors $Z^k; Z^{+k}$:
\begin{equation}\label{lindep}
  o_kZ^k = 0,\ \  o^+_kZ^{+k} = 0, \  \mathrm{for} \  Z^k= \beta(1,\,-1,\,\imath),\ Z^{+k}= \beta(1,\,-1,\,-\imath),\ \ \forall \beta \in \mathbb{R}\setminus \{0\},
\end{equation}
whose set is linear independent. Third, because of the elements $\partial_m^\omega, \omega_{m}$ transfer the fields $\Phi^{s,0}(x,\omega)$, $s\geq 0$ (see (\ref{npvec1})) into the fields
$\theta_{s,0}\Phi^{s-1,0}_m (x,\omega)$, $\Phi^{s+1,0}_m$, the elements $o_I$ from $\mathcal{A}(\Xi;
\mathbb{R}^{1,d-1})$ have the same property and  for their  actions on the fields $\Phi^{s,l}$, $l> 0$ both operators $\partial_m^\omega$, $\omega_{m}$ obey by the similar property:
\begin{eqnarray}\label{aa+vec0}
&& \partial_m^\omega \Phi^{s,l} =  s \Phi^{s-1,l}_m-l\Phi^{s,l+1}_m, \ \ \ \omega_{m} \Phi^{s,l} =  \Phi^{s+1,l}_m ,   \label{aa+vec1}
\end{eqnarray}
Thus, all operators $o_I$ when acting on $\Phi(x,\omega)$, preserve the  grading in $\mathcal{V}$, induced by the decomposition of $\Phi(x,\omega)$ by $g_0$: (\ref{npvec1}).

The algebra $\mathcal{A}(\Xi;
\mathbb{R}^{1,d-1})$ contains the subalgebra $\mathcal{A}_{BW}(\Xi;
\mathbb{R}^{1,d-1})$ = $\{l_0, l_1, m_1^+, m_{11}\}$ being closed with respect to the $[\ ,\ ]$-multiplication. This algebra does not closed with respect to the formal  Hermitian conjugation in $\mathcal{V}$, but maybe effectively used to formulate (un)constrained BRST-BFV description of dynamics for the CS field $\Phi(x,\omega)$.

\section{Constrained  BRST-BFV descriptions}
\label{BRSTBFV} %%%%%%%%%%%%%%%%%%%%%%%%%%%%%%%%%%%%%%%%%%%%%%%%
\setcounter{equation}{0}

To construct constrained formulations we extend the results of general research \cite{Reshetnyak_con}, realized there for HS fields with generalized integer and half-integer spins on $\mathbb{R}^{1,d-1}$, for CS case, however on the level of  the equations of motion.

\subsection{Constrained BRST operators, BRST-extended constraints}\label{BRSTBFV1}
There are two ways to develop constrained BRST-BFV descriptions for CS field, based respectively on the HS symmetry algebras $\mathcal{A}(\Xi;
\mathbb{R}^{1,d-1})$ and $\mathcal{A}_{BW}(\Xi;
\mathbb{R}^{1,d-1})$.
 \subsubsection{Case of $\mathcal{A}(\Xi;
\mathbb{R}^{1,d-1})$}\label{BRSTBFV11}
 We consider  the set of the first-class constraints $\{o_A\}$ as the dynamical one  with the element  $\Xi$, and the off-shell  algebraic constraint (one from the second-class constraints) $m_{11}$.  Due to the fact that the operator $g_0$  does not now relate to  CS value $\Xi$, as it was for the case of discrete spin \cite{Reshetnyak_con},  we introduce   \emph{generating equation} for superalgebra of the Grassmann-odd constrained BRST operator, $Q_C$,  and extended in  the  space $\mathcal{V}_C$: $\mathcal{V}_C =\mathcal{V}\otimes \mathcal{V}^{o_A}_{gh}$,    off-shell constraint $\widehat{M}_{11}$  in the form:
\begin{align}\label{eqQctot}
  & [Q_C,\, Q_C\}    = 0, &   [Q_C,\,  \widehat{M}_{11} \}= 0,  \ \, \mathrm{for } \  gh_H(Q_C,\, \widehat{M}_{11})=\epsilon\big(Q_C,\,  \widehat{M}_{11}\big) = (1,0),
  \end{align}
 with boundary conditions for $Q_C, \widehat{M}_{11}$:
 \begin{equation}\label{boudcond}\hspace{-0.7em}
\left( \frac{\overrightarrow{\delta}}{\delta  \mathcal{C}^A}, \frac{\overrightarrow{\delta}}{\delta  \eta_\Xi}, \frac{\overrightarrow{\delta}}{\delta  \eta^+_Z}, \frac{\overrightarrow{\delta}}{\delta  \eta_Z}\right)Q_C\big|_{\mathcal{C}=0} =\left(o_A ,\Xi  , \sum_kZ^k\mathcal{P}_k,\, \sum_kZ^{+k}\mathcal{P}^+_k\right), \ \  \widehat{M}_{11}\big|_{\mathcal{C}=\mathcal{P}=0} = {m}_{11},
 \end{equation}
 when vanishing ghost coordinates,  momenta  $\big(\mathcal{C}^A, \mathcal{P}_A;\eta_\Xi,  \mathcal{P}_\Xi\big)$ for constraints $(o_A, \Xi)$ and ones for  eigen-vectors $Z^k, Z^{+k}$: $\big(\eta^{(+)}_Z, \mathcal{P}^{(+)}_Z\big)$,  being by the generating  elements for the space $\mathcal{V}^{o_A}_{gh}$.

 The solution for the system (\ref{eqQctot}) is sought  in powers series in ghost operators  with choice of some $(\mathcal{C}\mathcal{P})$-ordering for $[{C}^A, {\mathcal{P}}_B\}=\delta^A_B$, which satisfy
 to the Grassmann, ghost number distributions and respective non-vanishing (anti)commutator relations:
\begin{eqnarray}
&& \qquad \begin{array}{|c|cccccc|} \hline
                   &   \mathcal{C}^{A} & {\mathcal{P}}_{A
} &  \eta_\Xi & \mathcal{P}_\Xi  & \eta^{(+)}_Z  & \mathcal{P}^{(+)}_Z  \\
                    \hline
                   \epsilon  &1  & 1 & 1 & 1 & 0 & 0 \\
                   gh_H  & 1 & -1 & 1 & -1 & 2 & -2\\
                  \hline \end{array}  \label{grassgh}\\
  &&  \{\eta_1,\, \mathcal{P}_1^+\}=   \{\eta_1^+,\, \mathcal{P}_1\}= 1, \qquad  \{\eta^m_1,\, \mathcal{P}_1^{m+}\}=   \{\eta_1^{m+},\, \mathcal{P}^m_1\}= 1,  \label{commrelgh}  \\
  && \{\eta_0,\, \mathcal{P}_0\}=\{\eta_\Xi,\, \mathcal{P}_\Xi\}=  \imath , \qquad   [\eta_Z,\, \mathcal{P}^{+}_Z] =   [\mathcal{P}_Z,\,\eta^{+}_Z] =1.  \nonumber
\end{eqnarray}
 In (\ref{grassgh}), (\ref{commrelgh}) the formal Hermitian conjugation for  the zero-mode ghosts is determined by the rule: $\big(\eta_0,\, \mathcal{P}_0,\,\eta_\Xi,\, \mathcal{P}_\Xi\big)^+$ = $\big(\eta_0,\, -\mathcal{P}_0,\,\eta_\Xi,\, -\mathcal{P}_\Xi\big)$ for formal hermitian operators $l_0, \Xi$ from the center of  $\mathcal{A}(\Xi;
\mathbb{R}^{1,d-1})$ with the rest ghost operators, which form the Wick pairs.

 The BRST operator, $Q'$, for the Lie algebra $\mathcal{A}(\Xi;
\mathbb{R}^{1,d-1})$ of the  constraints $o_I$ (\ref{geninalg}), whose linear dependence means the presence of  proper zero eigen-vectors $Z^I_{I_1}$: $o_IZ^I_{I_1} =0$, $\epsilon(Z^I_{I_1})=0$, such that they supercommute with $o_I$: $\big[o_I,\,Z^J_{I_1}\big\}=0$, should be found from  the equation: $[Q',\,Q'\}$ = $2(Q')^2=0$,  and has the form:
\begin{equation}\label{generalQ'}
    Q'  = {\mathcal{C}}^I{o}_I + \frac{1}{2}
    {\mathcal{C}}^I {\mathcal{C}}^Jf^K_{JI}{\mathcal{P}}_K  + {\mathcal{C}}^{I_1} Z^I_{I_1}{\mathcal{P}}_I .
\end{equation}
Here the set of  fermionic ghost operators $({\mathcal{C}}^I, {\mathcal{P}}_J)$  for the bosonic constraints ${o}_I$  and bosonic ghost coordinates and momenta $({\mathcal{C}}^{I_1}, {\mathcal{P}}_{J_1})$  for $Z^J_{I_1}$  corresponds to the minimal sector of BRST-BFV method \cite{BFV1}  for the topological (i.e. without Hamiltonian) first-stage reducible dyna\-mi\-cal system with the first-class constraints.

  For the case of the constraints $o_A$  (\ref{inconstraintsd}) whose algebra is subject to the  table~\ref{table in} and constant proper zero eigen-vectors $Z^I_{I_1} =(Z^k, Z^{+k})$,  the solution  for $Q_C$ in (\ref{eqQctot}) follows from the general anzatz  (\ref{generalQ'}) in the form:
\begin{equation}\label{generalQC}
    \widetilde{Q}_C  = {\mathcal{C}}^A\Big({o}_A + \frac{1}{2}
     {\mathcal{C}}^Bf^D_{BA}{\mathcal{P}}_D\Big)+ \eta_\Xi \Xi  + \eta_Z \sum_k Z^{+k} {\mathcal{P}}^+_k+ \eta^+_Z \sum_k Z^k {\mathcal{P}}_k ,
     \end{equation}
  for ${\mathcal{P}}_k = \big({\mathcal{P}}^m_1,\,{\mathcal{P}}_1,\,{\mathcal{P}}_\Xi\big)$ and ${\mathcal{P}}^+_k = \big({\mathcal{P}}^{m+}_1,\,{\mathcal{P}}^+_1,\,-{\mathcal{P}}_\Xi\big)$.
  Explicitly, we have
\begin{eqnarray}
   &&  \widetilde{Q}_C  = \eta_0l_0+\eta_1^+l_1+l^{+}_1\eta_1+m_1\eta_1^{m+}+\eta^m_1m^{+}_1
  +{\imath}\bigl(\eta_1^+\eta_1+\eta_1^{m+}\eta^m_1+\eta_1^+\eta^m_1+\eta^{m+}_1\eta_1\bigr){\cal{}P}_0  \nonumber \\
   && \phantom{Q_C  =}+\eta_\Xi \Xi + \eta_Z \big({\mathcal{P}}^{m+}_1-{\mathcal{P}}^{+}_1 - {\mathcal{P}}_\Xi\big) + \eta^+_Z \big({\mathcal{P}}^m_1-{\mathcal{P}}_1 + {\mathcal{P}}_\Xi\big) . \label{generalQCexp}
\end{eqnarray}
 Because of, first, the physical space (as the set of states being equivalent to one described by the equations (\ref{Eqb01}) and the first from (\ref{Eqb23})), in fact, should be extracted by imposing of  linear in ghost $\mathcal{C}^A, \Xi$  terms from $\widetilde{Q}_C$ (see, e.g.  Statement 2 in \cite{Reshetnyak_con}), second,  the operator  $\Xi$ can not be imposed as the constraint on the vectors from $\mathcal{V}$,  we instead consider another variant of inclusion of the term, $\eta_\Xi \Xi$, when calculating of zero ghost number cohomology of $\widetilde{Q}_C$ in $\mathcal{V}_C$.

 To do so, we define the representation in $\mathcal{V}_C$:
 \begin{eqnarray} \label{represvac}
\Big(\eta_1, \eta^{m+}_{1},  \mathcal{P}_1,
\mathcal{P}^{m+}_{1},  \mathcal{P}_{0}, \eta_{\Xi},
\eta_Z, \mathcal{P}_Z\Big) = \Big(\frac{\partial}{\partial \mathcal{P}_1^+}, \frac{\partial}{\partial \mathcal{P}_1^m},  \frac{\partial}{\partial \eta_1^+},
\frac{\partial}{\partial \eta_1^m},  \imath \frac{\partial}{\partial \eta_0}, \imath \frac{\partial}{\partial \mathcal{P}_{\Xi}}, \frac{\partial}{\partial \mathcal{P}_Z^+}
, \frac{\partial}{\partial \eta_Z^+}\Big) ,
\end{eqnarray}
such that the requirement  $(\eta_\Xi) \Xi \chi_C = 0$, for arbitrary physical  vector $ {\chi}_C $: $ \chi_C  \in \mathcal{V}_C$, $gh_{H}( \chi_C)=0$  to be  not depending on
 $\mathcal{P}_\Xi$  means that we, in fact, extract only linear independent constraints, when acting on  arbitrary $ \widetilde{\chi}_C =  \widetilde{\chi}_C(x,\omega, \mathcal{C}, \mathcal{P})$:
\begin{eqnarray}
\widetilde{\chi}_C  &=& \sum_n \eta_0
^{n_{f 0}} ( \eta_1^+ )^{n_{f 1}}( \eta_1^{m} )^{n_{f m}} (
\mathcal{P}_1^+ )^{n_{p 1}}(\mathcal{P}_1^{m} )^{n_{p m}}(\eta_Z^+)^{n_{f z}} (\mathcal{P}^+_Z)^{n_{p z}}(\mathcal{P}_\Xi)^{n_{p \Xi}} \label{chifconst1}
\\
&&{}\times  \Phi(x,\omega)_{ n_{f 0};   n_{f 1}, n_{f m},n_{p 1}, n_{p m},n_{f z}, n_{p z}, n_{p \Xi}} \,, \nonumber
\end{eqnarray}
where ${n_{f z}}, {n_{p z}}$ are running from $0$ to $\infty$, whereas the  rest $n's$  from $0$ to $1$. The ghost-independent vectors $\Phi(\omega)_{ n_{f 0};   ...}$ have the dependence in $\omega$ according to (\ref{presentaPhi}).  Thus, we  resolved the  linear dependence problem
for the sets:
$\{l_1,m_1,\Xi\}$ and $\{l_1^+,m_1^+,\Xi\}$ on the space of $\mathcal{P}_\Xi$-independent vectors (\ref{chifconst1})  and should  remove dependence on proper zero  eigen-vectors and respective ghosts $\eta^{(+)}_Z, \mathcal{P}^{(+)}_Z$ in $\widetilde{Q}_C$ and $\widetilde{\chi}_C $ turning to
\begin{equation}\label{Qcchic}
  \big({Q}_C, \, {\chi}_C \big) \ = \ \big(\widetilde{Q}_C,\, \widetilde{\chi}_C \big)\big|_{(\eta^{(+)}_Z=\mathcal{P}^{(+)}_Z=\mathcal{P}_\Xi=0)}.
\end{equation}
It provides that from ${Q}_C{\chi}_C  = 0$ it follows,  due to the choice of (\ref{represvac}), the equations in power of ghosts: $\big(l_0+O(\mathcal{C}^A), l_1+O(\mathcal{C}^A), m_1^{+}+O(\mathcal{C}^A) \big){\chi}_C  =0$, being compatible with (\ref{irrepconts}). It is in the agreement with the observation  that the operators $l_1^+, m_1$
 can not be imposed as the constraints on $\Phi(x,\omega)$, so that among the operators $l_1^+, m^+_1$ ($l_1, m_1$) the only $m^+_1$ ($l_1$) is the constraint.
The solution for the second equation in (\ref{eqQctot})  can  be found in the form
\begin{equation}
\widehat{M}_{11}\  =\ m_{11} + 2\eta_{1} \mathcal{P}_{1}+ 2\eta_{1}^m \mathcal{P}_{1}.
  \label{constralgB}
\end{equation}
The respective BRST-extended number particle operator $\widehat{\sigma}_C(g)$, $\epsilon\big(\widehat{\sigma}{}_C(g)\big)=0 $  (known for the discrete spin as the spin operator \cite{Reshetnyak_con}),
which should satisfy to the additional equations
\begin{equation}\label{gspin}
   [Q_C,\, \widehat{\sigma}_C(g)\}= 0, \qquad  [\widehat{M}_{11},\, \widehat{\sigma}_C(g)\}= 2 \big(\widehat{M}_{11}-\nu \big),
\end{equation}
is uniquely determined in the form\footnote{The operator $\widehat{\sigma}_C(g)$ (\ref{constralgsp}) is differed from the standard spin operator by the latter two terms due to $[g_0,m_1^+] = l_1^+ \ne m_1^+$.}
\begin{equation}
\widehat{\sigma}_C(g)\  =\ g_0 +\eta_{1}^+ \mathcal{P}_{1}-\eta_{1} \mathcal{P}^+_{1}+\eta_{1}^{m+} \mathcal{P}_{1}-\eta^m_{1} \mathcal{P}^{+}_{1} .  \label{constralgsp}
\end{equation}

\subsubsection{Case of $\mathcal{A}_{BW}(\Xi;
\mathbb{R}^{1,d-1})$}\label{BRSTBFV123}
  The nilpotent  unconstrained $Q'_{BW}$ and constrained $Q_{BW|C}$ BRST operators  for the algebra $\mathcal{A}_{BW}(\Xi;
\mathbb{R}^{1,d-1})$, which contain only linear independent  first-class constraints  subsystems by the respective numbers 4 and 3 looks as  (for the Grassman-odd ghost coordinate and momenta  $\eta_{11}^+, {\cal{}P}_{11}$: $\{\eta_{11}^+,, {\cal{}P}_{11}\}=1$ corresponding to $m_{11}$)
\begin{eqnarray}
   && Q'_{BW} =  Q_{BW|C}+ \eta_{11}^+\widehat{M}^{BW}_{11}  , \quad     Q_{BW|C} =  \eta_0l_0+\eta_1^+l_1+\eta^m_1m^{+}_1
  +{\imath}\eta_1^+\eta^m_1{\cal{}P}_0 ,  \label{generalQBW} \\
   && Q_{BW|C} = \widetilde{Q}_C\vert_{\big(\eta^{m+}_1=\eta_1=  {\cal{}P}^+_1= {\cal{}P}^m_1= \eta_\Xi = \mathcal{P}_\Xi  = \eta^{(+)}_Z  = \mathcal{P}^{(+)}_Z =0\big)} . \label{generalQBWexp}
\end{eqnarray}
The BRST-extended constraint $\widehat{M}^{BW}_{11}$  is determined  from the generating equation, $[Q_{BW|C},$ $\, \widehat{M}^{BW}_{11}\} =0$,   by the  expression:
\begin{equation}
\widehat{M}^{BW}_{11}\  =\ m_{11} + 2\eta_{1}^m \mathcal{P}_{1}, \qquad  \widehat{M}^{BW}_{11}\  =\ \widehat{M}_{11}- 2\eta_{1} \mathcal{P}_{1}.
  \label{constralgBW}
\end{equation}
The  extended vector ${\chi}_{BW|C}$ from  the  space $\mathcal{V}_{BW|C}$:  $\mathcal{V}_{BW|C} = \mathcal{V}\otimes \mathcal{V}^{o_{BW}}_{gh}$ for the primary constraints $o_{BW}=\{ {l}_0, l_1, m^+_1\}$ has the representation
\begin{eqnarray}
{\chi}_{BW|C}  &=& \sum_n \eta_0
^{n_{f 0}} ( \eta_1^+ )^{n_{f 1}}( \eta_1^{m} )^{n_{f m}} (
\mathcal{P}_1^+ )^{n_{p 1}}(\mathcal{P}_1^{m} )^{n_{p m}}
 \Phi(x,\omega)_{ n_{f 0};   n_{f 1}, n_{f m},n_{p 1}, n_{p m}}\,, \label{chifconstBW}
\end{eqnarray}
that implies ${\chi}_{BW|C}=\widetilde{\chi}_{C}\vert_{\big( \eta_\Xi = \mathcal{P}_\Xi  = \eta^{+}_Z  = \mathcal{P}^{+}_Z =0\big)}$.

\subsection{Constrained  dynamics}\label{BRSTBFV2}

To derive BRST constrained  dynamics we should solve spectral problem for  the vectors $\chi^l_C \in  \mathcal{V}^l_C$ due to existence of  $\mathbb{Z}$-grading in  $\mathcal{V}_{C}$: $\mathcal{V}_{C} = \bigoplus_{k}\mathcal{V}^k_{C}$   for $gh_H(\chi^k_C)=-k$, $k \in \mathbb{N}_0$ for the case of $\mathcal{A}(\Xi; \mathbb{R}^{1,d-1})$ algebra:
\begin{align}
\label{Qchi} & Q_C\chi^0_C=0, && \widehat{M}_{11}\chi^0_C=0 ,&& \left(\epsilon, {gh}_H\right)(\chi^0_C)=(0,0),
\\
& \delta \chi^0_C=Q_C\chi^1_C, && \widehat{M}_{11}\chi^1_C=0 ,&& \left(\epsilon,
{gh}_H\right)(\chi^1_C)=(1,-1), \label{Qchi1}
\\
& \delta \chi^1_C =Q_C \chi^2_C , && \widehat{M}_{11}\chi^2_C=0 , && \left(\epsilon,
{gh}_H\right)(\chi^2_C)=(0,-2) . \label{Qchi2}
\end{align}
The closedness of the superalgebra of
$Q_C,  \widehat{M}_{11}$ guarantees ,the joint set of solution for the system  (\ref{Qchi})-- (\ref{Qchi2}).
Thus, the physical state  $\chi_C \equiv  \chi^0_C$  for the vanishing of all  ghost variables $\eta_0,
\eta^+_1,  \eta^{m}_1,  \mathcal{P}^+_1, \mathcal{P}^{m}_1$,
 contains only the physical  vector $\Phi
= \Phi(x,\omega)_{ 0_{f 0};   0_{f 1},0_{f m},0_{p 1},}$ ${}_{0_{p m},0_{f z}, 0_{p z}}$ (\ref{presentaPhi}), so that
\begin{eqnarray}\label{decomptot}
\chi^0_C&=&\Phi+  \Phi_{aux} ,\qquad \Phi_{aux}\big|_{\textstyle  (\eta_0, \eta^+_1, \eta^{m}_1,  \mathcal{P}^+_1, \mathcal{P}^{m}_1)=0}=0.
\end{eqnarray}
 The vectors $\chi^k_C$ inherit by the construction the decomposition (\ref{presentaPhi}): $\chi^k_C = \chi^{(0)k}_C + \sum_{l\geq 1}\chi^{(l)k}_C$, in the sum of the vectors with only positive and mixed  degrees in $\omega_m$.  The equations of motion: $Q_C\chi_C=0$ ($\chi_C \equiv \chi^0_C$) in (\ref{Qchi})  obtained at independent degrees in powers of the ghost oscillators are yet non-Lagrangian (due to absence of finite scalar product definition in $\mathcal{V}_{C}$)   to be invariant with respect to reducible gauge transformations with off-shell constraints
 \begin{eqnarray}
 &&  Q_{C} \chi^0_{C}=0, \ \  \delta\chi^k_{C}
= \theta_{2,k}Q_{C}\chi^{k+1}_{C} , \ \ \widehat{M}_{11}\chi^k_C=0,  \ \ k=0,1,2. \label{LcontinB}
\end{eqnarray}
The vanishing of all  $\chi^{l}_{C} $, for $l \geq 3$ is  due to the   possible maximal ghost momenta degree: $\mathcal{P}^+_1\mathcal{P}^m_1$  to be realized  for only  $\chi^{2}_{C} $:
\begin{eqnarray}\label{chifconst2} {\chi}^2_C  &=&
\mathcal{P}_1^+ \mathcal{P}_1^{m}\varpi (x,\omega)  \ \  \mathrm{for} \ \  \varpi (x,\omega) \equiv \Phi(x,\omega)_{ 0_{f 0};   0_{f 1}, 0_{f m},1_{p 1}, 1_{p m}} .
\end{eqnarray}
 Thus,  we constructed the \emph{constrained gauge-invariant non-Lagrangian formulation of the first-stage reducibility}  for the massless scalar bosonic  field
with CS $\Xi$ for $\nu=1$.

Having the decomposition in ghost oscillators for the field and first level gauge parameters
$\chi^{l}_{C} $, $l=0,1$  with $\mathbb{R}$-valued coefficient functions (as well as for $ \varpi (x,\omega)$):
\begin{eqnarray}\label{chifconst0}\hspace{-0.5em} {\chi}^0_C  &\hspace{-0.5em}=& \hspace{-0.5em}\Phi(\omega) +\eta_1^+\Big(\mathcal{P}_1^+ \chi_1 (\omega)+\mathcal{P}_1^{m}\chi_1^m (\omega) \Big)+  \eta_1^{m}\Big(\mathcal{P}_1^+ \chi_{2} (\omega) +\mathcal{P}_1^{m}\chi_{2}^m (\omega) \Big)\\
 &\hspace{-0.5em}& \hspace{-0.5em}
 + \eta_0\Big(\mathcal{P}_1^+ \chi_0 (\omega) +\mathcal{P}_1^{m} \chi_0^m (\omega) + \eta_1^+\mathcal{P}_1^+ \mathcal{P}_1^{m}\chi_{01} (\omega)+\eta_1^{m}\mathcal{P}_1^+ \mathcal{P}_1^{m}\chi_{01}^m (\omega)\Big) \nonumber  \\
&& + \eta^+_1 \eta_1^{m}\mathcal{P}_1^+ \mathcal{P}_1^{m}\chi_{11}^m (\omega)\,, \nonumber \\
\label{chifconst11} \hspace{-0.5em} {\chi}^1_C   &\hspace{-0.5em}=& \hspace{-0.5em} \mathcal{P}_1^+ \varsigma (\omega)+\mathcal{P}_1^{m}\varsigma^m (\omega)+ \mathcal{P}_1^+ \mathcal{P}_1^{m}\Big(\eta_1^+\varsigma_{01} (\omega)+\eta_1^m\varsigma_{11} (\omega)+\eta_0\varsigma_0 (\omega) \Big),
\end{eqnarray}
from the BRST-extended constraints (\ref{LcontinB}) and structure of operator $\widehat{M}_{11}$ (\ref{constralgB})  it follows the constraints in powers of independent ghost monomials for the gauge parameters and field vectors:
 \begin{align}\label{constrcontinB2}
 & l=2:\  m_{11}\varpi = 0,  \hspace{-0.5em}&& \hspace{-0.5em} &&  \\
  &l=1:\    m_{11}\big( \varsigma,  \varsigma_0 , \varsigma_{01}\big) = 0,  \hspace{-0.5em}&& m_{11}\varsigma^m  + 2 \varsigma_{01}  =0,  \hspace{-0.5em}&& m_{11}\varsigma_{11}  + 2 \varsigma_{01} =0, \label{constrcontinB1}\\
  &l=0: \    m_{11}\big( \chi_0,  \chi_{01}\big) = 0,  \hspace{-0.5em} &&  m_{11}\big( \chi_{1},  \chi^m_{1},   \chi_{11}^m\big) = 0,  \hspace{-0.5em} &&
  m_{11}\chi_{0}^m  + 2 \chi_{01} =0,  \label{constrcontinB0}\\
  &\phantom{l=0: \ }    m_{11}\chi_{01}^m  + 2 \chi_{01} =0, \hspace{-0.5em}&&   m_{11}\chi_{2}^m  -\hspace{-0.15em}2 \chi_{11}^m + \hspace{-0.15em}2 \chi_{1}^m \hspace{-0.15em}=\hspace{-0.15em}0,  \hspace{-0.5em}&&
  m_{11}\chi_{2}  + 2 \chi_{1} =0,  \label{constrcontinB01}\\
  &\phantom{l=0: \ }    m_{11}\Phi + 2 \chi_{1}=0.  \hspace{-0.5em}&&   \hspace{-0.5em} &&
   \label{constrcontinB02}
 \end{align}
  The $\eta_0$-independent equivalent representation  for the  equations of motion and gauge transformations (\ref{LcontinB}) in the supermatrix form look:
 \begin{eqnarray}\hspace{-0.5em}&\hspace{-0.5em}&\hspace{-0.5em}   \left(\begin{array}{cc}
                                                                                          l_0 & - \Delta Q_C \\
                                                                                          -\Delta Q_C  &  (\eta_1^++\eta_1^{m+})(\eta_1+\eta_1^{m})
                                                                                        \end{array}\right)\left(\begin{array}{c}
                                                                                          S^{0}_C (\omega) \\
                                                                                          B^{0}_C (\omega)
                                                                                        \end{array}\right)=\left(\begin{array}{c}
                                                                                          0 \\
                                                                                          0
                                                                                        \end{array}\right)
 ,\label{Sclfvectf}\\
 \hspace{-0.8em}&\hspace{-0.8em}&\hspace{-0.8em} \label{gtranind} \delta\left(\begin{array}{c}
                                                                                          S^{l}_C(\omega) \\
                                                                                          B^{l}_C (\omega)
                                                                                        \end{array}\right)\hspace{-0.2em} = \hspace{-0.2em}
\left(\begin{array}{cc}\Delta Q_C                                            & -(\eta_1^++\eta_1^{m+})(\eta_1+\eta_1^{m})   \\
                                                                                          l_0  &  - \Delta Q_C
                                                                                        \end{array}\right)\hspace{-0.2em}\left(\begin{array}{c}
                                                                                          S^{l+1}_C (\omega) \\
                                                                                          B^{l+1}_C (\omega)
                                                                                        \end{array}\right)\hspace{-0.2em} \theta_{2,l},\ l= 0,1,2, \\
                                                                                        &&  \Delta Q_C \ = \  \eta_1^+l_1+\eta^m_1m^{+}_1+l_1^+\eta_1+m_1\eta^{m+}_1 \label{deltaqc}
\end{eqnarray}
  for $\chi^l_C(\omega) = S^l_C(\omega)+\eta_0 B^l_C(\omega)$,  $\chi^{-1}_C(\omega) = B^{2}_C (\omega)\equiv 0$.
 The respective gauge transformations in the ghost independent  form follow from (\ref{gtranind})\ for the zeroth level  gauge parameters:
 \begin{equation}\label{gtrgh11}
     \delta \Big( \varsigma ,\, \varsigma^m  ,\, \varsigma_{01} ,\,  \varsigma_{11}, \,  \varsigma_{0}  \Big)(\omega) \ =\ \Big(-m_1, \,l_1^+,\, l_1,\, m_1^+,\, l_0\Big)  \varpi(\omega),
 \end{equation}
 and for the field vectors (omitting explicit $\omega$-dependence)
 \begin{align}\label{gtrgh01}
     & \delta  \Phi  =  l_1^+  \varsigma    +m_1  \varsigma^m -   \varsigma_0 , \hspace{-0.5em}&&  \delta  \chi_{1}  = l_1  \varsigma +m_1  \varsigma_{01} , \hspace{-0.5em}&& \delta  \chi_{1}^m  = l_1  \varsigma^m  - l_1^+  \varsigma_{01} -  \varsigma_{0},\\
     \label{gtrgh02}
     & \delta  \chi_{2}  = m_1^+  \varsigma   + m_1  \varsigma_{11} - \varsigma_{0} , \hspace{-0.5em} &&  \delta  \chi_{2}^m  = m_1^+  \varsigma^m  - l_1^+  \varsigma_{11} , \hspace{-0.5em} &&  \delta \chi_{11}^m = l_1  \varsigma_{11} - m_1^+ \varsigma_{01} - \varsigma_{0}  ,\\
     \label{gtrgh03}
     & \delta  \chi_{0} =  m_1  \varsigma_0   +l_0 \varsigma ,  &&  \delta  \chi_{0}^m  =-l_1^+  \varsigma_0  + l_0 \varsigma^{m} , \hspace{-0.5em} &&  \\
     \label{gtrgh04}
     & \delta  \chi_{01}  = -l_1  \varsigma_0  + l_0  \varsigma_{01} , \hspace{-0.5em} &&  \delta  \chi_{01}^m  = - m_1^+  \varsigma_0  +l_0 \varsigma_{11} . \hspace{-0.5em} &&
 \end{align}
 The respective  ghost-independent equations of motion from  (\ref{Sclfvectf})  take the form
  in powers of ghost monomials $C(C\mathcal{P})^k$ for $k=0$:
\begin{eqnarray}
  & \eta_0: & l_0\Phi - l_1^+\chi_0 - m_1\chi_0^m =0,  \label{eta0} \\
  &\eta_1^+:& l_1\Phi  - l_1^+\chi_1 - m_1 \chi_1^m  - \chi_0  - \chi_{01}  =0, \label{eta1} \\
  &\eta_1^m:&  m_1^+ \Phi  - l_1^+ \chi_2 - m_1 \chi_2^m  + \chi_0^m - \chi_{01}^m  =0, \label{eta1m}\\
  &&   m_{11}\Phi + 2 \chi_{1} =0, \quad  m_{11} \chi_{1} =0,\label{m11}
\end{eqnarray}   as well as for $k = 1,2$:
\begin{eqnarray}
  &\eta_0\eta_1^+\mathcal{P}_1^+: & l_0\chi_1 - l_1 \chi_0 - m_1 \chi_{01} =0,   \label{eta0eta1P}\\
  &\eta_0\eta_1^+\mathcal{P}_1^m : & l_0\chi^m_1 - l_1 \chi_0^m  + l_1^+\chi_{01} =0, \label{eta0eta1Pm} \\
  &\eta_0\eta_1^m\mathcal{P}_1^+:&   l_0\chi_2 - m^+_1 \chi_0 - m_1\chi_{01}^m =0, \label{eta0etamP} \\
  &\eta_0\eta_1^m\mathcal{P}_1^m: &   l_0\chi_2^m - m^+_1 \chi_0^m + l^+_1\chi_{01}^m =0 , \label{eta0etamPm}\\
  &\eta_0\eta_1^+\eta_1^m\mathcal{P}_1^+\mathcal{P}_1^m: &   l_0\chi_{11}^m + m^+_1\chi_{01} - l_1\chi_{01}^m =0 ,\label{eta0eta1etamPPm}
\\
  &\eta_1^+\eta_1^m\mathcal{P}_1^+: & - m^+_1\chi_1 + l_1\chi_2 - m_1\chi_{11}^m  -  \chi_{0} -\chi_{01}   =0,  \label{eta1etamP} \\
  &\eta_1^+\eta_1^m\mathcal{P}_1^m : & -m^+_1\chi^m_1 +l_1\chi_2^m + l_1^+\chi_{11}^m -\chi_{0}^m +\chi_{01}^m =0.\label{eta1etamPm}
\end{eqnarray}
 Thus, the relations (\ref{gtrgh11})--(\ref{eta1etamPm}) determine the constrained gauge theory of the first stage  reducibility for the massless free  field $\Phi (x,\omega)$ of CS $\Xi$ in $\mathbb{R}^{1,d-1}$ subject to the constraints (\ref{constrcontinB2})--(\ref{constrcontinB02}), (\ref{m11})  with 9 auxiliary tensor fields. This theory is non-Lagrangian with off-shell holonomic constraint (\ref{m11}) and corresponds to the algebra $\mathcal{A}(\Xi;
\mathbb{R}^{1,d-1})$.

The special structure of the constraints permits to make gauge-fixing procedure starting from the  lowest  gauge  parameter $\varpi$ which together with the linear combination
of the gauge parameters $(\varsigma^m  -\varsigma_{11} )$ (that follows from (\ref{constrcontinB1}))  belongs to the set of  $\ker m_{11}$. After invertible change of the basis  of the zeroth-level gauge parameters:
  \begin{equation}\label{newbas}
  \big\{ \varsigma ,\,  \varsigma^m ,\,  \varsigma_{01},\,  \varsigma_{11}, \,  \varsigma_{0} \big\}\to  \big\{ \varsigma ,\,  \tilde{\varsigma}^m  ,\,  \varsigma_{01} ,\,  \tilde{\varsigma}_{11}, \,  \varsigma_{0} \big\} \  \mathrm{for}  \ \big(\tilde{\varsigma}^m, \tilde{\varsigma}_{11}\big)= \frac{1}{2}\big(\varsigma^m \mp \varsigma_{11}\big)
\end{equation}
we  may gauge away the parameter $ \tilde{\varsigma}^m $ from  $\delta \tilde{\varsigma}^m  = - \frac{\imath}{2}\Xi \varpi$ by means of complete using of $\varpi$. Now, the  theory becomes by irreducible gauge theory with independent gauge-invariant parameters $\varsigma ,\,  \varsigma_{01} ,\,  \tilde{\varsigma}_{11}, \,  \varsigma_{0} $ for $m_{11}^2 \tilde{\varsigma}_{11}=0$ and  with the rest parameters satisfying to the first constraints in  (\ref{constrcontinB1}).

Turning to the  field vectors we replace  the parameters $\varsigma^m, \varsigma_{11}$ in (\ref{gtrgh01})--(\ref{gtrgh04})  on $\tilde{\varsigma}_{11}$ and   see, that two pairs of the fields $\chi_{0}^m $, $ \chi_{01}^m $ and $\chi_{1}^m $, $ \chi_{11}^m $ obey to similar constraints in (\ref{constrcontinB0})--(\ref{constrcontinB02}) as the parameters $(\varsigma^m  -\varsigma_{11} )=2\tilde{\varsigma}_{11}$.
 Making  invertible change of the basis  of the fields:
  \begin{eqnarray}\label{newbasf}
 && \big\{ \chi_{0}^m , \chi_{01}^m , \chi_{1}^m ,  \chi_{11}^m \big\}\to  \big\{ \tilde{\chi}_{0}^m , \tilde{\chi}_{01}^m, \tilde{\chi}_{1}^m,  \tilde{\chi}_{11}^m \big\} \\
 &&\ \  \mathrm{for}  \ \big(   \tilde{\chi}_{0}^m, \tilde{\chi}_{01}^m;   \tilde{\chi}_{1}^m,  \tilde{\chi}_{11}^m \big)=\frac{1}{2} \big({\chi}_{0}^m \mp {\chi}_{01}^m;   {\chi}_{1}^m \mp {\chi}_{11}^m \big),\nonumber
\end{eqnarray}
with untouched rest fields: $\Phi,  \chi_0 ,  \chi_{01} ,  \chi_{1} ,   \chi_{2} ,
{\chi}_2^m,$
we  may gauge away the fields  $ \tilde{\chi}_{0}^m $  from  $\delta  \tilde{\chi}_{0}^m =\frac{\imath}{2} \Xi {\varsigma}_{0}$ and $ \tilde{\chi}_{1}^m $ from  $\delta \tilde{\chi}_{1}^m =  \frac{\imath}{2}\Xi {\varsigma}_{01}$  by means of complete using of ${\varsigma}_{0}$ and
$ {\varsigma}_{01}$ respectively, in view of theirs satisfaction to the same constraint.  Then, from the gauge transformation, $ \delta  \chi_{2}^m  =  \imath \Xi \tilde{\varsigma}_{11}$ we gauge away the field $  \chi_{2}^m $, which now obeys, together with $\tilde{\varsigma}_{11}$, to the constraints: $m_{11} | \chi_{2}^m \rangle$ =$m_{11}\tilde{\varsigma}_{11} = 0$ (\ref{constrcontinB01}) after using ${\varsigma}_{01}$.
Thus, the following 7 fields with gauge transformations  survive after partial gauge-fixing with unique gauge parameter $\varsigma$, $m_{11}\varsigma =0$:
\begin{align}\label{gtrgh0f}
     & \delta \Big( \Phi ,   \chi_{1},  \chi_{2},   \tilde{\chi}_{11}^m ,  \chi_{0} ,  \chi_{01},  \tilde{\chi}_{01}^m  \Big)=  \Big(l_1^+, l_1, m_1^+,  0, l_0, 0, 0\Big) \varsigma.
 \end{align}
From the transformed equations of motion (\ref{eta0etamPm}), (\ref{eta1etamPm}): $\imath\Xi \big(\tilde{\chi}_{01}^m , \tilde{\chi}_{11}^m\big) = 0$ follow vanishing of the fields  $\tilde{\chi}_{01}^m , \tilde{\chi}_{11}^m$. Therefore,  the remaining equations  (\ref{eta0})--(\ref{eta1etamPm})   transform as follows except for  (\ref{eta0eta1P}):
\begin{align}\label{eta0r} & l_0\Phi - l_1^+\chi_0 = 0, &   l_1\Phi  - l_1^+\chi_1   - \chi_0  - \chi_{01}  =0,
  &&  m_1^+ \Phi  - l_1^+ \chi_2     =0, \\
&   l_1^+\chi_{01} =0,  &  m^+_1\chi_{01}  =0 ,   &&   l_0\chi_2 - m^+_1 \chi_0 =0,  \label{eta0eta1etamPPmr}
\\
&{}  & - m^+_1\chi_1 + l_1\chi_2  -  \chi_{0} -\chi_{01}   =0. && \label{eta1etamPr}
 \end{align}
Two first equations in (\ref{eta0eta1etamPPmr}) have unique general solution:  $\chi_{01} = 0$.
The resulting equations of motion for the rest 4 fields take the form
\begin{align}& l_0\Phi - l_1^+\chi_0 = 0,  &
l_1\Phi  - l_1^+\chi_1   - \chi_0  =0,
  && l_0\chi_1 - l_1 \chi_0  =0, \label{eta1mrr}\\
  &
   l_0\chi_2 - m^+_1 \chi_0 =0,  & l_1\chi_2- m^+_1\chi_1   -  \chi_{0}  =0, &&  m_1^+ \Phi  - l_1^+ \chi_2     =0 ,  \label{eta1etamPrr}
 \end{align}
which may be considered with account of algebraic traceless constraints (\ref{constrcontinB0})--(\ref{constrcontinB02})  as the  \emph{triplet-like non-Lagrangian formulation} for scalar bosonic field with CS in the Bargmann--Wigner form, due to presence of the field ${\chi}_{0}(\omega)$  by analogy with case of HS fields with integer spin \cite{triplet}. Indeed, the equations (\ref{eta1mrr}) do not contain the field $\chi_2(\omega)$ and coincide  with  the conditions which determine the {triplet formulation} for the case of integer spin, but with more involved structure of the fields. Almost the same triplet interpretation (if makes the change $l^{(+)}_1 \to m^{(+)}_1$) valid for the latter equation in (\ref{eta1mrr}) and two first equations in (\ref{eta1etamPrr}) for the triplet: $\chi_2(\omega)$, $ \chi_0(\omega)$, $\chi_1(\omega)$. The latter equation in (\ref{eta1etamPrr}) entangles the basic fields: $\Phi$, $\chi_2$ from both triplet equations.

 After  resolution of the second   algebraic equation of  motion in (\ref{eta1mrr})     with respect to ${\chi}_{0}$   we get the system:
\begin{align}& (l_0 - l_1^+l_1)\Phi  + (l_1^+)^2\chi_1 = 0,
& (l_0+l_1 l_1^+)\chi_1 - (l_1)^2\Phi   =0, && \label{eta1mrrd}\\
  &
   l_0\chi_2 - m^+_1l_1\Phi  + m^+_1 l_1^+\chi_1 = 0,  & l_1\big(\chi_2 -  \Phi\big) - \imath\Xi\chi_1     =0, &&  l_1^+ \big( \chi_2 -\Phi \big) - \imath\Xi \Phi    = 0 ,  \label{eta1etamPrrd}
 \end{align}
 given in the configuration space $\mathcal{M}_{cl}$ parameterized by  $\Phi(\omega), \chi_1(\omega), \chi_2(\omega) $ subject to the gauge transformations (\ref{gtrgh0f}) and to the independent holonomic constraints
  \begin{align}
      &  m_{11}\chi_{1} (\omega) = m_{11} \varsigma(\omega)   = 0,
       &
  m_{11}\chi_{2}(\omega)  + 2 \chi_{1}(\omega)=0,  &&    m_{11}\Phi(\omega) + 2\chi_{1} (\omega) = 0.
   \label{constrfin}
 \end{align}
  The equations (\ref{eta1mrrd}), (\ref{eta1etamPrrd}) may be interpreted as the  \emph{duplet-like non-Lagrangian formulation} for scalar bosonic field with CS in the Bargmann--Wigner form by analogy with case of HS fields with integer spin \cite{triplet}. The equations (\ref{eta1mrrd}) coincide  by the form with  the conditions which determine the {duplet formulation} for discrete spin.

 If we gauge away the field $\chi_{1}(\omega)$ by using  the part of the gauge parameter $\varsigma(\omega)$, which therefore should be divergentless,  $l_1\varsigma(\omega)=0$,    (see for details, subsection~\ref{reductionC}),   the same gauge-fixing procedure may be applied  for now  $m_{11}$-traceless field $\chi_2$ so that the field  $\chi_2$ may be removed completely by means of using of the gauge transformations with remaining degrees of freedom in  $\varsigma(\omega)$ so that the remaining from the systems (\ref{eta1mrrd}), (\ref{eta1etamPrrd}), (\ref{constrfin})  equations on the initial field  $\Phi(\omega)$ coincide with the IR conditions (\ref{Eqb23m}), (\ref{Eqb23}).
 %%%%%%%%%%%%%%%%%%%%%%%%

 Expressing the field  $\chi_{1}(\omega)$ as generalized trace of the basic field $\Phi(\omega)$: $\chi_{1}$= $-\frac{1}{2}m_{11}\Phi$,  according to(\ref{constrfin}), we derive from the duplet-like formulation (\ref{eta1mrr}), (\ref{eta1etamPrrd}) and (\ref{constrfin}) the system:
  \begin{align}& \Big\{l_0 \hspace{-0.1em}- l_1^+l_1  \hspace{-0.1em} - \textstyle\frac{1}{2} (l_1^+)^2m_{11}\Big\} \Phi \hspace{-0.1em} \equiv \mathcal{L}_0\Phi\hspace{-0.1em} = \hspace{-0.1em}0,
\hspace{-0.5em}&&  \textstyle\Big\{\frac{1}{2} (l_0\hspace{-0.1em}+l_1 l_1^+)m_{11}+ \hspace{-0.1em} (l_1)^2\Big\}\Phi   = \hspace{-0.1em} \widehat{\mathcal{L}}_0\Phi \hspace{-0.1em}= 0,   \label{eta1mrrdd}\\
  &
   l_0\chi_2 - m^+_1\Big(l_1 +\frac{1}{2}l_1^+m_{11}\Big)\Phi  \equiv \mathcal{F}(\Phi, \chi_2) = 0,  \hspace{-0.5em}&&  \ \ l_1^+ \big( \chi_2 - \Phi  \big) - \imath\Xi \Phi    \equiv \mathcal{L}_1^+(\Phi, \chi_2)= 0 ,   \label{eta1etamPrrdd}
\\
 & l_1\big(\chi_2 -  \Phi\big) + \frac{1}{2}\imath\Xi m_{11}\Phi     = \mathcal{L}_1(\Phi, \chi_2) = 0,  \hspace{-0.5em} && \quad   m_{11}\big(\chi_{2}  - \Phi\big)= m^2_{11}\Phi = 0. \label{constrchi2phi}
 \end{align}
The second equation in (\ref{eta1mrrdd}) and the first one in  (\ref{constrchi2phi}) are the respective algebraic  consequences of the first equation  in (\ref{eta1mrrdd})  and the second one   (\ref{eta1etamPrrdd}), when applying to the latter ones of the trace operator $m_{11}$:
\begin{equation}\label{algconseq}
  \widehat{\mathcal{L}}_0\Phi  \ =\  m_{11}\mathcal{L}_0\Phi , \qquad \mathcal{L}_1(\Phi, \chi_2) \ =\  m_{11}\mathcal{L}^+_1(\Phi, \chi_2).
\end{equation}
 Thus, the independent system of equations  consists from 3 differential equations and 2  $m_{11}$-traceless (holonomic) constraints to be invariant with respect to gauge transformations, $\delta(\Phi(\omega), \chi_2(\omega)) = (l_1^+, m_1^+)\varsigma(\omega)$ for $m_{11}\varsigma(\omega)=0$. Note, the imposing of the above described  gauge-fixing procedure   on  the  auxiliary field $\chi_2$ turns the system  (\ref{eta1mrrdd})--(\ref{constrchi2phi}) to the Bargmann-Wigner equations  (\ref{Eqb23m}), (\ref{Eqb23}).

Presenting the independent gauge-invariant  equations  (\ref{eta1mrrd})--(\ref{constrchi2phi})  in powers of $\omega^{(m)_k}\times$ $\times\frac{\omega^{(n)_l}}{\omega^{2l}}$, $k, l \in \mathbb{N}_0$ similar to the Eqs. (\ref{Eqb01q}), (\ref{Eqb2q}):
 \begin{equation}\label{unfoldfll}
\omega^{(m)_k}\frac{\omega^{(n)_l}}{\omega^{2l}}:    \big(\mathcal{L}_0\Phi\big)^{{l}}_{{l}|(m)_k,(n)_l} =0 , \  \   \big(\mathcal{L}^+_1(\Phi, \chi_2)\big)^{{l}}_{{l}|(m)_k,(n)_l} = 0, \  \  \big(\mathcal{F}(\Phi, \chi_2)\big)^{{l}}_{{l}|(m)_k,(n)_l} =  0,
 \end{equation}
 we get with help of the real-valued Lagrangian multipliers $\hat{\lambda}^{{l}|(m)_k,(n)_l}_i$, $i=1,2,3$, $k,l\in \mathbb{N}_0$ (being different with ones used in (\ref{costrlagr})) the constrained gauge-invariant LF for CSR scalar (real-valued) field $\Phi$ of CS $\Xi$ in the tensor form
\begin{eqnarray}
 \hspace{-0.65em}&\hspace{-0.65em}&\hspace{-0.65em}   \mathcal{S}_{C|\Xi} \left(\Phi, \chi_2, \hat{\lambda}_i\right) =  \int d^dx \hspace{-0.15em} \sum_{k,l\geq 0}\hspace{-0.15em}\bigg\{\hspace{-0.1em} \hat{\lambda}^{{l}|(m)_k,(n)_l}_1 \big(\mathcal{L}_0\Phi\big)^{{l}}_{(m)_k,(n)_l}\hspace{-0.1em} +  \hspace{-0.1em}\imath \hat{\lambda}^{{l}|(m)_k,(n)_l}_2 \big(\mathcal{L}^+_1(\Phi, \chi_2)\big)^{{l}}_{{l}|(m)_k,(n)_l}\nonumber\\
 \hspace{-0.65em}&\hspace{-0.65em}&\hspace{-0.65em} \phantom{\mathcal{S}_{C|\Xi} (\Phi, \chi_2) =} + \hat{\lambda}^{{l}|(m)_k,(n)_l}_3 \big(\mathcal{F}(\Phi, \chi_2)\big)^{{l}}_{{l}|(m)_k,(n)_l}\bigg\}, \label{costrlagrBRST}\\
 \hspace{-0.65em}&\hspace{-0.65em}&\hspace{-0.65em}   \delta\hspace{-0.15em}\left(\hspace{-0.15em}\Phi^{{l}}, \chi^{{l}}_2\right)_{\hspace{-0.15em}(m)_k,(n)_l}\hspace{-0.55em} = \hspace{-0.1em} -\hspace{-0.1em}\Big({\partial_{\{m_{k}}}{\varsigma}^{{l}}_{(m)_{k-1}\},(n)_l},{\partial_{\{m_{k}}}{\varsigma}^{{l}}_{(m)_{k-1}\},(n)_l}\hspace{-0.15em}-\imath\Xi \eta_{\{m_{k}\{n_{l}} {\varsigma}^{{l-1}}_{(m)_{k-1}\},(n)_{l-1}\}}\Big),  \label{gaugetransfconstrBRST}\\
  \hspace{-0.65em}&\hspace{-0.65em}&\hspace{-0.65em}   \delta \hat{\lambda}^{l}_i{}^{(m)_k,(n)_l} (x)\ = \ \sum_{k',l'}\hat{R}^{l{}}_i{}^{(m)_k,(n)_l}_{l'{}(m_1)_{k'}(n_1)_{l'}}(x,\partial_x) \sigma^{l' (m_1)_{k'}(n_1)_{l'}}(x) \Leftrightarrow \delta \hat{\lambda}^{M_i}_{i} = \hat{R}{}^{N_i}_\alpha \sigma^\alpha \label{gaugetransfconstrDUAL}
\end{eqnarray}
with certain generators
$\hat{R}{}^{l{}}_i{}^{(m)_k,(n)_l}_{l'{}(m_1)_{k'}(n_1)_{l'}}$ of
gauge transformations for $\hat{\lambda}^{l}_i{}^{(m)_k,(n)_l}$,
$\delta_\sigma \mathcal{S}_{C|\Xi} = 0$, being  dual (see
Footnote~4) to those for the fields, whose specific form may be
determined explicitly. Here, we should impose the double (single) $m_{11}$-traceless  holonomic constraints on the auxiliary fields $\hat{\lambda}^{{l}|(m)_k,(n)_l}_i$ (on dual gauge parameters: $(m_{11}\sigma)^\alpha =0$)  due to structure of the action and the constraints  (\ref{constrchi2phi}) on
the fields $\Phi^{{l}|(m)_k,(n)_l}, \chi^{{l}|(m)_k,(n)_l}_2$ and gauge parameters ${\varsigma}^{{l}}_{p(m)_{k},(n)_l}$ according to the last representation in (\ref{Eqb2q}):
\begin{equation}\label{fullconstr}
\big(m^2_{11}\hat{\lambda}_i\big)^{{l}|(m)_k,(n)_l} = \big(m^2_{11}\Phi\big)^{l}_{(m)_k,(n)_l} = \big(m_{11}\big(\chi_{2}  - \Phi\big)\big)^{l}_{(m)_k,(n)_l} = \big(m_{11}\varsigma\big)^{l}_{(m)_k,(n)_l} =  0.
\end{equation}

The analogous forms of the constrained gauge-invariant  LFs may be formulated with help of respective sets of the Lagrangian multipliers for the {triplet-like non-Lagrangian formulation} (\ref{eta1mrr}), (\ref{eta1etamPrr}) (see, Eq. (\ref{conSghindf}) in the  Section~\ref{tripBRSTBFV}) and for  {duplet-like non-Lagrangian formulation} (\ref{eta1mrrd}), (\ref{eta1etamPrrd}) within Bargmann-Wigner form of the CSR equations.

One should stress, that an analysis of the form for the non-scalar field  $\Psi(x,\omega)_{(\mu^1)_{s_1}...(\mu^k)_{s_k}}$ with CS, $\Xi$, and integer generalized spin, $\mathbf{s}=(s_1,...,s_k)$, $k\leq [(d-4)/2]$ \cite{BrinkRamondKhan}, should necessary be by a gauge-invariant theory with reducible gauge symmetry, and based on the respective HCS symmetry algebra.

\subsubsection{On Higher Continuous Spin symmetry algebra  $\mathcal{A}(\Xi; Y(k),
\mathbb{R}^{1,d-1})$}\label{BRSTBFV12}

The most general  massless non-scalar CS one-valued  irreducible
representation of Poincare group in a  Minkowski space
$\mathbb{R}^{1,d-1}$  is described by a tensor field
$\Psi(\omega)_{(\mu^1)_{s_1}...(\mu^k)_{s_k}}
$ $\equiv$  $
\Psi(x,\omega)_{\mu^1_1\ldots}$ ${}_{\ldots\mu^1_{s_1},\mu^2_1\ldots\mu^2_{s_2},...,
 \mu^k_1\ldots \mu^k_{s_k}}$
 of rank $\sum_{i\geq 1}^k s_i$  to be corresponding to a Young tableaux with $k$ rows
of length $s_1, s_2, ..., s_k$, respectively
\begin{equation}\label{Young k}
\Psi(\omega)_{(\mu^1)_{s_1},(\mu^2)_{s_2},...,(\mu^k)_{s_k}}
\hspace{-0.3em}\longleftrightarrow \hspace{-0.3em}
\begin{array}{|c|c|c c c|c|c|c|c|c| c|}\hline%\vphantom{\biggm|}
  \!\mu^1_1 \!&\! \mu^1_2\! & \cdot \ & \cdot \ & \cdot \ & \cdot\  & \cdot\  & \cdot\ &
  \cdot\    &\!\! \mu^1_{s_1}\!\! \\
   \hline%\vphantom{\biggm|}
    \! \mu^2_1\! &\! \mu^2_2\! & \cdot\
   & \cdot\ & \cdot  & \cdot &  \cdot & \!\!\mu^2_{s_2}\!\!   \\
  \cline{1-8} \!\!\cdots\!\!   \\
   \cline{1-7}
    \! \mu^k_1\! &\! \mu^k_2\! & \cdot\
   & \cdot\ & \cdot  & \cdot &   \!\!\mu^k_{s_k}\!\!   \\
   \cline{1-7}%\vphantom{\biggm|}
\end{array}\ ,
\end{equation}
This field is symmetric with respect to the permutations of each
type of Lorentz indices
 $\mu^i$,
  and
obeys in addition to (\ref{Eqb01})--(\ref{Eqb23}) to the Klein-Gordon, divergentless
(\ref{Eq-1b}), traceless (\ref{Eq-2b}) and mixed-symmetry
equations (\ref{Eq-3b}) [for $i,j=1,...,k;\, l_i,m_i=1,...,s_i$]:
\begin{eqnarray}
 &&
\partial^\mu\partial_\mu\Psi(\omega)_{(\mu^1)_{s_1},(\mu^2)_{s_2},...,(\mu^k)_{s_k}}
 =0,\qquad \partial^{\mu^i_{l_i}}\Psi(\omega)_{
(\mu^1)_{s_1},(\mu^2)_{s_2},...,(\mu^k)_{s_k}} =0,  \label{Eq-1b}
\\
&& \eta^{\mu^i_{l_i}\mu^i_{m_i}}\Psi(\omega)_{
(\mu^1)_{s_1},(\mu^2)_{s_2},...,(\mu^k)_{s_k}}=
\eta^{\mu^i_{l_i}\mu^j_{m_j}}\Psi(\omega)_{
(\mu^1)_{s_1},(\mu^2)_{s_2},...,(\mu^k)_{s_k}} =0, \quad
 l_i<m_i,  \label{Eq-2b}\\
&& \Psi(\omega)_{
(\mu^1)_{s_1},...,\{(\mu^i)_{s_i}\underbrace{,...,\mu^j_{1}...}\mu^j_{l_j}\}...\mu^j_{s_j},...(\mu^k)_{s_k}}=0,\quad
i<j,\ 1\leq l_j\leq s_j, \label{Eq-3b}
\end{eqnarray}
where the  bracket below denote that the indices  in it do not
include in  symmetrization, i.e. the symmetrization concerns only
indices $(\mu^i)_{s_i}, \mu^j_{l_j} $ in
$\{(\mu^i)_{s_i}\underbrace{,...,\mu^j_{1}...}\mu^j_{l_j}\}$.

For a joint description of all the CSRs with given CS, but different spin $\mathbf{s}$
 we following to the case of  HS fields with only integer spin \cite{BuchbResh} introduce an auxiliary vector
space $\mathcal{V}_k$,  ($\mathcal{V}_0 \equiv \mathcal{V}$) generated in addition to $\omega$  by $k$ sets of bosonic  variables $\vec{\omega} = (\omega^1_{\mu^1}, ...,\omega^k_{\mu^k})$  $i,j =1,...,k; \mu^i,\nu^j
=0,1...,d-1$,
 and a set of constraints for an arbitrary string-like vector
$\Psi(x,\omega, \vec{\omega}) \in \mathcal{V}_k$,
\begin{eqnarray}
\label{PhysState}  \hspace{-2ex}&\hspace{-2ex}& \hspace{-2ex} \Psi(x, \omega, \vec{\omega})  =
\sum_{s_1=0}^{\infty}\sum_{s_2=0}^{s_1}\cdots\sum_{s_k=0}^{s_{k-1}}
\Psi(x, \omega)_{(\mu^1)_{s_1},(\mu^2)_{s_2},...,(\mu^k)_{s_k}}\,
\prod_{i=1}^k\prod_{l_i=1}^{s_i} \omega^{\mu^i_{l_i}}_i,\\
\label{lilijt} \hspace{-2ex} &\hspace{-2ex}& \hspace{-2ex} \bigl(l_0, {l}^i, l^{ij},
t^{i_1j_1} \bigr)\Psi(x, \omega, \vec{\omega}) \hspace{-0.1em} = \hspace{-0.1em}\Big(\hspace{-0.1em}\partial^2, -\frac{\partial}{\partial \omega_i^\mu} \partial^\mu,
\hspace{-0.1em}-\frac{1}{2} \eta^{mn}\frac{\partial}{\partial \omega_i^m}\frac{\partial}{\partial \omega_j^n}, \omega^{i_1}_\mu
\frac{\partial}{\partial \omega_{j_1\mu}}\hspace{-0.1em}\Big)\hspace{-0.1em} \Psi(x, \omega, \vec{\omega})\hspace{-0.1em}=\hspace{-0.1em}0,
\end{eqnarray}
(for $i\leq j;\, i_1 < j_1$).  The set of $(k(k+1)+1)$
primary constraints  (\ref{lilijt}) with $\{o^{\mathbf{s}}_\alpha\}$
= $\bigl\{{{l}}_0, {l}^i, l^{ij}, t^{i_1j_1} \bigr\}$ describes all CSR with given CS $\Xi$, if, in addition, the constraints $o_{\hat{\alpha}} $ hold
\begin{equation}\label{irrepcontgen}
\big(l_0,\, l_1,\,m^+_{1},\, m_{11} \big)\Psi(x,\omega,\vec{\omega})=0.
\end{equation}
 In turn, if we impose
 to the Eqs.  (\ref{lilijt}) the additional
constraints with number particles operators, $g_0^i$,
\begin{eqnarray}\label{g0iphys}
g_0^i\Psi(x, \omega, \vec{\omega}) = \big(s_i+{d}/{2}\big) \Psi(x, \omega, \vec{\omega}), \qquad
 g_0^i = \frac{1}{2}\big\{\omega^i_\mu ,\,\frac{\partial}{\partial \omega^i_\mu}\big\},
\end{eqnarray}
  then these
combined conditions
[which are reduced to ones (\ref{irrepconts}) for scalar CS field, $\Phi(x,\omega)$ = $\Psi(x,\omega,\vec{0})$]
 are equivalent to Eqs.
(\ref{Eq-1b})--(\ref{Eq-3b}) for the field
$\Psi(\omega)_{(\mu^1)_{s_1},(\mu^2)_{s_2},...,(\mu^k)_{s_k}}$ with
given spin $(\Xi, \mathbf{s})$.

The procedure of LF implies  the property of
BRST--BFV operator $Q$, $Q = C^\alpha o^{\mathbf{s}}_\alpha +C^{\hat{\alpha}} o_{\hat{\alpha}} + more$,  to be
Hermitian, that is equivalent to the formal requirements: $\{o^{\mathbf{s}}_\alpha\}^+
= \{o^{\mathbf{s}}_\alpha\}$, $\{o_{\hat{\alpha}}\}^+
= \{o_{\hat{\alpha}}\}$ and closedness for $\{o_{\hat{\alpha}},  o^{\mathbf{s}}_\alpha\}$ with respect to
the commutator multiplication $[\ ,\ ]$.  To provide these conditions
we consider an quasi-scalar product,  $\big(\ , \ \big)$, on
$\mathcal{V}_k$:
\begin{eqnarray}
\label{sproduct} \Big({\Omega}(\omega, \partial / \partial\vec{\omega}), \Psi(\omega,  \vec{\omega})\Big) \hspace{-0.2em} &\hspace{-0.2em} =  \hspace{-0.2em}& \hspace{-0.2em}\int\hspace{-0.2em}
d^dx\hspace{-0.15em}\sum_{s_1=0}^{\infty}\sum_{s_2=0}^{s_1} \hspace{-0.1em}\cdots\hspace{-0.1em}\sum_{s_k=0}^{s_{k-1}}
         \sum_{p_1=0}^{\infty}\sum_{p_2=0}^{p_1}\hspace{-0.1em}\cdots\hspace{-0.1em}\sum_{p_k=0}^{p_{k-1}}\Omega^*(x,\omega)_{(\nu^1)_{p_1},(\nu^2)_{p_2},...,(\nu^k)_{p_k}}
\nonumber\\
\hspace{-0.2em} &\hspace{-0.2em} & \hspace{-0.2em} \times \prod_{j=1}^k\prod_{m_j=1}^{p_j}
\frac{\partial}{\partial \omega^{\nu^j_{m_j}}_j}
\Psi(x,\omega)_{(\mu^1)_{s_1},(\mu^2)_{s_2},...,(\mu^k)_{s_k}}\,
\prod_{i=1}^k\prod_{l_i=1}^{s_i} \omega^{\mu^i_{l_i}}_i .
\end{eqnarray}
 The quasi-scalar  product $\big(\ , \ \big)$ presents the $\omega$-dependent bilinear operation on $\mathcal{V}^*_k \bigotimes \mathcal{V}_k$ being differed from the standard scalar product on the Fock space $\mathcal{H}$ used for HS fields with only integer spin $\mathbf{s}$ \cite{BuchbResh} [under identification of the oscillators $\big(a^i_{\mu^i}, a_{\nu^j}^{j+}\big) \equiv $ $-\imath\big({\partial}/\partial \omega_i^{\mu^i}, \omega_{\nu^j}^{j+}\big)$ with commutation relations,  $[a^i_{\mu^i}, a_{\nu^j}^{j+}]=-\eta_{\mu^i\nu^j}\delta^{ij}$].
As the result, the set of $\{o_{\hat{\alpha}}\}$ and $\{o^{\mathbf{s}}_\alpha\}$ are extended respectively formally to $o_I$ (\ref{geninalg}) and to $\{o^{\mathbf{s}}_P\} = \{o^{\mathbf{s}}_\alpha, (o^{\mathbf{s}}_\alpha)^+, g_0^i\}$ extended by means of the
operators $(o^{\mathbf{s}}_\alpha)^+$,
\begin{eqnarray} \label{lilijt+} \hspace{-2ex} && \hspace{-2ex} \bigl({l}^{i+},\
l^{ij+},\ t^{i_1j_1+} \bigr)  = \bigl(-\omega^{i}_\mu
\partial^\mu,\ -\textstyle\frac{1}{2}\omega^{i}_\mu \omega^{j\mu},\
\omega^{j_1}_{\mu}\frac{\partial}{\partial \omega_{i_1\mu}} \bigr) ,\ i\leq j;\ i_1 < j_1,
\end{eqnarray}
 with taken into account of formally self-conjugated
operators, $(l_0^+,\ {g_0^i}^+) = (l_0,\ {g_0^i})$.

The set of all operators $o_W$ has therefore the structure,
\begin{eqnarray}
\{o_W\} =  \{o_I; o^{\mathbf{s}}_P\} = \{o_{\hat{\alpha}}, o^+_{\hat{\alpha}}, g_0, \Xi, \nu ; o^{\mathbf{s}}_\alpha, o^{\mathbf{s}+}_\alpha;\ g_0^i\}; \ \  [o_I, o^{\mathbf{s}}_P] = 0. \label{inconstraints}
\end{eqnarray}
Explicitly, operators $o_W$ satisfy to the Lie-algebra commutation
relations,
\begin{equation}\label{geninalg1}
    [o_W,\ o_X]= f^Y_{WX}o_Y, \  f^Y_{WX}= \big(f^K_{IJ}, f^Q_{PR} \big) = - f^Y_{XW},
\end{equation}
where the structure constants $f^Y_{WX}$ with non-vanishing components $\big(f^K_{IJ}, f^Q_{PR} \big)$ are determined from the
Multiplication Tables~\ref{table in} and~\ref{table in1}. \hspace{-1ex}{\begin{table}
{{\footnotesize
\begin{center}
\begin{tabular}{||c||c|c|c|c|c|c|c||c||}\hline\hline
$\hspace{-0.2em}[\; \downarrow, \rightarrow
\}\hspace{-0.5em}$\hspace{-0.7em}&
 $t^{i_1j_1}$ & $t^+_{i_1j_1}$ &
$l_0$ & $l^i$ &$l^{i{}+}$ & $l^{i_1j_1}$ &$l^{i_1j_1{}+}$ &
$g^i_0$ \\
\hline\hline $t^{i_2j_2}$
    & $A^{i_2j_2, i_1j_1}$ & $B^{i_2j_2}{}_{i_1j_1}$
   & $0$&\hspace{-0.3em}
    $\hspace{-0.2em}l^{j_2}\delta^{i_2i}$\hspace{-0.5em} &
    \hspace{-0.3em}
    $-l^{i_2+}\delta^{j_2 i}$\hspace{-0.3em}
    &\hspace{-0.7em} $\hspace{-0.7em}l^{\{j_1j_2}\delta^{i_1\}i_2}
    \hspace{-0.9em}$ \hspace{-1.2em}& \hspace{-1.2em}$
    -l^{i_2\{i_1+}\delta^{j_1\}j_2}\hspace{-0.9em}$\hspace{-1.2em}& $F^{i_2j_2,i}$ \\
\hline $t^+_{i_2j_2}$
    & $-B^{i_1j_1}{}_{i_2j_2}$ & $A^+_{i_1j_1, i_2j_2}$
&$0$   & \hspace{-0.3em}
    $\hspace{-0.2em} l_{i_2}\delta^{i}_{j_2}$\hspace{-0.5em} &
    \hspace{-0.3em}
    $-l^+_{j_2}\delta^{i}_{i_2}$\hspace{-0.3em}
    & $l_{i_2}{}^{\{j_1}\delta^{i_1\}}_{j_2}$ & $-l_{j_2}{}^{\{j_1+}
    \delta^{i_1\}}_{i_2}$ & $-F_{i_2j_2}{}^{i+}$\\
\hline $l_0$
    & $0$ & $0$
& $0$   &
    $0$\hspace{-0.5em} & \hspace{-0.3em}
    $0$\hspace{-0.3em}
    & $0$ & $0$ & $0$ \\
\hline $l^j$
   & \hspace{-0.5em}$- l^{j_1}\delta^{i_1j}$ \hspace{-0.5em} &
   \hspace{-0.5em}$
   -l_{i_1}\delta_{j_1}^{j}$ \hspace{-0.9em}  & \hspace{-0.3em}$0$ \hspace{-0.3em} & $0$&
   \hspace{-0.3em}
   $l_0\delta^{ji}$\hspace{-0.3em}
    & $0$ & \hspace{-0.5em}$- \textstyle\frac{1}{2}l^{\{i_1+}\delta^{j_1\}j}$
    \hspace{-0.9em}&$l^j\delta^{ij}$  \\
\hline $l^{j+}$ & \hspace{-0.5em}$l^{i_1+}
   \delta^{j_1j}$\hspace{-0.7em} & \hspace{-0.7em}
   $l_{j_1}^+\delta_{i_1}^{j}$ \hspace{-1.0em} &
   $0$&\hspace{-0.3em}
      \hspace{-0.3em}
   $-l_0\delta^{ji}$\hspace{-0.3em}
    \hspace{-0.3em}
   &\hspace{-0.5em} $0$\hspace{-0.5em}
    &\hspace{-0.7em} $ \textstyle\frac{1}{2}l^{\{i_1}\delta^{j_1\}j}
    $\hspace{-0.7em} & $0$ &$-l^{j+}\delta^{ij}$  \\
\hline $l^{i_2j_2}$
    & \hspace{-0.3em}$\hspace{-0.4em}-l^{j_1\{j_2}\delta^{i_2\}i_1}\hspace{-0.5em}$
    \hspace{-0.5em} &\hspace{-0.5em} $\hspace{-0.4em}
    -l_{i_1}{}^{\{i_2+}\delta^{j_2\}}_{j_1}\hspace{-0.3em}$\hspace{-0.3em}
   & $0$&\hspace{-0.3em}
    $0$\hspace{-0.5em} & \hspace{-0.3em}
    $ \hspace{-0.7em}-\textstyle\frac{1}{2}l^{\{i_2}\delta^{j_2\}i}
    \hspace{-0.5em}$\hspace{-0.3em}
    & $0$ & \hspace{-0.7em}$\hspace{-0.3em}
L^{i_2j_2,i_1j_1}\hspace{-0.3em}$\hspace{-0.7em}& $\hspace{-0.7em}  l^{i\{i_2}\delta^{j_2\}i}\hspace{-0.7em}$\hspace{-0.7em} \\
\hline $l^{i_2j_2+}$
    & $ l^{i_1 \{i_2+}\delta^{j_2\}j_1}$ & $ l_{j_1}{}^{\{j_2+}
    \delta^{i_2\}}_{i_1}$
   & $0$&\hspace{-0.3em}
    $\hspace{-0.2em} \textstyle\frac{1}{2}l^{\{i_2+}\delta^{ij_2\}}$\hspace{-0.5em} & \hspace{-0.3em}
    $0$\hspace{-0.3em}
    & $-L^{i_1j_1,i_2j_2}$ & $0$ &$\hspace{-0.5em}  -l^{i\{i_2+}\delta^{j_2\}i}\hspace{-0.3em}$\hspace{-0.2em} \\
\hline\hline $g^j_0$
    & $-F^{i_1j_1,j}$ & $F_{i_1j_1}{}^{j+}$
   &$0$& \hspace{-0.3em}
    $\hspace{-0.2em}-l^i\delta^{ij}$\hspace{-0.5em} & \hspace{-0.3em}
    $l^{i+}\delta^{ij}$\hspace{-0.3em}
    & \hspace{-0.7em}$\hspace{-0.7em}  -l^{j\{i_1}\delta^{j_1\}j}\hspace{-0.7em}$\hspace{-0.7em} & $ l^{j\{i_1+}\delta^{j_1\}j}$&$0$ \\
   \hline\hline
\end{tabular}
\end{center}}} \vspace{-2ex}\caption{Higher Continuous Spin symmetry  algebra  $\mathcal{A}(\Xi;Y(k),
\mathbb{R}^{1,d-1})$.\label{table in1} }\end{table}

Note that,  in the table~\ref{table in1}, the operators
$t^{i_2j_2}, t^+_{i_2j_2}$ satisfy  the
properties
\begin{equation} \label{thetasymb} (t^{i_2j_2}, t^+_{i_2j_2}) \equiv
(t^{i_2j_2},t^+_{i_2j_2})\theta^{j_2i_2},
\end{equation}
 the products $B^{i_2j_2}_{i_1j_1}, A^{i_2j_2, i_1j_1},
F^{i_1j_1,i}, L^{i_2j_2,i_1j_1}$ are determined by the explicit
relations \cite{BuchbResh},
\begin{eqnarray}
 \hspace{-0.2em}B^{i_2j_2}{}_{i_1j_1}\hspace{-0.2em} &\hspace{-0.2em}=\hspace{-0.2em}&\hspace{-0.2em}
  (g_0^{i_2}-g_0^{j_2})\delta^{i_2}_{i_1}\delta^{j_2}_{j_1} +
  (t_{j_1}{}^{j_2}\theta^{j_2}{}_{j_1} + t^{j_2}{}^+_{j_1}\theta_{
  j_1}{}^{j_2})\delta^{i_2}_{i_1}
  -(t^+_{i_1}{}^{i_2}\theta^{i_2}{}_{i_1} + t^{i_2}{}_{i_1}\theta_{i_1}{}^{
  i_2})
  \delta^{j_2}_{j_1}
\,,\nonumber
\\
   A^{i_2j_2, i_1j_1} \hspace{-0.2em} &\hspace{-0.2em}=\hspace{-0.2em}&\hspace{-0.2em}  t^{i_1j_2}\delta^{i_2j_1}-
  t^{i_2j_1}\delta^{i_1j_2}  ,   \nonumber\\
   {} F^{i_2j_2,i} \hspace{-0.2em} &\hspace{-0.2em}=\hspace{-0.2em}&\hspace{-0.2em}
   t^{i_2j_2}(\delta^{j_2i}-\delta^{i_2i}),\label{Fijk} \\
  L^{i_2j_2,i_1j_1} \hspace{-0.2em} &\hspace{-0.2em}=\hspace{-0.2em}&\hspace{-0.2em}   \textstyle\frac{1}{4}\Bigl\{\delta^{i_2i_1}
\delta^{j_2j_1}\Bigl[2g_0^{i_2}\delta^{i_2j_2} + g_0^{i_2} +
g_0^{j_2}\Bigr]  - \delta^{j_2\{i_1}\Bigl[t^{j_1\}i_2}\theta^{i_2j_1\}} +t^{i_2j_1\}+}\theta^{j_1\}i_2}\Bigr] \nonumber \\
&& - \delta^{i_2\{i_1}\Bigl[t^{j_1\}j_2}\theta^{j_2j_1\}}
+t^{j_2j_1\}+}\theta^{j_1\}j_2}\Bigr] \Bigr\}
 \,,\nonumber
\end{eqnarray}
with known properties  of theirs antisymmetry and
Hermitian conjugation \cite{BuchbResh}.

We call the algebra of the  operators $o_W$ the \emph{higher continuous spin symmetry algebra in Minkow\-ski space with a Young
tableaux having $k$ rows} with notation $\mathcal{A}(\Xi; Y(k),
\mathbb{R}^{1,d-1})$. Note, that $\mathcal{A}(\Xi; Y(0),
\mathbb{R}^{1,d-1})$ = $\mathcal{A}(\Xi;
\mathbb{R}^{1,d-1})$. The subalgebra of $\{o^{\mathbf{s}}_P\}$ without space-time derivatives, i.e. $\big(l^{ij},\ t^{i_1j_1}, l^{ij+},\ t^{i_1j_1+}, g_0^i\big)$ is isomorphic to the symplectic algebra $sp(2k)$.  The Howe dual algebra \cite{Howe1} to the algebra
$so(1,d-1)$ is $sp(2k)$  if $k=\left[(d-4)/{2}\right]$.

The algebra $\mathcal{A}(\Xi; Y(k),
\mathbb{R}^{1,d-1})$ provides a basis to construct BRST-BFV gauge-invariant descriptions for the equations of motion and for the (un)constrained LFs following to the proposed above receipt for scalar CSR  in the Bargmann--Wigner form.

  The situation with $ISO(1,d-1)$ representations  with integer spin looks  another \cite{Pashnev1}, \cite{reviews3}, \cite{Reshetnyak_con}.

\subsection{Comparison with dynamics  for higher integer spin fields}\label{BRSTBFV3}

First of all, let us present the result for the BRST-BFV descriptions for scalar CS field obtained in the Section~\ref{BRSTBFV2}  in terms of the bosonic fields being subject to the usual traceless or double traceless condition, generated by the operator $l_{11}=m_{11}|_{\textstyle \nu=0}$, following to Fronsdal proposal  for (half-)integer HS fields \cite{Fronsdalfint, Fronsdalhalfint}.
  To this end we will use the representation (\ref{presentaPhi}), (\ref{npvec1}) for the fields  $\Phi(x,\omega)$ and  for $\chi^k_C$:
\begin{eqnarray}\label{npvec1g}
&& \chi^k_C = \sum_{s}\Big\{\chi^{(s,0)k} + \sum_{l>0} \chi^{(s,l)k}\Big\}, \ \ \chi^{(s,l)k}(x,\omega,\mathcal{C}, \mathcal{P}) = \mathcal{C}^{p} \mathcal{P}^{p+k}\varphi^{(s,l)k}(x,\omega), \\
  &&  gh_H(\chi^k_C)=gh_H(\chi^{(s,l)k}) = -k,\ \ \big(\mathrm{deg}_{\omega}, \mathrm{deg}_{\omega/\omega^2}\big)\chi^{(s, l)k} = (s, l), \label{npvec1g1}
\end{eqnarray}
so that the generalized-traceless constraints  for the field and gauge parameter satisfying to:
    \begin{equation}\label{gtreqs}
    m_{11}^2\Phi(\omega) =0, \qquad m_{11}\varsigma(\omega) =0
\end{equation}
should be rewritten in terms of Fronsdal-like double  traceless $\phi^{s,0}(\omega)$  and traceless $\psi^{s,0}(\omega)$ fields and new fields double traceless $\phi^{s,l}(\omega)$ and traceless $\psi^{s,l}(\omega)$, $\forall s, (l+1) \in \mathbb{N}_0$,  (for fixed degree in powers of $\omega^{m}$ according to [\ref{npvec1g} and due to relation: $\Phi^{s,l}(\omega) =  (-1)^l\widetilde{\Phi}^{s+l,0}(\omega)(l_{11}^+)^{-l}$]  as\footnote{The presentation  of (double) traceless condition, $l^{(2)}_{11}\psi^{s,l}(\omega) = 0$ is as usual   for  $l=0$, whereas for  $l>0$  should be  understood according to the latter algebraic equations respectively in  (\ref{Eqb01q}) , (\ref{Eqb2q}), but for $\nu=0$.}:
  \begin{eqnarray}\label{fronsdd}
 \hspace{-0.7em}  &\hspace{-0.7em}& \hspace{-0.7em} \Phi(\omega) =  \hspace{-0.2em}\sum_{n\geq 0}  \sum_{k\geq 0}^{\left[n/2\right]}\Big(\gamma_{k,n}(l_{11}^+)^{k} \phi^{n-2k,0}+\sum_{l>1}\widetilde{\gamma}^l_{k,n}(l_{11}^+)^k \phi^{n-2k,l}\Big),   \\
  \hspace{-0.7em} &\hspace{-0.7em}&\hspace{-0.7em} \varsigma(\omega) = \hspace{-0.2em}\sum_{n\geq 0} \sum_{k\geq 0}^{\left[n/2\right]} \Big( \delta_{k,n}(l_{11}^+)^k \psi^{n-2k,0} + \sum_{l>1}\widetilde{\delta}^l_{k,n}(l_{11}^+)^k \psi^{n-2k,l}\Big), \label{fronsdo}
  \end{eqnarray}
  with untouched negative-like part of  $\Phi(\omega)$ (as for the standard  Fronsdal-like fields) and in more general form
  \begin{eqnarray} \label{fronsddm}
 \hspace{-0.7em}  &\hspace{-0.7em}& \hspace{-0.7em} \Phi(\omega) =  \hspace{-0.2em}\sum_{n\geq 0}  \sum_{k\geq 0}^{\left[n/2\right]}\sum_{s\geq 0}^{\left[n/2\right]-k}\Big(\gamma^s_{k,n}(l_{11}^+)^{k+s} \phi^{n-2k-s,s}+\sum_{l>1}\widetilde{\gamma}^{l,s}_{k,n}(l_{11}^+)^{k+s} \phi^{n-2k-s,l+s}\Big),   \\
  \hspace{-0.7em} &\hspace{-0.7em}&\hspace{-0.7em} \varsigma(\omega) = \hspace{-0.2em}\sum_{n\geq 0}\sum_{k\geq 0}^{\left[n/2\right]} \sum_{s\geq 0}^{\left[n/2\right]-k}\Big(\delta^s_{k,n}(l_{11}^+)^{k+s} \psi^{n-2k-s,s} + \sum_{l>1}\widetilde{\delta}^{l,s}_{k,n}(l_{11}^+)^{k+s} \psi^{n-2k-s,l+s}\Big), \label{fronsdog}\\
  \hspace{-0.7em}  &\hspace{-0.7em}& \hspace{-0.7em} \vartheta(\omega) = \big(\vartheta^{0} +\sum_{l\geq 1}\vartheta^{l}\big)(\omega) \ \mathrm{for} \ \vartheta \in \{\phi, \psi \}, \  \ l^2_{11}\phi^{n,l} =0,\ l_{11}\psi^{n,l} =0, \label{fronsdoom}
  \end{eqnarray}
  where the fields $\phi(\omega), \psi(\omega)$ have the decomposition (\ref{presentaPhi}) according to  (\ref{npvec1}).
Here,  the unknown rational coefficients  $\widetilde{\gamma}^l_{k,n}$, $\widetilde{\delta}^l_{k,n}$   for the decomposition (\ref{fronsdd}), (\ref{fronsdo}) are determined from (\ref{gtreqs}) as the solutions of the system of recursive  equations
 at each fixed monomial $(l_{11}^+)^k\vartheta^{n-2k,l}$, $(l_{11}^+)^{k+1}l_{11}\phi^{n-2k,l}$  and $(l_{11}^+)^k\psi^{n-2k,l}$ for $k=0,...,\left[n/2\right]$; $n,l\in \mathbb{N}_0$:
\begin{eqnarray}
   &&    \left\{\nu^2 \widetilde{\gamma}^l_{k,n}(l_{11}^+)^k  + 2\nu \widetilde{\gamma}^l_{k+1,n+2} l_{11}(l_{11}^+)^{k+1} + \widetilde{\gamma}^l_{k+2,n+4}\big[l_{11}^2,\,(l_{11}^+)^{k+2}\big] \right\}\phi^{n-2k,l}=0,
\label{recurseq2}\\
\label{recurseq1}
 &&  \left\{\nu \widetilde{\delta}^l_{k,n}(l_{11}^+)^k  + \widetilde{\delta}^l_{k+1,n+2}\big[l_{11},\,(l_{11}^+)^{k+1}\big]\right\}\psi^{n-2k,l} =0\end{eqnarray}
(for $\gamma_{k,n}\equiv  \widetilde{\gamma}^0_{k,n}$, ${\delta}_{k,n}\equiv \widetilde{\delta}^0_{k,n}$).
In case of decomposition (\ref{fronsddm}), (\ref{fronsdog}) the  coefficients  $\widetilde{\gamma}^{l,s}_{k,n}$, $\widetilde{\delta}^{l,s}_{k,n}$ should be determined from more general system at each fixed monomial $(l_{11}^+)^{k+s}\vartheta^{n-2k-s,l+s}$, $(l_{11}^+)^{k+s}l_{11}\phi^{n-2k-s,l+s}$  and $(l_{11}^+)^{k+s}\psi^{n-2k-s,l+s}$ for $k=0,...,\left[n/2\right]$, $s=0,...,\left[n/2\right]-k$; $n, l \in \mathbb{N}_0$:
\begin{eqnarray}
\hspace{-0.85em}&\hspace{-0.85em}& \hspace{-0.85em}   \left\{\hspace{-0.2em}\nu^2 \widetilde{\gamma}^{l,s}_{k,n}(l_{11}^+)^{k+s}  \hspace{-0.1em}+ 2\nu \widetilde{\gamma}^{l,s}_{k+1,n+2} l_{11}(l_{11}^+)^{k+s+1}\hspace{-0.1em} + \widetilde{\gamma}^{l,s}_{k+2,n+4}\big[l_{11}^2,\,(l_{11}^+)^{k+s+2}\big] \hspace{-0.2em}\right\}\hspace{-0.2em}\phi^{n-2k-s,l+s}\hspace{-0.2em}=0,
\label{recurseq22}\\
\hspace{-0.5em} &\hspace{-0.5em}& \hspace{-0.5em} \left\{\nu \widetilde{\delta}^{l,s}_{k,n}(l_{11}^+)^{k+s}  + \widetilde{\delta}^{l,s}_{k+1,n+2}\big[l_{11},\,(l_{11}^+)^{k+s+1}\big]\right\}\psi^{n-2k-s,l+s} =0, \label{recurseq33}
\end{eqnarray}
(for $\gamma^s_{k,n}\equiv  \widetilde{\gamma}^{0,s}_{k,n}$, ${\delta}^s_{k,n}\equiv \widetilde{\delta}^{0,s}_{k,n}$).
The solution for the system (\ref{recurseq2}), (\ref{recurseq1}) is found in the form
\begin{eqnarray}
  \hspace{-0.75em} &\hspace{-0.75em}& \hspace{-0.75em} \widetilde{\gamma}^l_{k,n}  =  \frac{(-\nu)^k\ \widetilde{\gamma}^l_{0,n-2k}}{4^k k! \prod\limits_{i=1}^k\big[(n-2k-l+ d/2 -1\big)+i-1\big] } =  \frac{(-\nu)^k\ \widetilde{\gamma}^l_{0,n-2k}}{4^k k!\big(n-2k-l+ d/2-1  \big)_k }  \label{gammas}
      ,\\
  \hspace{-0.75em} &\hspace{-0.75em}& \hspace{-0.75em} \widetilde{\delta}^l_{k,n}  =  \frac{(-\nu)^k\ \widetilde{\delta}^l_{0,n-2k}}{4^k k!\prod\limits_{i=1}^k\big[(n-2k-l+ d/2  \big)+i-1\big] } =  \frac{(-\nu)^k\ \widetilde{\delta}^l_{0,n-2k}}{4^k k!\big(n-2k-l+ d/2  \big)_k }
       \label{deltas}
\end{eqnarray}
with arbitrary constants   $\widetilde{\delta}^l_{0,n}$, $\widetilde{\gamma}^l_{0,n}$ concrete choice of which depends on $n, l$, $d$ and with $(x)_n$ being by the Pochhammer symbol. The coefficients related as: $\big({\widetilde{\delta}^l_{k,n-1}}/{\widetilde{\delta}^l_{0,n-1-2k}}\big)$ = $\big({\widetilde{\gamma}^l_{k,n}}/{\widetilde{\gamma}^l_{0,n-2k}}\big)$.  The solution (\ref{gammas}) for (\ref{recurseq2}) follows from the recursive relations \begin{eqnarray}
  &&  \nu^2\widetilde{\gamma}^l_{k,n}\hspace{-0.15em}  + \hspace{-0.15em}8 (k+\hspace{-0.15em}1) \nu \widetilde{\gamma}^l_{k+1,n+2} \big[n-k-l+d/2 \big]  \nonumber \\
  && \qquad + 4^2\widetilde{\gamma}^l_{k+2,n+4}\prod_{i=k}^{k+1}(i+1)\big[i+ n-2k-l +d/2\big] =0,
\label{solrecurseq2}\\
\label{solrecurseq1}
&&  \left\{ 2\nu\widetilde{\gamma}^l_{k+1,n+2}  + 8(k+2)\widetilde{\gamma}^l_{k+2,n+4}\big[k+ n-2k-l+d/2 \big] \right\} l_{11}  =0,
\end{eqnarray}
with account for:
\begin{equation}\label{auxg0}
g_0\vartheta^{n-2k,l} = \big( (n-2k-l)+d/2 \big) \vartheta^{n-2k,l}, \ \ \vartheta \in \{\phi, \psi \}
.
\end{equation}
Substituting $\widetilde{\gamma}^l_{k+2,n+4}$ expressed from (\ref{solrecurseq1}) in terms of $\widetilde{\gamma}^l_{k+1,n+2}$ in (\ref{solrecurseq2}) we get (\ref{gammas}).

Note, as to the general systems (\ref{recurseq22}), (\ref{recurseq33}), that due to the ambiguity in the definition of the monomial
\begin{equation}\label{ambdef}
  (l_{11}^+)^{k+s}\vartheta^{n-2k-s,l+s} = (l_{11}^+)^{(k+m)+(s-m)}\vartheta^{(n+m)-2(k+m)-(s-m),(l+m)+(s-m)}, \ \ m=1,\ldots, s,
\end{equation}
the respective coefficients $\widetilde{\gamma}^{l,s}_{k,n}$, $\widetilde{\delta}^{l,s}_{k,n}$ should satisfy to the relations:
\begin{equation}\label{ambdef2}
\widetilde{\gamma}^{l,s}_{k,n}   = \widetilde{\gamma}^{l+m,s-m}_{k+m,n+m}, \ \ \widetilde{\delta}^{l,s}_{k,n} = \widetilde{\delta}^{l+m,s-m}_{k+m,n+m}, \ m=1,...,s.
\end{equation}
One can easily  see that the solutions for (\ref{recurseq22}), (\ref{recurseq33})  can be found analogously to (\ref{gammas}), (\ref{deltas})  in the form:
\begin{eqnarray}
 \hspace{-0.9em}  &\hspace{-0.9em}&\hspace{-0.9em} \Big( \hspace{-0.1em} \widetilde{\gamma}^{l,s}_{k,n},\, \widetilde{\delta}^{l,s}_{k,n} \hspace{-0.1em}\Big)  \hspace{-0.1em} =   \hspace{-0.1em} \frac{(-\nu)^k}{4^k k!}\Big(\hspace{-0.1em}\frac{ \ \ \widetilde{\gamma}^{l,s}_{0,n-2k}}{\big(n-2(k+s)-l+ d/2-1  \big)_k },\,     \frac{\ \ \widetilde{\delta}^{l,s}_{0,n-2k}}{\big(n-2(k+s)-l+ d/2  \big)_k }\hspace{-0.1em}\Big) \hspace{-0.1em},     \label{deltasg}
\end{eqnarray}
so that $\widetilde{\gamma}^{l,0}_{k,n}\equiv \widetilde{\gamma}^{l}_{k,n}$, $\widetilde{\delta}^{l,0}_{k,n}\equiv \widetilde{\delta}^{l}_{k,n}$.

 Now, substituting, instead of  $\Phi, \chi_j$, $ \varsigma $, for  $j=1,2$ theirs presentations in terms of series of respective  traceless: $\chi_{F|1}^{n,l} $,  $n,l\in \mathbb{N}_0$  and double traceless tensor fields: $w_i^{n,l}$, $w_i \in \big\{\phi, \chi_{F|2} \big\} $, $i=1,2$,   as well as the gauge parameters $\epsilon^{n,l}$:
   \begin{eqnarray}\label{fronsdd0}
 \hspace{-0.7em}  &\hspace{-0.7em}& \hspace{-0.7em} W_i(\omega) = \hspace{-0.15em} \sum_{n\geq 0} \sum_{k\geq 0}^{\left[n/2\right]}\hspace{-0.15em} \sum_{l\geq 0}\frac{(-\nu)^k\ \widetilde{\gamma}^l_{0,n-2k}}{4^k k!\big(n\hspace{-0.15em}-\hspace{-0.15em}2k\hspace{-0.15em}-l+ \hspace{-0.15em}d/2\hspace{-0.15em}-\hspace{-0.15em}1  \big)_k }(l_{11}^+)^k w_i^{n-2k,l}, \   l^2_{11}w_i^{n,l} =0, \\
  \hspace{-0.5em} &\hspace{-0.5em}&\hspace{-0.5em} \chi_1(\omega) = \hspace{-0.15em} \sum_{n\geq 0} \sum_{k\geq 0}^{\left[n/2\right]} \sum_{l\geq 0}\frac{(-\nu)^k\ \widetilde{\delta}^l_{0,n-2k}}{4^k k!\big(n-2k-l+ d/2  \big)_k } (l_{11}^+)^k \chi_{F|1}^{n-2k,l}, \   \ l_{11}\chi_{F|1}^{n-2k,l} =0,\, \label{fronsdo1}
 \\
 \hspace{-0.5em} &\hspace{-0.5em}&\hspace{-0.5em} \varsigma(\omega) = \hspace{-0.15em} \sum_{n\geq 0} \sum_{k\geq 0}^{\left[n/2\right]}\sum_{l\geq 0} \frac{(-\nu)^k\ \widetilde{\delta}^l_{0,n-2k}}{4^k k!\big(n-2k-l+ d/2  \big)_k } (l_{11}^+)^k \epsilon^{n-2k,l}, \   \ l_{11}\epsilon^{n,l} =0, \label{fronsdo2}
    \end{eqnarray}
(for $W_i\in\big\{\Phi, \chi_2 \big\}$)   we get gauge-invariant non-Lagrangian duplet-like  (\ref{eta1mrrd}), (\ref{eta1etamPrrd})  and with expressed  field  $\chi_1$, $\chi_1=1/2 m_{11}\Phi$, (\ref{eta1mrrdd}) (\ref{eta1etamPrrdd}), (\ref{constrchi2phi})  (with the  fields ${\Phi}$, $\chi_2$ and ${\phi}^{n,l}$,  ${\chi}_{F|2}^{n,l}$) formulations  in terms of Fronsdal-like totally-symmetric standard (for $l=0$) and new (for $l>0$) fields. Representing   the expression (\ref{fronsdd0}) in powers of  $\omega^{(m)_k}\frac{\omega^{(n)_l}}{\omega^{2l}}$  we may derive the latter gauge-invariant EoM from the action $\mathcal{S}_{C|\Xi} \left(\Phi, \chi_2\right)$  (\ref{costrlagrBRST})  with help of the  Lagrangian multipliers $\hat{\lambda}^{{l}|(m)_k,(n)_l}_i$, $i=1,2,3$, $k,l\in \mathbb{N}_0$. Because of the   multipliers are  double $m_{11}$-traceless  as well (\ref{fullconstr}) they should be expressed according to the above receipt (\ref{fronsdd0}) in terms of  double $l_{11}$-traceless Fronsdall-like multipliers $\hat{\lambda}^{{l}|(m)_k,(n)_l}_{F|i}$. The constrained gauge-invariant action (\ref{costrlagrBRST}) for CSR scalar field  of CS $\Xi$  presenting in terms of Fronsdal-like fields  in the tensor form  (but in  Bargmann--Wigner approach) in this case  looks as
   \begin{eqnarray}
 \hspace{-0.65em}&\hspace{-0.65em}&\hspace{-0.65em}   \mathcal{S}_{C|\Xi} \left(\hspace{-0.15em}\phi^{n,l}, \chi_{F|2}^{n,l}, \hat{\lambda}^{n,l}_{F|i}\hspace{-0.15em}\right) =   \mathcal{S}_{C|\Xi} \left(\hspace{-0.15em}\Phi, \chi_2, \hat{\lambda}_i\right)\vert_{\left(\Phi, \chi_2, \hat{\lambda}^{{l}|(m)_k,(n)_l}_{i} \hspace{-0.15em}\right) = \left(\hspace{-0.15em}\Phi(\phi^{n,l}) , \chi_2(\chi_{F|2}^{n,l} ), \hat{\lambda}^{n,l}_{i} (\hat{\lambda}^{n,l}_{F|i})\hspace{-0.15em}\right)}\hspace{-0.15em} . \label{costrlagrBRSTFr}
 \end{eqnarray}
   To compare these results with triplet, duplet and Fronsdal LFs for the fields with all integer spins  we remind that the latters are encoded by only the D'Alambert,  divergentless   and usual  traceless (for $\nu=0$)   equations (\ref{Eqb01q}) for the basic field $\phi^{s}\equiv \phi^{s,0}$  of any integer spin $s$, $s \in \mathbb{N}_0$: $g_0\phi^s = (s+ \frac{d}{2}) \phi^s$, without presence of new fields $\phi^{\bullet, l}$, $l>0$.

The respective constrained gauge-invariant LFs for the massless field, $\phi^s$ of integer spin $s$ in terms, first,  of triplet: $\phi^s$,
$\chi_0^{s-1}$, $\chi_1^{s-2}$,  second, of  duplet  $\phi^s$, $\chi_1^{s-2}$  (having expressed of  $\chi_0^{s-1}$, being similar to  $\chi_0(\omega)$  (\ref{chifconst0}), from triplet formulation through  algebraic EoM) with indices $s$, $s-1$, $s-2$  meaning  the rank of  the component Lorentz tensors, i.e. $\mathrm{deg}_{\omega}\phi(\chi)^{s} = s$ (\ref{fronsdd}),  and, third, in terms of unique field $\phi^s$  look as
 \begin{eqnarray}
\hspace{-0.7em}  &\hspace{-0.7em}& \hspace{-0.7em} \label{Sclsrtri}  \mathcal{S}_{C|s}\left({\phi}, {\chi_0}, {\chi_1}\right)  =  \left( \phi^s(\partial_\omega)     \chi_0^{s-1}(\partial_\omega)\  \chi_1^{s-2}(\partial_\omega)   \right)\hspace{-0.3em}\left(\hspace{-0.5em}\begin{array}{ccc}
  l_0 & - l_1^+ & 0 \\
 -l_1 & 1  & l^+_1\\
 0 & l_1  & -l_0  \end{array}\hspace{-0.5em}\right)\hspace{-0.3em}
       \left( \hspace{-0.3em} \begin{array}{c}{\phi}^s(\omega)\\ {\chi_0}^{s-1}(\omega) \\ {\chi_1}^{s-2}(\omega)  \end{array} \hspace{-0.3em}\right) \hspace{-0.15em}, \\
            &&   \quad   \delta \left( \phi^{s}(\omega) , \chi_0^{s-1}(\omega),     \chi_1^{s-2}(\omega)\right) = \left( l_1^+ ,l_0, l_1\right) \epsilon^{s-1}(\omega), \nonumber \\
           &&  \quad   \mathrm{and}\ \
            l_{11}\left({\phi}(\omega), {\chi_0}(\omega), {\chi_1}(\omega),  \epsilon(\omega)\right) = (-2{\chi_1}(\omega),0,0,0 ) ; \label{gaugetripletp}\\
\hspace{-0.5em}  &\hspace{-0.5em}& \hspace{-0.5em} \label{Sclsrdou}  \mathcal{\hat{S}}_{C|s}\left({\phi}, {\chi_1}\right)  =  \left( \phi^s (\partial_\omega)  \  \chi_1^{s-2}(\partial_\omega)    \right)\left(\begin{array}{cc}
  l_0-l_1^+l_1 & (l_1^+)^2 \\
 l_1^2 & -l_0 -  l_1l_1^+     \end{array}\right)
       \left(  \begin{array}{c}{\phi}^s(\omega)\\ \chi_1^{s-2}(\omega)   \end{array} \right);\\
\hspace{-0.5em}  &\hspace{-0.5em}& \hspace{-0.5em} \label{Sclsrsingl} \mathcal{\hat{S}}^0_{C|s}\left({\phi}\right)  =  \phi^s(\partial_\omega)   \left(  l_0-l_1^+l_1  -\frac{1}{2}(l_1^+)^2l_{11}
 -\frac{1}{2}l^+_{11}l_1^2  -\frac{1}{4}l^+_{11}(l_0 + l_1l_1^+)l_{11}    \right)
       {\phi}^s(\omega), \\
            &&   \quad   \delta  \phi^{s}(\omega)  =  l_1^+  \epsilon^{s-1}(\omega)\ \   \mathrm{and}\ \
            l^2_{11}{\phi}^s(\omega) = l_{11}\epsilon^{s-1}(\omega) = 0, \label{gaugesingle}
            \end{eqnarray}
            for $\mathcal{\hat{S}}_{C|s} = \mathcal{{S}}_{C|s}\vert_{\chi_0=\chi_0(\phi,\chi_1)}$ and $\mathcal{\hat{S}}^0_{C|s} = \mathcal{\hat{S}}_{C|s}
            \vert_{\chi_1=-(1/2)l_{11}\phi}$.
Thus,  the gauge-invariant actions
\begin{eqnarray}\label{infintactions}
&& \Big(\mathcal{S}_{C|\infty}(\phi^0,\chi^0_0, \chi^0_1),\ \mathcal{\hat{S}}_{C|\infty}(\phi^0, \chi^0_1),\ \mathcal{{\hat{S}}}^0_{C|\infty}(\phi^0)\Big) \  =  \ \sum_{s\geq 0}\Big(\mathcal{S}_{C|s}, \mathcal{\hat{S}}_{C|s}, \mathcal{{\hat{S}}}^0_{C|s}\Big),\\
 && \qquad \mathrm{with} \  \big(\phi^0, \chi^0_0,  \chi^0_1\big)(\omega) =\Big(\sum_{s\geq 0}\phi^s,\sum_{s\geq 1}\chi_0^{s-1},\sum_{s\geq 2} \chi_1^{s-2}\Big)(\omega)  \nonumber
\end{eqnarray}
according to the rules (\ref{presentaPhi}), (\ref{fronsdoom})   for massless fields of all spins $s=0$, $1$, $2$,$\ldots$  take  in the ghost-independent  vector-like notations  the respective forms: (\ref{Sclsrtri}), (\ref{Sclsrdou}), (\ref{Sclsrsingl}) with allowance made for the changes $\big(\phi^s , \chi_0^{s-1}$, $\chi_1^{s-2}\big)$ $\to  \big(\phi^0, \chi^0_0,  \chi^0_1\big)$.  The corresponding gauge transformations (\ref{gaugetripletp}),  (\ref{gaugesingle})  are now written for the fields of all integer spins with  the gauge parameter: $\epsilon^0 =\sum_{s\geq 1}\epsilon^{s-1} $, with the same forms of the traceless constraints.

In the  tensor  notations the latter duplet and Fronsdal LFs read (up to the common factor $(1/2)$):
 \begin{eqnarray}
   \hspace{-0.7em}&\hspace{-0.7em}&\hspace{-0.7em} \mathcal{\hat{S}}^0_{C|\infty}(\phi^0, \hspace{-0.1em}\chi^0_1)   =\hspace{-0.2em} \sum_{s\geq 0}\hspace{-0.2em}\frac{(-1)^s}{s!}\hspace{-0.2em} \int \hspace{-0.15em}d^dx  \hspace{-0.1em}\Bigg\{\hspace{-0.15em}\phi_{(m)_s}\hspace{-0.1em}\Big(\hspace{-0.15em}\partial^2\phi^{(m)_s}\hspace{-0.15em} -
  s \partial^{m_s}\partial^{n}\phi^{(m)_{s-1}}{\hspace{-0.2em}}_{n}\hspace{-0.1em}  + \hspace{-0.1em} {s(s-1)}\partial^{m_{s-1}}\partial^{m_s}\chi_1^{\hspace{-0.1em}(m)_{s-2}}\hspace{-0.15em}\Big)  \nonumber\\
   \hspace{-0.5em}&\hspace{-0.5em}&\hspace{-0.5em} \ \ -
   {s(s-1)} \chi_{1\,(m)_{s-2}}\Big(2\partial^2\chi_1^{(m)_{s-2}} +  (s-2)\partial^{m_{s-2}}\partial^{m}\chi_1^{(m)_{s-3}}{\hspace{-0.2em}}_{m}- \partial_{m_{s-1}}\partial_{m_s}\phi^{(m)_{s}} \Big)  \Bigg\}  \label{doubl},\\
 \hspace{-0.7em}&\hspace{-0.7em}&\hspace{-0.7em} \mathcal{\hat{S}}^0_{C|\infty}(\phi^0)    =\hspace{-0.2em} \sum_{s\geq 0}\hspace{-0.2em}\frac{(-1)^s}{s!}\hspace{-0.2em} \int \hspace{-0.15em}d^dx  \hspace{-0.1em} \Bigg\{\hspace{-0.15em}\phi_{(m)_s}\Big(\hspace{-0.15em}\partial^2\phi^{(m)_s}\hspace{-0.15em} -
  s \partial^{m_s}\partial^{n}\phi^{(m)_{s-1}}{\hspace{-0.2em}}_{n}\hspace{-0.1em}+\hspace{-0.1em} {s(s-1)}\partial^{m_{s-1}}\partial^{m_s}\phi^{(m)_{s-2}m}{}_{m}\hspace{-0.15em}\Big)   \nonumber\\
   \hspace{-0.5em}&\hspace{-0.5em}&\hspace{-0.5em} \  -
   \frac{1}{2}{s(s-1)}\phi_{(m)_{s-2}m}{}^{m}\Big(\partial^2\phi^{(m)_{s-2}n}{}_{n} + \frac{1}{2} (s-2)\partial^{m_{s-2}}\partial^{m}\phi^{(m)_{s-3}}{}_{mn}{}^{n}  \Big) \Bigg\}   \label{Fronsdal},\end{eqnarray}
with  traceless field: $ \sum_{s\geq 2}{\chi_1}^{(m)_{s-2}}$: ${\chi_1}^{(m)_{s-4}m}{}_m =0$,  gauge parameter $ \sum_{s\geq 1}{\epsilon}^{(m)_{s-1}}$: ${\epsilon}^{(m)_{s-3}m}{}_m$ $=0$, and double traceless basic field: $ \sum_{s\geq 0}{\phi}^{(m)_{s}}$: ${\phi}^{(m)_{s-2}m}{}_m = 2{\chi_1}^{(m)_{s-2}} $,  providing the standard form of the gauge transformations:
\begin{equation}\label{gtranintinf}
\delta \Big(\sum_{s\geq 0}\phi_{(m)_{s}}, \sum_{s\geq 2}\chi_{1\,(m)_{s-2}}\Big)= - \sum_{s\geq 0}\Big(\partial_{\{m_s}{\epsilon}_{(m)_{s-1}\}},\, \partial^{m_{s-1}}{\epsilon}_{(m)_{s-1}}\Big),
\end{equation}
 from (\ref{gaugetripletp}), (\ref{gaugesingle}).
To be complete, note the constrained BRST-BFV LF for HS field, $\phi^s$, of integer spin $s$ are given by the relations:
 \begin{eqnarray}
 \hspace{-0.6em}&\hspace{-0.7em}& \hspace{-0.7em} {\cal{}S}_{C|s}\left({\phi}^0, \hspace{-0.1em}\chi_0^0, \hspace{-0.1em}\chi_1^0\right)\hspace{-0.1em}= \hspace{-0.1em}\int \hspace{-0.1em} d \eta_0 \chi^{0{}s}_{C}(\partial_\omega) Q_{C|\mathrm{int}} \chi^{0s}_{C}(\omega) ,\, \  \delta\big(\chi^{0s}_{C}, \chi^{1s}_{C}\big)(\omega)
= \big(Q_{C|\mathrm{int}}\chi^{1s}_{C}(\omega)
 ,0\big), \label{LintB}\\
 \hspace{-0.6em}&\hspace{-0.7em}& \hspace{-0.7em} \widehat{L}_{11}\chi^{k{}s}_C = \Big({l}_{11}+ 2\eta_{1} \mathcal{P}_{1} \Big)\chi^{k{}s}_C = 0, \  \widehat{\sigma}_{C|\mathrm{int}}(g)\chi^{ks}_C = \left(s+ \frac{d}{2}\right) \chi^{k{}s}_C,\ \ k=0,1, \label{constrintB}
\end{eqnarray}
which are related with ones (\ref{LcontinB})  for  CS field $\Phi(\omega) $ as follows:
\begin{equation}\label{relcontint}\hspace{-0.7em}
\left(Q_{C|\mathrm{int}},\, \sum_{s\geq 0}\chi^{k{}s}_C,\, \widehat{L}_{11}+ \nu, \widehat{\sigma}_{C|\mathrm{int}}(g)\right)  =  \left(Q_{C},\, {\chi^{0 k}_C}^0,\, \widehat{M}_{11},\, \widehat{\sigma}_C(g)\right)\vert_{ (\eta_1^m = \eta_1^{m+}=\mathcal{P}_1^{m+}=\mathcal{P}_1^m = 0 )},  \end{equation}
where  ${\chi^{0 k}_C} = {\chi^{0 k}_C}^0+\sum_{l>0}{\chi^{0 k}_C}^l$ and $\chi^0_{C}(\partial_\omega)$ is dual for $\chi^0_{C}(\omega)$ with respect to the natural  scalar product in the respective Hilbert space $\mathcal{H}_C$.  Explicit comparison of the  duplet LF (\ref{doubl}), (\ref{gtranintinf}) for HS fields of all integer spins and duplet-like LF (\ref{costrlagrBRST}), (\ref{gaugetransfconstrBRST}), (\ref{fullconstr}) for CS field shows its  difference both by the contents of the configuration spaces, due to presence, first, of "negative spin value" fields: $\big(\Phi^{\bullet,l}, \chi^{\bullet,l}_1, \chi^{\bullet,l}_2\big)$, second, by     the field $\big(\sum_{s,l\geq 0} \chi^{s,l}(\omega) \big)$ in the latter, third, by the Lagrangian multipliers presence for CS field, by the structure of the constraints among the fields. The only EoM (\ref{eta1mrrd})  for the standard fields $\Phi^0(\omega),  \chi_1^0(\omega)$ have  the  Lagrangian form (without using   Lagrangian multipliers) and when rewritten in terms of Fronsdal-like fields $\phi^0(\omega),  \chi_{F|1}^0(\omega)$  may be derived from the functional ${\mathcal{S}}^{0}_{C|\Xi}\big(\Phi^0, \chi_1^0\big)$:
\begin{equation}\label{compdupISCSD}
  {\mathcal{S}}^{0}_{C|\Xi}\big(\Phi^0, \chi_1^0\big)\vert_{\left(\Phi^0, \chi^0_1 \right) = \left(\Phi^0(\phi^{n,0}) , \chi^0_1(\chi_{F|1}^{n.0} )\right)}  = \mathcal{\hat{S}}^0_{C|\infty}(\phi^0, \chi^0_1).
\end{equation}
  We stress, that the main difference concerns the presence of infinite number of new tensor fields with "negative spin values", that, in turn, follows from the Bargmann-Wigner and Fronsdal forms of the equations selecting respectively the CSR and the integer spin representations.
  In case of Fronsdal-like form of the equations (suggested by Shuster and Toro  \cite{ShusterToro})  they can be derived from the Lagrangian BRST-BFV EoM  (without new fields presence) with the constrained Lagrangian formulation  closely related with one for massless totally-symmetric fields for all integer spins (see footnote 3).

\subsection{Equivalence  to Initial Irreducible Relations}\label{reductionC}

Let us preliminarily consider the problem of establishing of the equivalence of the Lagrangian EoM for massless  totally-symmetric field $\phi_{(m)_s}(x)$ with integer spin $s$, in the triplet formulation (\ref{Sclsrtri}), (\ref{gaugetripletp}), which have the form,  when expanding the EoM:  $Q_{C|\mathrm{int}} \chi^{0{}s}_{C}(\omega)  = 0$, in powers of   ghost coordinates (together with  respective traceless constraints (\ref{gaugetripletp})):
\begin{align}
  & \eta_0: \ \  l_0\phi^{s,0}  - l_1^+\chi^{s-1,0}_0=0,
  &&\eta_1^+: \ \ l_1\phi^{s,0} - l_1^+\chi_1^{s-2,0} - \chi_0^{s-1,0}  =0,  \label{trip1}\\
&\eta_0\eta_1^+\mathcal{P}_1^+: \  l_0\chi_1^{s-2,0} - l_1\chi_0^{s-1,0}  =0,&&  l_{11}\hspace{-0.15em}\left(\phi^{s,0}, \chi_0^{s-1,0}, \chi_1^{s-2,0},  \epsilon^{s-1,0}\right) = \hspace{-0.15em}(-\hspace{-0.15em}2\chi_1^{s-2,0},0,0,0 ) \label{trip2}
\end{align}
with non-Lagrangian conditions which should extract the massless UIR  of the Poincare group $ISO(1,d-1)$ with discrete spin $s$ in terms of tensor fields:
\begin{equation}\label{irrepint}
  \big(l_0,\, l_1,\, l_{11}\big)\phi^{s,0} =(0,0,0).
\end{equation}
The conditions (\ref{irrepint}) do not fix completely an ambiguity  in the definition of $\phi^{s,0}$ as a representative of the UIR space of  $ISO(1,d-1)$ group  due to existence of a residual gauge symmetry, which we intend to determine. First of all, we use part of the degrees of freedom from the gauge parameter $\epsilon^{s-1,0}$ to gauge away the field $\chi_1^{s-2,0}$ by means of the gauge transformations (\ref{gaugetripletp}). For $s=0$ the equivalence is trivial, whereas for $s=1$ there is no field $\chi_1^{-1,0}\equiv 0$.  To do so, we expand $\epsilon^{s-1,0}$ into sum of  longitudinal, $\epsilon_L^{s-1,0}$, and transverse, $\epsilon_{\bot}^{s-1,0}$, components:
\begin{equation}\label{decomplotr}
 \epsilon^{s-1,0} = \epsilon_L^{s-1,0}+\epsilon_{\bot}^{s-1,0} \equiv \sum_{k=1}^{s-1} (-1)^{k-1}\frac{(l_{1}^+)^k l_1^k}{k! (l_0)^k}\epsilon^{s-1,0} + \left( 1+ \sum_{k=1}^{s-1} (-1)^{k}\frac{(l_{1}^+)^k l_1^k}{k! (l_0)^k}\right)\epsilon^{s-1,0},
\end{equation}
so that $l_1\epsilon_{\bot}^{s-1,0} \equiv 0$ and both of the components are traceless: $l_{11} \epsilon_L = l_{11}\epsilon_{\bot} =0$. Thus, first, we use only part: $\epsilon_L^{\chi}$  from the parameter  $\epsilon_L^{s-1,0} = \epsilon_L^{\chi}+ \epsilon_L^{\phi}$ for $s\geq 2$ to gauge away the field $\chi_1^{s-2,0}$ completely. So we have, from the stability of the solution  $\chi_1^{s-2,0}=0$ under the gauge symmetry, that $\delta\chi_1^{s-2,0}=0 \Leftrightarrow  l_1\epsilon_L^{s-1,0}=0$ $\Rightarrow  l_1\epsilon_L^{\phi {}s-1,0} = - l_1\epsilon_L^{\chi{}s-1,0}$\footnote{The realization of the first step allows one to get Maxwell-like LF with traceless field $\phi^{s,0}$, when having substituted  $\chi_0^{s-1,0}$ being expressed from the second equation in (\ref{trip1}) into the first one, as follows: $\big(l_0 - l_1^+l_1\big)\phi^{s,0}=0$, so that $\mathcal{S}^0_{C|s}\left({\phi^{0}}\right) =  \phi^{s,0}(\partial_\omega)    \left(  l_0-l_1^+l_1  \right)
       \phi^{s,0}$ with $\delta \phi^{s,0} = l_1^+ \epsilon^{s-1,0}$, $l_1\epsilon^{s-1,0} = 0$. The equivalent reducible  respective LF for the latter  with elimination of the differential constraint on  $\epsilon$ were considered  in \cite{franciaSharapov} among them for AdS space.}.

Second, from the first equation in (\ref{trip2}) we observe that the field $\chi_0^{s-1,0}$ is the transverse  one and we may therefore use the unused parameter $\epsilon_{\bot}^{s-1,0}$ choosing $\epsilon_{\bot}^{s-1,0}=-(l_0)^{-1}\chi_0^{s-1,0}$   to gauge away this field completely, so that the   stability of the solution  $\chi_0^{s-1,0}=0$ under the gauge transformations means, that $\delta\chi_0^{s-1,0}=0 \Leftrightarrow l_0\epsilon^{s-1,0} =0 $.

As the result, from the  equations  (\ref{trip1}), (\ref{trip2}) it follows the validity of the system (\ref{irrepint}) with residual gauge transformations determined by the  longitudinal gauge parameter, $\epsilon_L^{\phi{}s-1,0}$, which satisfy to the same restrictions as the field $\phi^{s,0}$ in (\ref{irrepint}).
Therefore, the conditions which should select the (tensor) field of any spin $s\in \mathbb{N}_0$ as the element of  irreducible massless unitary representation  must be determined as:
 \begin{equation}\label{irrepintnew}
  \big(l_0,\, l_1,\, l_{11}\big)\phi^{s,0} =(0,0,0),\   \, \delta\phi^{s,0} = l_1^+ \epsilon^{s-1,0}, \  \  \big(l_0,\, l_1,\, l_{11}\big)\epsilon^{s-1,0} = (0,0,0).
\end{equation}
 The latter equations on $\epsilon^{s-1,0}$ means that the parameter may be considered as the element of massless UIR of $ISO(1,d-1)$ of spin $s-1$, but without own gauge symmetry\footnote{For the case of mixed-symmetric massless HS field with generalized integer spin $\mathbf{s} = (s_1,...,s_k)$ given on $\mathbb{R}^{1,d-1}$ the conditions of extraction of only UIR of Poincare group $ISO(1,d-1)$ in the space of tensor fields $\phi_{(m^1)_{s_1}...(m^k)_k}(x) \in Y(s_1,...,s_k)$, $k\leq \left[d/2\right]$: $\big(l_0,\, l_i,\, l_{ij}, t_{rs}\big)|\phi\rangle_{(s)_k}$ $ = 0$ being initial in \cite{BuchbResh} (with use of the Fock space notations) should be  augmented according to (\ref{irrepintnew}) by adding the reducible gauge symmetry: $\delta \phi_{(m^1)_{s_1}...(m^k)_k} = l_1^+\epsilon_{(m^1)_{s_1-1}...(m^k)_k}$, ... ,  subject to the same requirements  as for the tensor field itself.  For the half-integer totally- and mixed-symmetric massless  HS fields the situation with the exact formulation of the UIR is the same, e.g. one can show that for totally-symmetric case it is necessary to add the gauge transformations of the same form with gauge spin-tensor of rank $(n-1)$,  but for basic spin-tensor field $\psi_{(m)_{n}}$ of spin, $n+1/2$, with suppressed Dirac indice and being subject to the same conditions: Dirac and $\gamma$-traceless constraints. Thus, the theorem in \cite{Reshetnyak_con} concerning the equivalence of the solutions of the equations of motion from the respective constrained BRST-BFV LF and ones for UIR conditions will be guaranteed, because of the latter solutions contains some gauge identities due to residual gauge symmetry presence.}. Note, first that the dimensional reduction procedure being applied to massless UIR conditions (\ref{irrepintnew}) of $ISO(1,d)$ in $\mathbb{R}^{1,d}$ space-time permits one explicitly derive the massive UIR conditions  of $ISO(1,d-1)$ in $\mathbb{R}^{1,d-1}$ with  the same spin, as follows:
 $\big(l_0 +m^2,\, l_1,\, l_{11}\big)\phi^{s,0} =(0,0,0)$ without any gauge symmetry. Second, the independent counting of the numbers of the physical degrees of freedom being extracted by (\ref{irrepintnew}) and by the equations (\ref{trip1}), (\ref{trip2}) with the gauge symmetry transformations (\ref{gaugetripletp}) shows  their coincidence.

Having in mind, the above analysis for HS field with integer spin,   let us consider
non-Lagrangian EoM for the basic field $\Phi(\omega)$ with CS, which follow from BRST-BFV equation (\ref{Qchi}) (or from (\ref{gtranind})), as well as the  holonomic  constraints.

Again,
it may turn out that the conditions (\ref{irrepconts}) do not fix completely an ambiguity  in the definition of $\Phi(\omega)$ as a representative of the CSR space of $ISO(1,d-1)$ group, due to existence of a residual gauge symmetry, which we should to determine. We will call the equations (\ref{eta0})--(\ref{eta1etamPm})  as the \emph{BRST-unfolded equations}, due to appearance of any field variable there with a coefficient being, at most the first degree in powers of the  symmetry algebra $\mathcal{A}(\Xi;
\mathbb{R}^{1,d-1})$ elements $o_I$\footnote{The analogous type of the BRST-unfolded equations were written in (\ref{trip1}), (\ref{trip2}) for totally-symmetric integer spin case.}.

First of all, we repeat the procedure from the
Section~\ref{BRSTBFV2} of gauge fixing up to surviving of only the   fields $\big( \Phi $,  $\chi_{1}$,  $\chi_{2} $,  $ \chi_{0}  \big)(\omega)$ with   equivalent transforming
of the equations (\ref{eta0})--(\ref{eta1etamPm})  into \emph{triplet-like non-Lagrangian formulation} (\ref{eta1mrr}), (\ref{eta1etamPrr})  with the gauge transformations (\ref{gtrgh0f})  with unique independent gauge parameter $\varsigma(\omega) $ with  account of algebraic traceless constraints (\ref{constrcontinB0})--(\ref{constrcontinB02}).
Second, we  expand $\varsigma $ into sum of  longitudinal, $\varsigma_L(\omega)$, and transverse, $\varsigma_{\bot}(\omega)$, components:
\begin{equation}\label{decompcs}
  \varsigma = \varsigma_L+\varsigma_{\bot} \equiv \sum_{k=1}^{\infty} (-1)^{k-1}\frac{(l_{1}^+)^k l_1^k}{k! (l_0)^k}\varsigma + \sum_{k=0}^{\infty} (-1)^{k}\frac{(l_{1}^+)^k l_1^k}{k! (l_0)^k}\varsigma,
\end{equation}
so that $l_1\varsigma_{\bot} \equiv 0$ and both of the components are generalized traceless: $m_{11} \varsigma_L = m_{11}\varsigma_{\bot} =0$. Then, we use a part $\varsigma_L^{\chi_1}(\omega)$ of the longitudinal gauge parameter:
\begin{equation}\label{decomplongvs}
\varsigma_L  = \varsigma_L^{\chi_1} + \varsigma_L^{\chi_2} + \varsigma_L^{\Phi}
\end{equation}
 to gauge away the field $\chi_1(\omega)$ completely. From the stability: $\delta \chi_1(\omega) =0$, of the solution  $\chi_1(\omega)=0$ under the gauge symmetry,  it follows the relations:
\begin{equation}\label{cont2}
  \delta\chi_1=0 \ \Leftrightarrow\  l_1 \varsigma_L = l_1\varsigma =0\ \Rightarrow \  l_1\varsigma_L^{\Phi} + l_1\varsigma_L^{\chi_2} = - l_1 \varsigma_L^{\chi_1} \  \mathrm{and} \ \ m_{11}\Phi =0.
  \end{equation}
Third,  from  the second  equation in  (\ref{eta1etamPrr}) we obtain that: $l_1\chi_2  = \chi_{0}$ and thus the field  $\chi_2$ is double transverse, due to $l_1\chi_{0}=0$ from the third equation in  (\ref{eta1mrr}). Then, we use the remaining degrees of freedom from the parameter $\varsigma$ (both $\varsigma_L^{\chi_2}\rangle$   and $\varsigma_{\bot}$)
to gauge away the field $\chi_2$ completely. Then, the requirement $\delta\chi_2 =0 $ leads to:
      \begin{equation}\label{cont3}
      m_1^+\big(\varsigma_L^{\chi_2}+\varsigma_L^{\Phi} +\varsigma_{\bot}\big) =  m_1^+\varsigma =0\ \Rightarrow \   m_1^+\varsigma_L^{\Phi} = - m_1^+\big(\varsigma_L^{\chi_2} +\varsigma_{\bot}\big) \  \mathrm{and} \ \ \chi_0  = 0.
 \end{equation}
 As the result, the only initial field  $\Phi$ survives after the  procedure above, satisfying to the relations (\ref{irrepconts}) without any residual gauge transformations due to using of the parameter $\varsigma$ completely. Indeed, from a possible expression $\delta\Phi = l_1^+ \varsigma$ and $\big(l_0,\, l_1,\,m_1^+,\, m_{11}\big)\varsigma = (0,0,0,0)$ it should be that $\delta\Phi = -\imath\Xi  \varsigma$ and it would mean that the field $\Phi$ does not contain any physical degrees of freedom.
It means, that in opposite to the case of integer massless UIR of $ISO(1,d-1)$ one-valued CSR conditions do not include residual gauge transformations. Again, we suppose, that the dimensional reduction when applied to massless CSR conditions (\ref{irrepconts}) in  $\mathbb{R}^{1,d}$ can be used to derive massive-like CSR relations in $\mathbb{R}^{1,d-1}$ for the same value of CS $\Xi$.

 Thus,  we show, that the
CSR equations (\ref{irrepconts})  [or, equivalently,
(\ref{Eqb01}), (\ref{Eqb23})],  can be
achieved by using the BRST-BFV  equations (\ref{LcontinB}) after gauge-fixing and
removing the auxiliary fields by using a total set  of the equations of
motion. We stress that these equations (\ref{LcontinB}) completely equivalent to the Shuster-Toro form \cite{ShusterToro} of the gauge-invariant equations of motion because of they both describe the same  field $\Phi(\omega)$ with CS $\Xi$ from CSR of the Poincare group.

\section{BRST-BV minimal descriptions}
\setcounter{equation}{0}

\label{mincBRSTBV} %%%%%%%%%%%%%%%%%%%%%%%%%%%%%%%%%%%%%%%%%%%%%%%%

To construct a quantum action being sufficient for determination of the non-degenerate path integral within conventional BV quantization  method \cite{BV}, one necessary to derive  preliminarily the so called BV action in the minimal sector of field and antifield variables organized in terms of respective vectors on a  space $\mathcal{V}_{g}$,
when considering instead of the  field vector $\chi^0_C  \in \mathcal{V}_C$ the \emph{generalized field-antifield vector} $\chi_{g|C}  \in  \mathcal{V}_{g|C}$:
\begin{equation}\label{genFock}
\mathcal{V}_{g|C}:
  =\mathcal{V}_g\otimes \mathcal{V}^{o_A}_{gh} \ \ \mathrm{ with \  \mathbb{Z}-grading} \ \   \mathcal{V}_{g|C}=\lim_{M\to \infty} \oplus^M_{l=-M}\mathcal{V}^l_{g|C}
\end{equation}
for $gh_{\mathrm{tot}}(\chi^l_{g|C} )=-l$, $\chi^l_{g|C}  \in \mathcal{V}^l_{g|C} $.
The total configuration space for initial first-stage  reducible gauge constrained LF  in the minimal sector, $\mathcal{M}_{\min}$ = $\{\Phi^A_{\min}(x,\omega)\}$,  contains,  in addition to the field   $\chi^0_C $,  the $0$-level  ghost  field  vector, $C^{0}_C$, and $1$-st level  ghost  field  one,  $C^{1}_C$,  introduced by the rule according to (\ref{presentaPhi}):
\begin{eqnarray}\label{corrchiC2tot}
 &&  \varpi(x,\omega) =C^1 (x,\omega) \mu_0\mu_1 \   \Longrightarrow  \   \chi^{2}_{C}  = C^{1}_C \mu_0\mu_1, \\
\label{corrchiC1tot}
 &&  \varsigma_k(x,\omega) =C^0_k (x,\omega) \mu_0 \ \   \Longrightarrow  \   \chi^{1}_{C}  = C^{0}_C \mu_0, \ \  C^{i}_C = C^{0|i}_C + \sum_{l\geq 1}C^{l| i}_C, \\
 && \begin{array}{|c|cccccc|} \hline
                   &   \mathcal{C}^{A} & {\mathcal{P}}_{A
} &  C^1 (x,\omega)& C^0_k (x,\omega) & C^{i}_C  & \mu_i  \\
                    \hline
                   \epsilon  &1  & 1 & 0 & 1 & 0 & 1 \\
                   gh_H  & 1 & -1 & 0 & 0 & -1-i & 0\\
                                      gh_L  & 0 & 0 & 2 & 1 & i+1 & -1 \\
                                                         gh_{\mathrm{tot}}  & 1 & -1 & 2 & 2  & 0 & -1\\
                  \hline \end{array},\ \  i=0,1;  \label{grassghtot}    \end{eqnarray}
(where under $\varsigma_k $ and $C^0_k$ we mean all component fields in $\chi^{1}_{C}$  (\ref{chifconst11}) and with constant $\mu_i$: $\{\mu_i,\,\mu_j\}=0$, $i,j=0,1$),  which due to the vanishing of the total ghost number and Grassmann parity may be combined with $ \chi^{0}_C$ into  \emph{generalized field vector}:
\begin{eqnarray}\label{genvectortot}
 \chi^{0}_{\mathrm{gen}|C} = \chi^{0|0}_{\mathrm{gen}|C}  +  \sum_{l\geq 1}\chi^{l|0}_{\mathrm{gen}|C}   =   \chi^{0}_C  + \sum_{i=0}^1 C^{i}_C ,  \ \    \big(\epsilon, gh_{\mathrm{tot}}\big)\chi^{0}_{\mathrm{gen}|C}=(0,0).
\end{eqnarray}

The corresponding (according to (\ref{chifconst1}))  antifields  $ \Phi^*_{A{}\min}(x,\omega)$ = $ \big(
\Phi^*_{ n_{f 0};   n_{f 1}, n_{f m},n_{p 1}, n_{p m}, 0, 0, 0}$; $C^{*0}_k, C^{*1} \big)(x,\omega)$ and respective  space vectors from $\mathcal{V}^0_{g|C}$ with the $\mathbb{Z}_2-$, and $\mathbb{Z}-$gradings
 \begin{eqnarray}
 && \begin{array}{|c|ccccc|} \hline
                   &   \Phi^*_{ n_{f 0}; ...}(x,\omega) & C^{*0}_k (x,\omega)  &   C^{*1} (x,\omega) & \chi^{*0}_C  & C^{*i}_C  \\
                    \hline
                   \epsilon  &1  & 0 & 1 & 0 & 0 \\
                   gh_H  & 0 &  0 & 0 & 1 & 2+i\\
                                      gh_L  & -1 & -2 & -3 &  -1 & -2-i \\
                                                         gh_{\mathrm{tot}}  & -1 & -2 & -3   & 0 & 0\\
                  \hline \end{array},\ \  i=0,1;  \label{grassghtotant}    \end{eqnarray}
are combined into \emph{generalized  antifield  vector}  as follows:
\begin{eqnarray}\label{genavectortot0}
  \chi^{*0}_{\mathrm{gen}|C} & =& \chi^{*0}_C +\sum_{i=0}^1C^{*i}_C  =  \left\{B^{*0}_C + B^{*0}_{c|C} \right\} + \eta_0 \left\{S^{*0}_C + \sum_{i\geq 0} S^{*i}_{c|C} \right\} ,    \\
 \chi^{*0}_C(\mathcal{C},\mathcal{ P}, \omega)  & =&\eta_1^+ \chi^*_0 (\omega) +\eta_1^{m}\chi_0^{*m} (\omega)  + \eta_1^+\mathcal{P}_1^+ \eta_1^{m}\chi_{01}^* (\omega) +\eta_1^{m}\eta_1^+ \mathcal{P}_1^{m}\chi_{01}^{*m} (\omega)  \label{afield0}\\
&&
 + \eta_0\hspace{-0.15em}\Big[\Phi^* +\hspace{-0.15em}\mathcal{P}_1^+\Big(\hspace{-0.15em}\eta_1^+ \chi^*_1 (\omega) +\hspace{-0.15em}\eta_1^{m}\chi_1^{*m} (\omega) \Big)\hspace{-0.15em}+  \hspace{-0.15em}\mathcal{P}_1^{m}\Big(\hspace{-0.15em}\eta_1^+ \chi_{2}^* (\omega) +\hspace{-0.15em}\eta_1^{m}\chi_{2}^{*m} (\omega) \hspace{-0.15em}\Big) \nonumber  \\
&& + \eta^+_1 \eta_1^{m}\mathcal{P}_1^+ \mathcal{P}_1^{m}\chi_{11}^{*m} (\omega) \Big]\,, \nonumber\\
\label{achifconst11} C^{*0}_C(\mathcal{C},\mathcal{ P}, \omega) & =& \eta_0\Big(\eta_1^+ C_\varsigma^* (\omega) +\eta_1^{m}C_\varsigma^{*m} (\omega) + \eta_1^+\mathcal{P}_1^+ \eta_1^{m}C_{\varsigma|01}^* (\omega)\\
&&+\mathcal{P}_1^m\eta_1^+ \eta_1^{m}C_{\varsigma|11}^* (\omega) \Big)+\eta_1^+ \eta_1^{m}C_{\varsigma|0}^* (\omega)  ,\nonumber \\
 C^{*1}_C(\mathcal{C},\mathcal{ P}, \omega)  & =& \eta_0
\eta_1^+ \eta_1^{m}C_\varpi^{*} (\omega) , \label{achifconst111}
\end{eqnarray}
for $ B^{*1}_{c|C} \equiv 0$ and $ \chi^{*0}_{\mathrm{gen}|C}$ = $ \chi^{0*0}_{\mathrm{gen}|C} +  \sum_{l\geq1}\chi^{l*0}_{\mathrm{gen}|C}$.   The ghost  independent antifield vectors
have the decompositions in powers of $\omega^m$ similar to (\ref{presentaPhi}) and (\ref{npvec1})  as for the respective field vectors.
The generalized field (\ref{genvectortot}) and antifield  (\ref{genavectortot0}) vectors  are uniquely written in terms of the  generalized field-antifield vector:
\begin{eqnarray}\label{genafvectorts}
 \chi^{0}_{\mathrm{g}|C} & =& \chi^{0}_{\mathrm{gen}|C}   + \chi^{*0}_{\mathrm{gen}|C}  \ = \   \chi^{0|0}_{\mathrm{g}|C} + \sum_{l\geq1}\chi^{l|0}_{\mathrm{g}|C} , \ \   \big(\epsilon, gh_{\mathrm{tot}}\big)\chi^{0}_{\mathrm{g}|C}=(0,0).
\end{eqnarray}
The presentation of the  constrained minimal BV  actions are different for the case of   Lagrangian form of BRST-BFV EoM, $Q_{C|\mathrm{int}} \chi^{0\infty}_{C}(\omega)=0$  as for the field  $\chi^{0\infty}_{C}=\sum_{s\geq 0}\chi^{0s}_{C}$,of all integer spin $s=0,1,...$ (\ref{LintB}) and for the non-Lagrangian BRST-BFV EoM  (\ref{LcontinB})  as for the free massless   field ${\Phi}(x,\omega)$ of CS $\Xi$ (for $\nu=1$) in  $\mathbb{R}^{1,d-1}$. In the former case the minimal action $S_{\min}\equiv S_{C|\infty} $  is  given according to the general prescription: $S_{\min}=\mathcal{S}_0+\Phi^*_{A{}\min}\overrightarrow{s}\Phi^{A}_{\min}$ \cite{BV},  with account for specific of the vector  space $\mathcal{V}^0_{g|C}$, now being endowed by Fock space structure, irreducibllity of the gauge theory   and reality of $S_{C|\infty}$:
\begin{eqnarray}\label{BVprescripint}
 S_{C|\infty} &=&  \mathcal{S}_{C|\infty} + \int d \eta_0 \left\{\chi^{*\infty}_{C}(\partial_\omega)\overrightarrow{s}_\infty C^{0\infty}_{C}+  C^{0\infty}_{C}(\partial_\omega)\overleftarrow{s}_\infty \chi^{*\infty}_{C}(\omega) \right\},\\
 && \chi^{*\infty}_{C}(\omega)\ = \
  \eta_0\big[\Phi^{*\infty} +\mathcal{P}_1^+\eta_1^+ \chi^{*\infty}_1 (\omega)\big]
+ \eta_1^+ \chi^{*\infty}_0 (\omega),  \label{vantfint}
\end{eqnarray}
 with right $\overleftarrow{s}_\infty$ (left $\overrightarrow{s}_\infty$) generator of \emph{ Lagrangian BRST-like transformations} in the minimal sector of the fields combined within the generalized  field $\chi^{0\infty}_{\mathrm{gen}|C}(\omega)$:
 \begin{equation}\label{BRSTLagrint}
 \chi^{0\infty}_{\mathrm{gen}|C} =    \chi^{0\infty}_C  +  C^{\infty}_C ,\ \
   \delta_B \chi^{0\infty}_{\mathrm{gen}|C} \ = \ \mu \overrightarrow{s}_\infty \chi^{0\infty}_{\mathrm{gen}|C} \ = \ \mu Q_{C|\mathrm{int}}  \chi^{0\infty}_{\mathrm{gen}|C}.
 \end{equation}
For dual vector, $\chi^{0\infty}_{\mathrm{gen}|C}(\partial_\omega) = \big(\chi^{0\infty}_{C}+ C^{\infty}_C\big)(\partial_\omega)$, the transformation (\ref{BRSTLagrint})  with account of hermiticity $Q_{C|\mathrm{int}}, \mu$ looks as:
\begin{equation}\label{dBRSTLagr}
\delta_B\chi^{0\infty}_{\mathrm{gen}|C}(\partial_\omega) = \big(\delta_B \chi^{0\infty}_{\mathrm{gen}|C}(\omega)\big)^+: \ \delta_B\chi^{0\infty}_{\mathrm{gen}|C}(\partial_\omega)= \chi^{0\infty}_{\mathrm{gen}|C}(\partial_\omega)\overleftarrow{s}_\infty\mu = \chi^{0\infty}_{\mathrm{gen}|C}(\partial_\omega)Q_{C|\mathrm{int}}\mu   .
\end{equation}
   Explicitly, the action $S_{C|\infty}$ and its BRST-like invariance transformations can be given  in the form
\begin{eqnarray} \label{Sminconfints}
  \hspace{-0.65em}&\hspace{-0.65em}&\hspace{-0.65em}S_{C|\infty}  =  \hspace{-0.25em} \int\hspace{-0.15em} d \eta_0 \chi^{0\infty}_{\mathrm{g}|C}(\partial_\omega)Q_{C|\mathrm{int}}\chi^{0\infty}_{\mathrm{g}|C}(\omega) ,\   \chi^{0\infty}_{\mathrm{g}|C}(\omega)  =  \big(\chi^{0\infty}_{\mathrm{gen}|C}   + \chi^{*0\infty}_{\mathrm{gen}|C}\big) (\omega) ,\ \delta_B S_{C|\infty} = 0, \end{eqnarray}
Here, both the generalized field, $\chi^{0\infty}_{\mathrm{gen}|C}$, and antifield, $\chi^{*0\infty}_{\mathrm{gen}|C}$,  vectors are subject to the off-shell BRST extended constraints  $\widehat{L}_{11}$ (\ref{constrintB}):
\begin{eqnarray}\label{constralgaftsint}
&& \widehat{L}_{11} \chi^{0\infty}_{\mathrm{g}|C} = 0 \  \Longleftrightarrow \  \widehat{L}_{11}\big(\chi^{0\infty}_{\mathrm{gen}|C},\, \chi^{*0\infty}_{\mathrm{gen}|C}\big) = (0,\, 0).
\end{eqnarray}

However for the non-Lagrangian BRST-BFV  description for free massless  CSR  field ${\Phi}(x,\omega)$ in the Bargmann-Wigner representation the minimal BV action, $  S_{\mathrm{min}\Xi}=S_{C|\Xi}$,  may be found with help of only Lagrangian multipliers procedure. We apply it for the duplet-like non-Lagrangian formulation with expressed field $\chi_{1}(\omega)$:  $\chi_{1}(\omega)$= $-\frac{1}{2}m_{11}\Phi(\omega)$  (\ref{eta1mrrdd}) --(\ref{constrchi2phi})  with only  independent gauge-invariant equations of motion derived from the action $\mathcal{S}_{C|\Xi} \left(\Phi, \chi_2, \hat{\lambda}_i \right)$  (\ref{costrlagrBRST}).
The functional
 $S_{C|\Xi}=S_{C|\Xi}\left(\Phi, \chi_2, \hat{\lambda}_i, \Phi^*, \chi^*_2, \hat{\lambda}^*_i, C_\varsigma, C_\sigma \right)$ and its non-vanishing  global invariance transformation (following from (\ref{gaugetransfconstrBRST}), (\ref{gaugetransfconstrDUAL})) \begin{eqnarray}\label{BVprescrip}
\hspace{-0.85em} &\hspace{-0.85em}&\hspace{-0.85em} S_{C|\Xi} =  \mathcal{S}_{C|\Xi} - \int d^dx \sum_{k,l\geq 0} \left\{ \Phi^{*(m)_{k},(n)_l}{\partial_{\{m_{k}}}{C}^{{l}}_{(m)_{k-1}\},(n)_l} \right. \\
 \hspace{-0.65em}&\hspace{-0.65em}&\hspace{-0.65em} \phantom{S_{C|\Xi} =} \left.+
 \chi^{*(m)_{k},(n)_l}_2\Big({\partial_{\{m_{k}}}{C}^{{l}}_{(m)_{k-1}\},(n)_l}-\imath\Xi\eta_{\{m_{k}\{n_{l}} {C_\varsigma}^{{l-1}}_{(m)_{k-1}\},(n)_{l-1}\}}\Big) \right.\nonumber\\
 \hspace{-0.65em}&\hspace{-0.65em}&\hspace{-0.65em} \phantom{S_{C|\Xi} =} \left.+\sum_{k',l'}\hat{\lambda}^{*l}_{i{}(m)_k,(n)_l}\hat{R}^{l{}}_i{}^{(m)_k,(n)_l}_{l'{}(m_1)_{k'}(n_1)_{l'}}(x,\partial_x) C_{\sigma}^{l' (m_1)_{k'}(n_1)_{l'}}(x)  \right. \Big\}, \nonumber \\
\hspace{-1.45em} &\hspace{-1.45em}&\hspace{-1.45em} \delta_{\hspace{-0.1em}B}\hspace{-0.15em}\left(\hspace{-0.15em}\Phi^{{l}},\hspace{-0.15em} \chi^{{l}}_2\right)_{\hspace{-0.15em}(m)_k,(n)_l}\hspace{-0.35em} = \hspace{-0.1em} -\hspace{-0.15em}\Big(\hspace{-0.15em}{\partial_{\{m_{k}}}\hspace{-0.15em}{C_\varsigma}^{{l}}_{(m)_{k-1}\},(n)_l}, {\partial_{\{m_{k}}}\hspace{-0.15em}{C_\varsigma}^{{l}}_{(m)_{k-1}\},(n)_l}\hspace{-0.15em}-\hspace{-0.15em}\imath\Xi \eta_{\{m_{k}\{n_{l}} \hspace{-0.1em}{C_\varsigma}^{{l-1}}_{(m)_{k-1}\},(n)_{l-1}\}}\hspace{-0.2em}\Big)\hspace{-0.15em}\mu \label{BRSTphi}\\
%%%%%%%%%%%%%%%%%%%%%%%%%
\hspace{-1em} &\hspace{-1em}&\hspace{-1em} \delta_B\hat{\lambda}^{l}_{i}{}^{(m)_k,(n)_l}= \sum_{k',l'}\hat{R}^{l{}}_i{}^{(m)_k,(n)_l}_{l'{}(m_1)_{k'}(n_1)_{l'}}(x,\partial_x) C_{\sigma}^{l' (m_1)_{k'}(n_1)_{l'}}\mu ,\label{BRSTlambda}
\end{eqnarray}
solve the problem in the tensor form. Thus, we have derived the
constrained BRST-BV minimal action for an irreducible form of
constrained BRST-BFV LF (\ref{LcontinB}) for free CSR of the
$ISO(1,d-1)$ group described by the field $\Phi(x,\omega)$ and
auxiliary classical $\chi_2(x,\omega)$, the ghost
$C_\varsigma(x,\omega)$, the Lagrangian multipliers
$\hat{\lambda}_i(x)$, $i=1,2,3$, the ghost $C_\sigma$ fields and
theirs antifields subject to the generalized traceless constraints
(\ref{fullconstr}) where the substitution  $\varsigma
\leftrightarrow C_\varsigma$ should be made.

The difference of the BRST--BFV descriptions for the fields describing   all integer spin representations
and the fields for CSR presented in the Subsection~\ref{BRSTBFV3} is inherited for  BRST--BV descriptions as well.
Note, the constrained BRST--BV actions   for the CSR field   in case of Fronsdal-like form of the equations (following to Shuster and Toro  \cite{ShusterToro})   can be derived from the constrained Lagrangian BRST-BFV LF  (without the  fields with "negative spin values") which should be  closely related with the minimal BV action $S_{C|\infty}$  (\ref{Sminconfints})  for massless totally-symmetric fields for all integer spins.

Different BRST-BV minimal actions may be used as the starting points to construct a  LF for the CS field, being interacting both with itself, or with another scalar CS fields and with HS fields with integer spin in $R^{1,d-1}$ on a base of preservation underlying master equation.

\section{Generalized  quartet-like unconstrained   descriptions}
\setcounter{equation}{0}

\label{tripBRSTBFV}
To solve the problem, beyond  of the extension of the constrained BRST-BFV approach to unconstrained one, it is sufficient to start from the  triplet-like non-Lagrangian formulation (\ref{eta1mrr}), (\ref{eta1etamPrr}). We may obtain unconstrained quartet-like non-Lagrangian formulation (following, in part, to idea of \cite{quartmixbosemas} for the case of integer spin) by introducing a compensator field $\vartheta(\omega) $: $\delta\vartheta(\omega) = m_{11}\varsigma(\omega)$. Then we should enlarge the constraints (\ref{constrcontinB0})--(\ref{constrcontinB02})  on the fields $\big(\Phi,  \chi_{1},  \chi_{2} $, $ \chi_{0}  \big)$  with  nontrivial  gauge transformations (\ref{gtrgh0f}) leaving by invariant the EoM    (\ref{eta1mrr}), (\ref{eta1etamPrr}) up to the gauge-invariant equations as follows:
\begin{align}&\    m_{11}\chi_0  -l_0\vartheta  = 0,   &&  m_{11} \chi_{1}   -l_1 \vartheta = 0,    \label{quaconstrcontinB0}\\
  &  m_{11}\chi_{2}  + 2\chi_{1}  - m_1^+\vartheta=0,  &&    m_{11}\Phi + 2\chi_{1} -l_1^+ \vartheta=0.
   \label{quaconstrcontinB02}
 \end{align}
 Introducing four  new sets of  bosonic (real one-valued)  fields $\lambda_j$, $j=1,...,4$,  playing the role of the Lagrangian multipliers for the modified constraints (\ref{quaconstrcontinB0}), (\ref{quaconstrcontinB02})  in addition to ones ${\widetilde{\lambda}}_p$, $p=1,...,6$ for the equations of motion    (\ref{eta1mrr}), (\ref{eta1etamPrr}),  we get an unconstrained LF with the action in the tensor form:
\begin{eqnarray}\hspace{-0.3em}&\hspace{-0.3em}&\hspace{-0.3em} {\cal{}S}_{\Xi} =  {\cal{}S}_{C|\Xi}\big(\Phi,\chi_0, \chi_1, \chi_2;{\widetilde{\lambda}}_p\big)  +  \int d^dx  \sum_{k,l\geq 0}\bigg[{\lambda}^{{l}|(m)_k,(n)_l}_1\Big(m_{11}\Phi + 2 \chi_{1}-l_1^+ \vartheta\Big)^{l}_{(m)_k,(n)_l} \label{unSghindf}\\
 \hspace{-0.3em}&\hspace{-0.3em}&\hspace{-0.3em} \ \ + \lambda_2^{{l}|(m)_k,(n)_l}\Big(m_{11}\chi_{2}  + 2\chi_{1}  - m_1^+\vartheta \Big)^{l}_{(m)_k,(n)_l} +  \lambda_3^{{l}|(m)_k,(n)_l}\Big(m_{11} \chi_{1}  -l_1 \vartheta\Big)^{l}_{(m)_k,(n)_l}\nonumber\\
 \hspace{-0.3em}&\hspace{-0.3em}&\hspace{-0.3em}\ \  +  \lambda_4^{{l}|(m)_k,(n)_l}\Big(m_{11}\chi_0  -l_0\vartheta\Big)^{l}_{(m)_k,(n)_l}  \bigg] ,\nonumber\\
 \hspace{-0.3em}&\hspace{-0.3em}&\hspace{-0.3em} {\cal{}S}_{C|\Xi}\big(\Phi,\chi_0, \chi_1, \chi_2;{\widetilde{\lambda}}_p\big) =  \int d^dx  \sum_{k,l\geq 0}\bigg[\widetilde{\lambda}_1 ^{{l}|(m)_k,(n)_l}\Big(l_0\Phi - l_1^+\chi_0
  \Big)^{l}_{(m)_k,(n)_l}  \label{conSghindf}  \\
 && + \widetilde{\lambda}_2 ^{{l}|(m)_k,(n)_l}\Big(l_1\Phi  - l_1^+\chi_1   - \chi_0\Big)^{l}_{(m)_k,(n)_l}+\widetilde{\lambda}_3 ^{{l}|(m)_k,(n)_l}\Big( l_0\chi_1 - l_1 \chi_0\Big)^{l}_{(m)_k,(n)_l}    \nonumber\\
 &&+ \widetilde{\lambda}_4 ^{{l}|(m)_k,(n)_l}\Big(  l_0\chi_2 - m^+_1 \chi_0\Big)^{l}_{(m)_k,(n)_l} +\widetilde{\lambda}^{{l}|(m)_k,(n)_l}_5\Big(l_1\chi_2- m^+_1\chi_1   -  \chi_{0}\Big)^{l}_{(m)_k,(n)_l} \nonumber\\
 && + \widetilde{\lambda}_6 ^{{l}|(m)_k,(n)_l}\Big(m_1^+ \Phi  - l_1^+ \chi_2\Big)^{l}_{(m)_k,(n)_l}  \bigg] \nonumber
 \end{eqnarray}
 which is invariant with respect to the  gauge transformations with the unconstrained gauge parameter $ \varsigma (\omega)$ for the fields
\begin{align}\label{gtrgh0fquart}
     & \delta \big(\Phi ,  \chi_{1}, \chi_{2},     \chi_{0}, \vartheta   \big)(\omega) =  \big(l_1^+, l_1, m_1^+,   l_0, m_{11}\big) \varsigma(\omega).
     \end{align}
     and with respect to the dual gauge transformations for the Lagrangian multipliers ${\widetilde{\lambda}}_p, \lambda_j$ with additional  unconstrained gauge parameters $ \sigma^{{l}|(m)_k,(n)_l}$:
 \begin{eqnarray}
      \delta \big(\widetilde{\lambda}{}^{l}_p,\, {\lambda}{}^{l}_j\big){}^{(m)_k,(n)_l} (x)\ = \ \sum_{k',l'}\Big(\widetilde{R}_p,\,{R}_j\Big)^{l{}(m)_k,(n)_l}_{l'{}(m_1)_{k'}(n_1)_{l'}}(x,\partial_x) \sigma^{l' (m_1)_{k'}(n_1)_{l'}}(x)  \label{ungaugetransfDUAL}
\end{eqnarray}
with some local field-independent generators $\widetilde{R}_p{}^{l{}(m)_k,(n)_l}_{l'{}(m_1)_{k'}(n_1)_{l'}},\,{R}_j{}^{l{}(m)_k,(n)_l}_{l'{}(m_1)_{k'}(n_1)_{l'}}$ whose  specific form should be  derived from the gauge invariance for the EoM for the Lagrangian multipliers:
\begin{eqnarray}
 \frac{\delta {\cal{}S}_{\Xi}}{\delta \Phi}&=& l_0 \widetilde{\lambda}_1+ l^d_1\widetilde{\lambda}_2+ m_1^{+d}\widetilde{\lambda}_6  + m_{11}^{d}{\lambda}_1 =0, \nonumber  \\
 \frac{\delta {\cal{}S}_{\Xi}}{\delta \chi_0}  &=&- l^{+d}_1 \widetilde{\lambda}_1 - \widetilde{\lambda}_2 - l^d_1\widetilde{\lambda}_3 - m_1^{+d}\widetilde{\lambda}_4 - \widetilde{\lambda}_5 + m_{11}^{d}{\lambda}_4  =0, \nonumber\\
  \frac{\delta {\cal{}S}_{\Xi}}{\delta \chi_2} &=& l_0 \widetilde{\lambda}_4+ l^d_1\widetilde{\lambda}_5- l_1^{+d}\widetilde{\lambda}_6  + m_{11}^{d}{\lambda}_2 =0,  \label{EoMLm} \\
  \frac{\delta {\cal{}S}_{\Xi}}{\delta \chi_1} &=& l_0 \widetilde{\lambda}_3 - l^{+d}_1\widetilde{\lambda}_2 -  m_1^{+d}\widetilde{\lambda}_5  + m_{11}^{d}{\lambda}_3 +2({\lambda}_1+ {\lambda}_2) =0, \nonumber\\
  \frac{\delta {\cal{}S}_{\Xi}}{\delta \vartheta} &=& -l_0 {\lambda}_4  - l^d_1{\lambda}_1 - m_1^{+d}{\lambda}_2 - l_{1}^{d}{\lambda}_3   =0 .\nonumber
\end{eqnarray}
Here, the form of the dual operators $l^d_1, m_1^{+d}, l_1^{+d}, m_{11}^{d}$ are determined with help of  (\ref{Eqb2q}), (\ref{in-opersf}) according the rule for any tensors  $G^{l}_{(m)_k,(n)_l}(x)$, $F^{l}_{(m)_k,(n)_l}(x)$ with a compact support
\begin{eqnarray}\label{dualrule}
 \hspace{-0.5em}  &\hspace{-0.5em} &\hspace{-0.5em}  \int d^dx G^{l}_{(m)_k,(n)_l} (\hat{A} F)^{{l}|(m)_k,(n)_l} \ = \  \int d^dx (\hat{A}{}^d G)^{l}_{(m)_k,(n)_l}  F^{{l}|(m)_k,(n)_l},  \  \hat{A} \in \{l_1, m_1^{(+)}, l_1^{+}\},\\
 \hspace{-0.5em}  &\hspace{-0.5em} &\hspace{-0.5em} G^{l}_{(m)_k,(n)_l} (m_{11} F)^{{l}|(m)_k,(n)_l} \ = \  (m_{11}^d G)^{l}_{(m)_k,(n)_l} F^{{l}|(m)_k,(n)_l}.\nonumber
\end{eqnarray}
        Again, by the choice of appropriate initial and boundary conditions for ${\widetilde{\lambda}}{}^{{l}|(m)_k,(n)_l}_p$, $\lambda^{{l}|(m)_k,(n)_l}_j$   we always able to fix their unwanted degrees of freedom completely \cite{Sharapovaug}. We use again the  such form of the free actions  as the  auxiliary ones to derive preferably the EoM for the fields $\Phi,  \chi_{1}, \chi_{2},     \chi_{0}, \vartheta$.

Applying the terminology from the HS fields with discrete spin we will call the obtained irreducible gauge-invariant LF as the \emph{quartet-like unconstrained formulation} for scalar bosonic field with CS $\Xi$ on $\mathbb{R}^{1,d-1}$ within Bargmann--Wigner representation. In turn, the functional ${\cal{}S}_{C|\Xi}\big(\Phi$, $\chi_0, \chi_1, \chi_2;{\widetilde{\lambda}}_p\big)$ (\ref{conSghindf}) should describe the constrained irreducible gauge-invariant LF in the so-called  \emph{triplet-like  formulation} with constrained fields, Lagrangian multipliers and gauge parameters $\sigma^{l' (m_1)_{k'}(n_1)_{l'}}$.

The unconstrained  LF given by the relations (\ref{unSghindf}),  (\ref{conSghindf}),  (\ref{gtrgh0fquart}), (\ref{ungaugetransfDUAL}) presents the basic result of the section.

\section{Conclusion} \label{Conclus}

In this paper, we have developed a constrained BRST--BFV approach to a gauge-invariant  description of EoM and action of free scalar CSR for the Poincare group, with a fixed arbitrary CS $\Xi$ (when parameter  $\nu=1$) in Minkowski space-time $\mathbb{R}^{1,d-1}$ of an arbitrary dimension in a \textquotedblleft metric-like\textquotedblright\ formulation within Bargmann-Wigner representation. The final constrained BRST-BFV representation for EoM, given by (\ref{LcontinB}),  in fact, determined by Wigner fields of two space-time variables $x^m,\omega^m$, represents a first-stage reducible gauge theory and contains an auxiliary set of fields providing a BRST-unfolded form (in a ghost-independent representation), of the field equations (\ref{eta0})--(\ref{eta1etamPm}) and gauge transformations (\ref{gtrgh11}), (\ref{gtrgh01})--(\ref{gtrgh04}).

To present a constrained BRST--BFV gauge-invariant description of EoM, by transforming  the Bargmann--Wigner equations (\ref{Eqb01w}) into four constraints (\ref{Eqb01}), (\ref{Eqb23}) (equivalently, (\ref{irrepconts}))  imposed on the CS real-valued  field $\Phi(x,\omega)$ in the coordinate form.
 The decomposition (\ref{presentaPhi}) for the field $\Phi(x,\omega)$ presents an original ansatz for the non-trivial  solution for (\ref{Eqb01}), (\ref{Eqb23})  in powers  of direct and inverse degrees in the variables $\omega^m$ (\ref{presentaPhi}), using an infinite set of conventional independent $\Phi^0_{(m)_k}(x)$ and additional ${\Phi}^{l}_{(m)_k, (n)_l}(x)$ tensor fields. The vector $\Phi(x,\omega)$ (in the space $\mathcal{V}$ having no scalar product structure) contains a standard contribution with the usual  $\Phi^{{0}}_{(m)_k}(x)$ massless tensor fields of rank $k=0,1,2,...$ and a new one, $\Phi^{{l}}_{(m)_k,(n)_l}(x)$  from which the number particle operator extracts some vectors with "mixed positive and negative spin values" $(k,l)$: $k, l-1=0,1,2,... $.  CSR realizations  on (\ref{Eqb01q}) and (\ref{Eqb2q}) ones are different but not independent, due to  \emph{"coupling" equation} for $l=0$ in the second,  third  and fourth lines of (\ref{Eqb2q}), due to an ambiguity in the definitions of $\Phi^{{l}}_{(m)_k,(n)_l}(x)$ (\ref{unambiguity}), (\ref{decompmon}).    The closure of the constraint algebra (\ref{irrepconts}) under the commutator multiplication and a formal Hermitian conjugation generates a higher continuous spin symmetry algebra $\mathcal{A}(\Xi;\mathbb{R}^{1,d-1})$ given by Table~\ref{table in} with two center elements: the parameter $\nu$ and the value of CS, $\Xi$ for $\nu=1$, since any linear combination of constraints should also be a constraint.   Extracting a second-class constraint subsystem: the generalized trace, $m_{11}$, its dual, $m_{11}^+$, and the particle number,  $g_0$, operators from the remaining $(4+1)$ first-class differential constraints, i.e., the divergence, $l_1$, the generalized   divergence, $m_1$, their formal  duals, $l^+_1, m^+_1$, and the D'Alambert operator, we construct, with respect to a reducible set of first-class constraints (considering $m_1-l_1=-i\Xi $ as a constraint), a constrained BRST operator, $\widetilde{Q}_C$ (\ref{generalQCexp}),  and a BRST-extended off-shell constraint, $\widehat{M}_{11}=m_{11}+...$,  in an enlarged vector space,  $\mathcal{V}_C$. They are found as a solution of the generating equations (\ref{eqQctot}) with the boundary conditions (\ref{boudcond}). Calculating the $Q_C$-cohomology in the ghost number zero subspace of $\mathcal{V}_C$, which should lead to the Bargmann--Wigner equations fixes in a unique way, the representation (\ref{represvac}) in $\mathcal{V}_C$, which allows one to select an independent set of constraints and then to reduce $\widetilde{Q}_C$  to the constrained BRST operator ${Q}_C$ (\ref{Qcchic}), without first-stage reducible ghost operators, and to determine the off-shell constraint  $\widehat{M}_{11}$ (\ref{constralgB}). The familiar application of the spectral problem, with a BRST equation ${Q}_C \chi^0_C = 0$, (\ref{Qchi})--(\ref{Qchi2}), however with no spin condition, as in the case of HS fields with discrete spin \cite{Reshetnyak_con}, leads to the constrained BRST--BFV description of the first-stage reducible EoM (\ref{LcontinB}). In the ghost-independent form the latter problem is realized with EoM (\ref{eta0})--(\ref{eta1etamPm}) for one initial and 9 auxiliary fields (in Bargmann--Wigner form with 2 sets of variables $x^m, \omega^m$) invariant with respect to reducible gauge transformations (\ref{gtrgh01})--(\ref{gtrgh04}) with 5 gauge parameters, invariant under the transformations (\ref{gtrgh11}) with an independent gauge for gauge parameter $\varpi(x,\omega)$ and off-shell holonomic constraints (\ref{constrcontinB2})--(\ref{constrcontinB02}).

A specific structure of the constraints and gauge transformations has permitted one to realize a partial gauge-fixing, jointly with a resolution of some of the non-Lagrangian EoM to obtain from the constrained BRST--BFV description the \emph{triplet-like} (\ref{eta1mrr}), (\ref{eta1etamPrr})  and  then \emph{duplet-like} (\ref{eta1mrrd}), (\ref{eta1etamPrrd}), (\ref{constrfin})   formulations of EoM  for scalar  CSR  fields in Bargmann--Wigner representation. These formulations are classified as irreducible gauge theories, respectively, with constrained three and two additional auxiliary fields, by analogy with the triplet and doublet  desriptions for an HS field of an integer spin $s$ \cite{triplet}.   Expressing the field  $\chi_{1}(x,\omega)$ as a generalized trace of the basic CS field $\Phi(x,\omega)$, the non-Lagrangian gauge-invariant EoM (\ref{eta1mrrdd})--(\ref{constrchi2phi}) has also been derived with the help of an additional field $\chi_2(x,\omega)$. The respective constrained gauge-invariant LF  for triplet-like and latter formulations with the actions ${\cal{}S}_{C|\Xi}\big(\Phi,\chi_0, \chi_1, \chi_2;{\widetilde{\lambda}}_p\big)$, $\mathcal{S}_{C|\Xi} \left(\Phi, \chi_2, \hat{\lambda}_i\right)$  have been obtained with the help of some appropriate sets of real-valued  gauge Lagrangian multipliers  (following \cite{Sharapovaug})  for CSR scalar (real-valued) field $\Phi$ of CS $\Xi$ in the tensor form, respectively, in (\ref{costrlagrBRST})--(\ref{fullconstr})  and (\ref{conSghindf}), (\ref{gtrgh0fquart}), (\ref{ungaugetransfDUAL})  for only gauge  ${\widetilde{\lambda}}_p$.
 The fields and gauge parameter $\varsigma(\omega)$   satisfy the generalized traceless (simply, $m_{11}$-traceless) conditions (\ref{constrfin}). We stress that the dynamics of the fields and Lagrangian multipliers is completely decoupled in the presented LFs for free CS field  and the unwanted degrees of freedom for the Lagrangian multipliers can be accurately treated, e.g., it can be  removed by the appropriate choice of the initial conditions for the respective EoM.

The characteristic feature of the constrained BRST--BFV descriptions of EoM  and theirs derivative descriptions is the presence of respective sets of new infinite set of   tensor fields with so called "negative spin values".

We have found, first, the interrelations of the resulting BRST--BFV description of EoM for a scalar CSR field (in the Bargmann-Wigner form) given in the basis of $m_{11}$-traceless fields with those for totally-symmetric HS fields with any integer spin $s= 0,1,2,...$ in terms of Fronsdal-like (traceless) standard and new fields. Second, we have found the correspondence of the (double) $m_{11}$-traceless fields with the usual and new Fronsdal-like (double) traceless fields in (\ref{fronsdd0})--(\ref{fronsdo2}). The latter allows one to present  the parts of  all the constrained LFs which contain the usual tensor and auxiliary fields  for an CSR field entirely in terms of Fronsdal-like  fields. We have shown that the constrained LFs  for all integer spins do not coincide with respective ones for a scalar CSR field. However, for vanishing CS $\Xi=0$ the EoM (\ref{eta1mrrdd})–-(\ref{constrchi2phi})  under the identification $\Phi=\chi_2$ without new (negative spin values) tensors they coincide with ones for HS field $\phi^0(x,\omega)$ (\ref{infintactions}) with all integer spins. We stress that, first,  that the BRST EoM (\ref{LcontinB}) completely equivalent to the Shuster-Toro form \cite{ShusterToro} of the gauge-invariant equations of motion because  they both describe the same  field $\Phi(\omega)$ with CS $\Xi$ from CSR of the Poincare group. Second, in case of the  Shuster-Toro form of equations
\cite{ShusterToro}) which select the CSR field,  they can be derived from the Lagrangian BRST-BFV EoM (without
new fields presence) with the constrained LF closely related with that
for massless totally-symmetric fields for all integer spins. We intend
to sol\-ve this problem in a separate work based on the result
of Appendix~\ref{addalgebra3}, which justifies the presence
of a CSR field  in the spectrum  of an open bosonic string within a
special tensionless limit.

We have established an equivalence of non-Lagrangian EoM in the BRST unfolded form (\ref{eta0})--(\ref{eta1etamPm}) of the suggested constrained BRST-BFV description  with the irreducible CSR relations (\ref{Eqb01}), (\ref{Eqb23}). Incidentally, we have clarified the form of conditions necessary to select UIR of $ISO(1,d-1)$ with integer spin  (\ref{irrepintnew}) and with residual gauge transformations, thus determining a class of gauge equivalent configurations instead of its  unique representative.
Note that the constraints in the respective conditions that select massless UIR  both with CS and with integer spin are sufficient (without using the residual gauge transformations) to construct the constrained BRST operators and to derive the respective BRST--BFV description of EoM and  LFs.

We have developed a BRST--BV approach to the suggested constrained BRST--BFV gauge-invariant description of EoM for a CSR field in a $R^{1,d-1}$ space-time and explicitly constructed the BRST--BV action (\ref{BVprescrip})  for the classical action $\mathcal{S}_{C|\Xi} \left(\Phi, \chi_2, \hat{\lambda}_i\right)$  with 2 fields and  with a corresponding BRST-like invariance (\ref{BRSTphi}), (\ref{BRSTlambda})  in the minimal set of constrained field-antifield configurations both for the fields and for the Lagrangian multipliers.  The crucial point here is that all the fields, ghost fields and their antifields are combined within a unique generalized field-antifield vector (\ref{genafvectorts}) and contain new auxiliary (anti)field tensors as well.  The actions serve, first, to construct quantum actions under an appropriate choice of  gauge conditions, and second, to develop a construction of theories interacting with the CS field with accurate elaboration of the degrees of freedom for the set of Lagrangian multipliers.  We stress that the construction of the minimal BRST--BV actions is differed from the procedure of finding BRST-BV minimal and quantum actions developed in \cite{Metsaev3} for the scalar CS field.

An  unconstrained quartet-like LF (similar to the one for the integer spin case \cite{quartmixbosemas}) has also been found in (\ref{unSghindf}) by including a compensator field to remove the $m_{11}$-tracelessness  of the gauge parameter and by adding to the action for a triplet-like LF (\ref{conSghindf}) of the augmented gauge-invariant constraint conditions (\ref{quaconstrcontinB0}), (\ref{quaconstrcontinB02}) with $(4+6)$  unconstrained Lagrangian multipliers. These multipliers are subject to the attributed gauge transformations (\ref{ungaugetransfDUAL}) being dual for the gauge transformations  (\ref{gtrgh0fquart}) for the fields.

The higher continuous spin symmetry algebra $\mathcal{A}(\Xi; Y(k),
\mathbb{R}^{1,d-1})$  which corresponds to the
most general  massless non-scalar CS one-valued  irreducible
representation of Poincare group in Minkowski space
$\mathbb{R}^{1,d-1}$  for $k=0, 1, ..., k=\left[(d-4)/{2}\right]$ of the Bargmann--Wigner form  is suggested in the Subsection~\ref{BRSTBFV12} as well to be different from one in \cite{Alkalaev}.

We have presented in Appendix~\ref{addalgebra} another way of higher continuous spin symmetry algebra $\mathcal{A}(\Xi;\mathbb{R}^{1,d-1})$ realization by means of two sets of oscillator pairs corresponding to direct and inverse degrees in variables $\omega^m$  with endowing the Fock space $\mathcal{V}\equiv \mathcal{H}$ with a new scalar product.  It should serve   for further study of the algebra $\mathcal{A}(\Xi;\mathbb{R}^{1,d-1})$ and its application for BRST LFs for CSR field.

It has been shown  in Appendix~\ref{addalgebra2} that  there is no possibility to endow the vector space $\mathcal{V}$ with a Hilbert space structure with finite scalar product when explicitly working with the inverse degrees in powers of oscillators. This point proves an impossibility to use the latters  for the purpose of BRST-BFV Lagrangian formulation of the form $S_\Xi \sim  \langle\Phi \big| Q \big|\Phi\rangle^\prime $.

There are numerous ways to elaborate the suggested constrained BRST--BFV and BRST--BV approaches for CSR in the Bargmann-Wigner representation, so as to study the Lagrangian dynamics of CSR in $\mathbb{R}^{1,d-1}$ in the case of arbitrary one-valued mixed-symmetric UIR with CS  as well as to adapt the formalism to accomodate two-valued CSR in $\mathbb{R}^{1,d-1}$.

\vspace{-1ex}

\paragraph{Acknowledgements}

 A.R. is  grateful to I. Buchbinder,  A. Isaev, Yu. Zinoviev, A.
Sharapov, K. Stepanyantz, P. Moshin and to the participants of the
International Conference ``QFTG'2018'', which originated the idea
of this paper. He thanks G. Bonelli, R. Metsaev, M. Najafizadeh,
B. Mischuk, V. Krykhtin , for valuable discussions and important
clarifying comments as well as to the referees who indicated
explicitly the problem with the finiteness of  the scalar product,
the peculiarities with a tensionless limit in the strings for CSR
and with the Lagrangians augmented by Lagrangian multipliers. The paper was
supported by the Program of Fundamental Research sponsored by the
Russian Academy of Sciences, 2013-2020.

\appendix
\section*{Appendix}

\section{Higher Continuous Spin symmetry   algebra \\ $\mathcal{A}(\Xi;\mathbb{R}^{1,d-1})$ with two sets of oscillators }\label{addalgebra}
\renewcommand{\theequation}{\Alph{section}.\arabic{equation}}
\setcounter{equation}{0}

In this appendix, we describe another way to present the algebra $\mathcal{A}(\Xi;\mathbb{R}^{1,d-1})$ in the sector of new tensor fields ${\Phi}^l_{(m)_k,(n)_l}(x)$. To this end,  we endow  $\mathcal{V}$ by the Fock space structure  $\mathcal{V} \to \mathcal{H}$ with a new scalar product  by presenting $\mathcal{H}$ as $\mathcal{H}=\mathcal{H}^0+ \sum_{l>0}\mathcal{H}^l$, which is generated  by  two pairs of the Grassmann-even bosonic (dependent) oscillators   with help of translational invariant vacuum vector:
$|0\rangle$: ${\partial^{m}}|0\rangle =0 $:
\begin{equation}\label{intosc}
\big(a_m, a^{+n}\big)\equiv  -\imath\Big({\partial_\omega^m},\omega^n\Big),\ (b_m, b^{+n})\equiv - \Big(\frac{\partial}{\partial  b^{+m}}, \frac{\imath\omega^{n}}{\omega^2}\Big), \ \  \big(a_m,\, b_m\big)|0\rangle = (0, 0).
\end{equation}
which are  subject to the commutation relations:
\begin{equation}\label{commrelnew}
  [a^m, a^{+n}]=-\eta^{mn}, \ [b^m, b^{+n}]= -\eta^{mn}, \  \big[a^{+}_m,\, b^{+}_n\big] \ = \ 0,\ \ \big[a_m,\, b^{+}_n\big] \ = \ \eta_{mn}(b^+)^2-2b^+_mb^+_n.
\end{equation}
The validity of the latter commutator in (\ref{commrelnew})  follows from   (\ref{intosc}) and the explicit calcu\-la\-tion of $\big[\partial_\omega^m,\, {\imath\omega^{n}}/{\omega^2}\big]$, whereas the commutators $\big[a_m,\, b_n\big]$, $\big[a^{+}_m,\, b_n\big]$ are still remained undetermined.

The field $\Phi(x,\omega)$  (\ref{presentaPhi}) is presented as the  vector  from $\mathcal{H}$
\begin{eqnarray}
&& \big|\Phi\rangle = \sum_{k,l\geq 0}\frac{\imath^{k+l}}{k!l!}\Phi^l_{(m)_k,(n)_l}(x)\prod_{i=1}^k a^{+m_i}\prod_{j=1}^lb^{+n_j}|0\rangle  \equiv  \big[\Phi^{0}(x,\imath a^{+}) +\sum_{l\geq 1}\Phi^{l}(x,\imath a^{+}, \imath b^{+})\big]  |0\rangle \nonumber \\
&& \phantom{\big|\Phi\rangle}   \equiv \big|\Phi^{0}\rangle + \big|\Phi^{-}\rangle, \quad    \label{baspresentaPhi}
\end{eqnarray}
for  square integrable component functions in $\Phi^{0}(x,\imath a^{+})$ and  $\Phi^{l}(x,\imath a^{+},\imath b^{+})$ obtained from the decomposition (\ref{presentaPhi}).
The different pairs of the oscillators are not independent, in view of $\omega^{m_1}\frac{\omega^{n_1}}{\omega^{2}}= \omega^{n_1}\frac{\omega^{m_1}}{\omega^{2}}$, (\ref{unambiguity1}) and (\ref{intosc}):
  \begin{eqnarray}&&  \Big(a^{+m}b^{+n} = a^{+n}b^{+m} \ \mathrm{ and }\  a^{+m}b^+_{m} = -1\Big)  \Longrightarrow     \label{abrels0}\\
  \label{abrels}
&&     a^{+2} b^{+2} = 1 \Longrightarrow  \ C^{+m} = - D^{+m}/D^{+2} = - D^{+m}C^{+2}   \ \mathrm{for} \  C, D \in \{b, a\},
\end{eqnarray}
because of, $a^{+2} b^{+2} = (- \omega^m\omega_m)\cdot(- \omega^n/ \omega^2)(\omega_n/ \omega^2) =  (- \omega^n\{\omega_n/ \omega^2\})\cdot(-\omega_m \{\omega^m/ \omega^2\}) = (a^{+n}b^+_{n})\cdot (a^+_{m}b^{+m})$, so that they both  look as the inverse-like operators for each other, creating "particle" and "antiparticle" respectively.

An idea to consider the oscillators $a^{+m}$, $b^{+n} $ as independent ones but with additional  constraints imposed, explicitly on the vector $\big|\Phi\rangle$:
\begin{equation}\label{newconstraints}
  \big[F(a^{m}b_{m}), G(a^{2}b^{2})\big]|\Phi\rangle  = 0,\    \texttt{for} \ \big[F(0), G(0)\big] =0  \ \texttt{and}   \ \Big[\frac{d F(y)}{d y},\frac{d G(y)}{d y} \Big]\big|_{y=0}\hspace{-0.2em} = \big[C_F,  C_G\big],
\end{equation}
(for  unknown analytical functions $F, G$,) with some real constants $C_F$, $C_G$ leads to highly non-linear expressions for the operators $ F(a^{m}b_{m})$ and $G(a^{2}b^{2})$  generating the mixed traces for the component tensors in  $|\Phi\rangle$ as it was shown in the Section~\ref{HSsymm} .  Instead, we will explicitly resolve the oscillator constraints (\ref{abrels}) thus reducing the ambiguity (\ref{unambiguity}) in the choice of the component fields in $|\Phi\rangle$. To this end, we introduce system of projectors generated by the decomposition (\ref{decompmon}): $P_0, P_1; P_0+P_1 = 1$, such that $P_iP_j=\delta_{ij}P_i$, $i,j=0,1$, which are  associated with the  decomposition of any product
$a^{+m}b^{+ n}$ on trace and traceless parts (according the rules (\ref{unambiguity}), (\ref{decompmon})) when decompose of any product
$a^{+m}b^{+ n}$  in each monomial $(\prod a^{+})^{(k)_l}(\prod b^{+})^{(n)_l}$ for $k,l>0$ and the same for the product $a^{+m_1}a^{+m_2}b^{+n_1}b^{+n_2}$, but  for $k,l>1$  as follows:
\begin{eqnarray}
  a^{+m}b^{+ n} &=& d^{-1} \eta^{mn}\eta_{k_1k_2}  a^{+k_1}b^{+k_2} + \big(\delta^m_{k_1}\delta^n_{k_2}-d^{-1} \eta^{mn} \eta_{k_1k_2} \big) a^{+k_1}b^{+k_2}
  \nonumber\\
  &\equiv&  \big({P_0}^{mn}_{k_1k_2} + {P_1}^{mn}_{k_1k_2}\big) a^{+k_1}b^{+k_2}, \label{a+b+}
 \end{eqnarray}
 \vspace{-2ex}
 \begin{eqnarray} (\prod a^{+})^{(m)_2}(\prod b^{+})^{(n)_2} &=&  d^{-2} \eta^{m_1m_2}\eta^{n_1n_2}\eta_{k_1k_2}\eta_{l_1l_2}(\prod a^{+})^{(k)_2}(\prod b^{+})^{(l)_2} \nonumber\\
  && + d^{-1}\eta^{m_1m_2}\big(\delta^{n_1}_{l_1}\delta^{n_2}_{l_2}-d^{-1}\eta^{n_1n_2} \eta_{l_1l_2} \big) a^{+2}(\prod b^{+})^{(l)_2} \nonumber\\
  && + d^{-1}\eta^{n_1n_2}\big(\delta^{m_1}_{k_1}\delta^{m_2}_{k_2}-d^{-1}\eta^{m_1m_2} \eta_{k_1k_2} \big) (\prod a^{+})^{(k)_2}b^{+2} \nonumber\\
  && + \big(\delta^{m_1}_{k_1}\delta^{m_2}_{k_2}\delta^{n_1}_{l_1}\delta^{n_2}_{l_2}-d^{-2} \eta^{m_1m_2}\eta^{n_1n_2} \eta_{k_1k_2} \eta_{l_1l_2} \big) (\prod a^{+})^{(k)_2}(\prod b^{+})^{(l)_2} \nonumber\\
  & \equiv & \sum_{i=0}^1\sum_{j=0}^1 {P_i}^{m_1m_2}_{k_1k_2}{P_j}^{n_1n_2}_{l_1l_2}(\prod a^{+})^{(k)_2}(\prod b^{+})^{(l)_2}\label{a+2b+2}.
\end{eqnarray}
Due to the properties (\ref{abrels}) the first summands in (\ref{a+b+}), (\ref{a+2b+2}) are equal respectively to
\begin{equation}\label{red123}
 {P_0}^{mn}_{k_1k_2} a^{+k_1}b^{+k_2}  = - d^{-1} \eta^{mn}, \qquad  {P_0}^{m_1m_2}_{k_1k_2}{P_0}^{n_1n_2}_{l_1l_2}(\prod a^{+})^{(k)_2}(\prod b^{+})^{(l)_2} = d^{-2} \eta^{m_1m_2}\eta^{n_1n_2}.
\end{equation}
The projectors ${P_i}^{m_1m_2}_{k_1k_2}$ (because of the total symmetry  of all component tensor fields $\Phi^{{l}}_{(m)_k,(n)_l}$, and thus due to $a^{+m}b^{+n} = a^{+n}b^{+m}$ inside $\big|\Phi\rangle$) satisfy to the symmetry properties:
\begin{equation}\label{symmprojectors}
  {P_i}^{m_1m_2}_{k_1k_2}= {P_i}^{m_2m_1}_{k_1k_2}={P_i}^{m_1m_2}_{k_2k_1}, \ i=0,1.
\end{equation}
The last properties permits to find that the decomposition of the quartic term $(\prod a^{+})^{(m)_2}\times$ $\times(\prod b^{+})^{(n)_2} \equiv (a^+b^+)^{(mn)_2}$  into components generated by the relations  (\ref{a+b+}), (\ref{red123}):
 \begin{equation}\label{1decomp}
   (a^+b^+)^{(mn)_2} = d^{-2} \eta^{m_1\{n_1}\eta^{m_2\}n_2} + \Big({P_1}^{m_1\{n_1}_{k_1l_1}{P_0}^{m_2\}n_2}_{k_2l_2}+\delta^{m_1}_{k_1}\delta^{\{n_1}_{l_1}{P_1}^{m_2\}n_2}_{k_2l_2}\Big)(a^+b^+)^{(kl)_2}
 \end{equation}
coincides with  the decomposition (\ref{a+2b+2}) with account for Eqs. (\ref{abrels}), (\ref{red123}):
 \begin{equation}\label{2decomp}
(a^+b^+)^{(mn)_2} = d^{-2} \eta^{m_1\{m_2}\eta^{n_1\}n_2} + \Big({P_1}^{m_1\{m_2}_{k_1k_2}{P_0}^{n_1\}n_2}_{l_1l_2}+\delta^{m_1}_{k_1}\delta^{\{m_2}_{k_2}{P_1}^{n_1\}n_2}_{l_1l_2}\Big)(a^+b^+)^{(kl)_2}.
\end{equation}
Therefore,  the decomposition (\ref{a+b+}) is sufficient to reduce an ambiguity in the choice of the component tensors in $\big|\Phi\rangle$ (\ref{baspresentaPhi}), which should now be determined as
\begin{eqnarray}
\hspace{-0.35em}&\hspace{-0.35em}&\hspace{-0.35em} \big|\Phi\rangle = \sum_{k\geq 0}\bigg\{\sum_{l\geq 0}^{k}\frac{\imath^{k+l}}{k!l!}\Big(\hspace{-0.2em}\prod_{h=1}^{l}P_1{}_{m_hn_h}^{\sigma_h\rho_h} \hspace{-0.2em}\Big)\Phi^l_{(\sigma)_lm_{l+1}...m_k,(\rho)_l}+ \sum_{l>k}\frac{\imath^{k+l}}{k!l!}\Big(\hspace{-0.2em}\prod_{h=1}^{k}P_1{}_{m_hn_h}^{\sigma_h\rho_h} \hspace{-0.2em} \Big)\Phi^l_{(\sigma)_k,(\rho)_kn_{k+1}...n_l}\hspace{-0.2em}\bigg\} \nonumber \\
\hspace{-0.35em}&\hspace{-0.35em}&\hspace{-0.35em}  \times \prod_{i=1}^k a^{+m_i}\prod_{j=1}^lb^{+n_j}|0\rangle \equiv \big(\Phi^{0}(x,\imath a^{+}) +\sum_{l\geq 1}\widehat{\Phi}^{l}(x,\imath a^{+}, \imath b^{+})\big)  |0\rangle  \equiv \big|\Phi^{0}\rangle + \big|\widehat{\Phi}^{-}\rangle,  \label{baspresentaPhird}\\
\hspace{-0.35em}&\hspace{-0.35em}&\hspace{-0.35em} \widehat{\Phi}{}^l_{(m)_k,(n)_l} \equiv \Big(\prod_{h=1}^{l}P_1{}_{m_hn_h}^{\sigma_h\rho_h} \Big)\Phi^l_{(\sigma)_lm_{l+1}...m_k,(\rho)_l}\theta_{k,l} + \Big(\prod_{h=1}^{k}P_1{}_{m_hn_h}^{\sigma_h\rho_h} \Big)\Phi^l_{(\sigma)_k,(\rho)_kn_{k+1}...n_l}\theta_{l,k-1},\label{hatphi}
\end{eqnarray}
 where the operators $a^{+m},b^{+ n}$ are already considered on the set of such vectors  as independent oscillators. Note, the traceless projection concerned only new tensor fields $\Phi^l_{(m)_k,(n)_l}(x)$, $l>0$, whereas  the standard fields, $\Phi^{0}_{(m)_k}(x)$, have not been touched when resolving the operator identities. Because of, $P_0{}^{m_in_i}_{\sigma_i\rho_i}\widehat{\Phi}{}^l_{(m)_k,(n)_l}(x)=0 $ for $i=1,...,\min(k,l)$ the field $\widetilde{\Phi}{}^l_{(m)_k,(n)_l}$ given by the Eq.(\ref{unambiguity}) contains the field $\widehat{\Phi}{}^l_{(m)_k,(n)_l}$ as its traceless part without summands proportional  to   $P_0{}_{m_hn_h}^{\sigma_h\rho_h}$ projector according to (\ref{hatphi}), (\ref{difftrace}). It  means, in fact that all new  tensor fields
$\Phi^l_{(m)_k,(n)_l}(x)$, $l>0$ should be traceless when calculating of any traces:
\begin{equation}\label{anyrrless}
  \eta^{m_{k-i}m_k}\Phi^l_{(m)_k,(n)_l}\ = \ \eta^{m_{k-i}n_{l-j}}\Phi^l_{(m)_k,(n)_l} \ = \ \eta^{n_{l-j}n_l}\Phi^l_{(m)_k,(n)_l} \ =\ 0,
\end{equation}
for $ i=1,...,{k-1}$, $j=1,...,{l-1}$, when $l, k>0 $.
Therefore, instead of the tensors $\widehat{\Phi}{}^l_{(m)_k,(n)_l}$ we may equivalently write ${\Phi}{}^l_{(m)_k,(n)_l}$ in the decomposition (\ref{baspresentaPhird})  implying validity of the traceless condition (\ref{anyrrless}).

The  Poincare group IR relations (\ref{Eqb01}),  (\ref{Eqb23})   in the tensor form (\ref{Eqb01q}), (\ref{Eqb2q}) take the equivalent representation  in terms of the operators [confer with the Eqs. (\ref{irrepconts}), (\ref{in-opersf})]
\begin{align}
 &  \big(l_0,\, l_1,\,m^+_{1},\, m_{11} \big)\big|\Phi\rangle \ = \  0 \ ; &&\label{irrepcont}\\
&   l_1  =  -\imath a^m{\partial_m}  -\imath  b^{+n}\big[2b^{+m}b_{n}- b^{+}_nb^{m} \big]{\partial_m} ,
      &&  m^+_{1}  =   -\imath a^{+m}{\partial_m}+\imath\Xi \label{in-opers2},\end{align}
      \vspace{-2ex}
    \begin{equation}
           \widetilde{m}_{11} =  a^2 + \big\{ a^m,b^{+n}\big[2b^{+}_mb_{n}-  b^{+}_nb_{m} \big]  \big\}+{b^+}^2\big({b^{+}}^2b^2 -\hspace{-0.15em}2 (d-2) b^{+k}b_{k}\big)+ \nu  \label{in-opers3},
 \end{equation} where, first, the sign "$\big\{\ ,\ \}$" in (\ref{in-opers3}) is the anticommutator, second, we have used the rule to express the derivative ${\partial}/{\partial \omega^m}$ in terms of the oscillators without their negative degrees:
   \begin{eqnarray}\label{exprombb+}
  \frac{\partial}{\partial \omega^m} & = & \frac{\partial}{\partial \omega^m}\big|_{-\imath\frac{ \omega^{n}}{\omega^2} = b^{+n}=\mathrm{const} }+ \frac{\partial}{\partial \omega^m}\big|_{-\imath \omega^{n} = a^{+n}=\mathrm{const} } \\
 &=&  \imath a_m + \frac{\partial b^{+n}}{\partial \omega^m}\frac{\partial}{\partial  b^{+n}}=   \imath a_m - \imath \frac{\partial \big(\omega^{n}/{\omega^2}\big)}{\partial \omega^m}\frac{\partial}{\partial  b^{+n}}  \  = \  \imath a_m+  \imath \left(\frac{\delta^n_m}{\omega^2} - 2\frac{ \omega^{n} \omega_{m}}{\omega^4}\right) b_{n} \nonumber\\
 & =& \imath a_m+   \imath \left(\frac{\delta^n_m \omega^k \omega_k}{\omega^2\cdot\omega^2} - 2\frac{ \omega^{n} \omega_{m}}{\omega^4}\right) b_{n}  \ = \    \imath a_m+ \imath b^{+n}\big[2b^{+}_mb_{n}- b^{+}_nb_{m} \big], \label{exprombba+}
\end{eqnarray}

To get Lagrangian form of the equations (\ref{irrepcont})  (without Lagrangian multipliers) we need  $\mathbb{R}$-valued Lagrangian action within BRST--BFV approach. Therefore, the set of initial constraints $\{o_\alpha\}=\big(l_0,\, l_1,\,m^+_{1},\, m_{11} \big)$  (\ref{in-opers2}), (\ref{in-opers3}) should be
closed with respect to $[\ ,\ ]$-multiplication and Hermitian conjugation in $\mathcal{H}$. To do so we determine the scalar product on the space of the vectors $\ref{baspresentaPhird}$ and its   dual as follows
\begin{eqnarray}\label{sproductn1}
\hspace{-0.55em}&\hspace{-0.55em}&\hspace{-0.55em}  \langle{\Psi}\big|\Phi\rangle \hspace{-0.15em}  =  \hspace{-0.25em} \langle{\Psi}^{0}\big|\Phi^0\rangle + \langle{\Psi}^{-}\big|\Phi^{-}\rangle\ = \  \int
d^dx \Big\{\sum_{k_1,k_2=0}^{\infty}
         \hspace{-0.15em}\frac{\imath^{k_1}(-\imath)^{k_2}}{k_1!k_2!}
 \langle 0|\hspace{-0.15em}\prod_{j=1}^{k_2}\hspace{-0.15em} a^{m_{j}}\Psi^*_{(m)_{k_2}}\hspace{-0.15em}\times \\
 \hspace{-0.55em}&\hspace{-0.55em}&\hspace{-0.55em} \phantom{\big|\Phi\rangle \hspace{-0.15em}}\times
\Phi_{(n)_{k_1}}\hspace{-0.15em}
\prod_{i=1}^{k_1} a^{+n_i}|0\rangle + \sum_{k_1,k_2,l_1,l_2>0}^{\infty}
         \hspace{-0.15em}\frac{(-\imath)^{l_2+k_2}\imath^{l_1+k_1}}{k_1!l_1!k_2!l_2!}
 \langle 0|\hspace{-0.15em}\prod_{i_2,j_2=1}^{k_2,l_2} b^{n'_{j_2}}a^{m'_{i_2}}{\Psi}^{l_2*}_{(m')_{k_2},(n')_{l_2}}\hspace{-0.15em}
 \nonumber \\
 \hspace{-0.55em}&\hspace{-0.55em}&\hspace{-0.55em} \phantom{\big|\Phi\rangle \hspace{-0.15em}} \times \hspace{-0.15em}{\Phi}^{l_1}_{(m)_{k_1},(n)_{l_1}}\prod_{i_1,j_1=1}^{k_1,l_1} {a^{+m_{i_1}}}{b^{+n_{j_1}}}|0\rangle\Big\}\hspace{-0.15em} =  \sum_{k,l=0}^{\infty}\hspace{-0.15em}\frac{(-1)^{k+l}}{k!l!}\int d^dx {\Psi}{}^{l*}_{(m)_k,(n)_l}{\Phi}^{l{(m)_k,(n)_l}} , \nonumber
\end{eqnarray}
where the non-diagonal terms proportional to
\begin{eqnarray}
 \hspace{-0.5em}&\hspace{-0.5em}&\hspace{-0.5em} \Big(\sum_{p=1}z_{p,l}{\Psi}^{2p+l *}_{(m)_{p}(m^\prime)_{k},}{}_{,(n^\prime)_l(n)_p}{}^{(n)_p}\Big){\Phi}^{p+l}{}^{(m^\prime)_{k},(m)_{p}(n^\prime)_l}+ c.c. \label{nondiag}
\end{eqnarray}
(for some rationals $z_{p,l}$) arising in (\ref{sproductn1}) from the non-commutativity of  $a^m$ and $b^+_n$ (\ref{commrelnew})  and  its  (usual)  Hermitian conjugated for  $a^{+m}$ and $b_n$ should vanish due to the traceless condition (\ref{anyrrless}).  For instance, for $p=1$; $l,k=0$, we have,
\begin{equation}\label{exlub}
\langle0|{\Psi}^{2*}_{m,(n)_2} b^{n_1}b^{n_2}a^m {\Phi}^{1}_{k}b^{+k}|0\rangle =  2 {\Psi}^{2*}_{m,n}{}^n {\Phi}^{1m}-4{\Psi}^{2*}_{m,}{}^m{}_n {\Phi}^{1n}=-2 {\Psi}^{2*}_{m,n}{}^n {\Phi}^{1m}=0.
\end{equation}
The Hermitian conjugated for   (\ref{commrelnew}) mixed oscillator's commutators  take the form
\begin{equation}\label{noncommab}
  \big[a_m,\, b_n\big] \ = \ 0, \ \ \big[a^{+}_m,\, b_n\big] \ = \ -\eta_{mn}b^2+2b_mb_n.
\end{equation}
From the Hermitian conjugation of the identity (\ref{abrels}), $
 (a^{+m}b^+_{m})^+ = b_{m}a^{m}= -1 $,  the operator $ \widetilde{m}_{11}$ in (\ref{in-opers3}) is simplified to
 \begin{equation}
  m_{11} =  a^2 +\hspace{-0.15em} 2\big\{ a^m,b^{+n}b^{+}_mb_{n}  \big\} - a^m  {b^{+}}^2b_{m}  + {b^{+}}^2\big(1+{b^+}^2b^2 -2 (d-2) b^{+k}b_{k}\big)+ \nu  \label{in-opers3mod}.
 \end{equation}
The closedness of the set of operators (\ref{in-opers2}), (\ref{in-opers3mod}), first, with respect to the  Hermitian conjugation in $\mathcal{H}$ with a scalar product (\ref{sproductn1})  leads  to its augmentation by the operators
\begin{eqnarray}
\hspace{-0.55em} m_{1}\hspace{-0.15em}&\hspace{-0.15em} =\hspace{-0.15em} &\hspace{-0.15em}  -\imath a^{m}{\partial_m}-\imath\Xi, \label{add-opers0}\\
%%%%%%%%%%%%%
\hspace{-0.55em}  l^+_1 \hspace{-0.15em}&\hspace{-0.15em} =\hspace{-0.15em} &\hspace{-0.15em}  -\imath a^{+m}{\partial_m}  -\imath  \big[2b^+_{n}b^{m}- b^{+m}b_n \big]b^{n}{\partial_m}, \label{add-opers-}\\
\hspace{-055em} m^+_{11}\hspace{-0.35em}&\hspace{-0.15em} =\hspace{-0.15em} &\hspace{-0.15em} {a^{+}}^2 +\hspace{-0.15em} 2\big\{ a^{+m},b^{+n}b_mb_{n}  \big\} -   b^{+}_{m}b^2a^{+m}  + \big(1 + {b^+}^2b^2 -
\hspace{-0.15em}2 (d-2) b^{+l}b_{l}\big)b^2 + \nu . \label{add-opers1}
\end{eqnarray}
Second,  its closedness with respect to the $[\ ,\ ]$-multiplication for $l^+_1, l_1, m_1, m_1^+$:
\begin{eqnarray}
  \hspace{-0.3em} \big[l_{1},\,l^+_{1}\big]\hspace{-0.3em} &\hspace{-0.3em}= & (1+3{b^+}^2b^2- 4(b^{+l}b_{l})^2)l_0  -  2\Big\{{b^+}^2b_mb_k+b^+_mb^+_k{b}^2 \nonumber \\
\hspace{-0.3em} \phantom{\big[l_{1},\,l^+_{1}\big]}\hspace{-0.3em} &\hspace{-0.3em} &  -2 b^+_m\big(b^{+l}b_{l}+\textstyle\frac{1}{2}(d-2)\big)b_k\Big\}\partial^m\partial^k        \nonumber  \\
%%%%%%%%%%%%%%%%%%%%
  \hspace{-0.3em}&\hspace{-0.3em}=\hspace{-0.3em} &\hspace{-0.3em}  \big(1+3{b^+}^2b^2-4(b^{+l}b_{l})^2\big)l_0 - 2l_1^{b+}\big(l_1^{+}-m_1^{+}\big) - 2\big(l_1-m_1\big)l_1^{b}-4 l_1^{b+} \Big\{\textstyle\frac{1}{2}(d-2)\nonumber \\
   \hspace{-0.7em} \hspace{-0.3em} &\hspace{-0.3em}\hspace{-0.3em}&\hspace{-0.3em} - b^{+l}b_{l}\Big\}l_1^{b}  - 2 \imath\Xi  (l_1^{b+} - l_1^{b}) \label{comml1l1+}\\
   \hspace{-0.7em} \big[l_{1},\,m^+_{1}\big]\hspace{-0.3em} &\hspace{-0.3em}= & \hspace{-0.3em} (1+{b^+}^2b^2)l_0  -  2\big({b^+}^2b_mb_k+b^+_mb^+_k{b}^2\big)\partial^m\partial^k +4 b^+_mb^{+l}b_{l}b_k\partial^m\partial^k  \nonumber  \\
%%%%%%%%%%%%%%%%%%%%
  \hspace{-0.3em}&\hspace{-0.3em}=\hspace{-0.3em} &\hspace{-0.3em}  \big(1+{b^+}^2b^2\big)l_0 - 2l_1^{b+}\big(l_1^{+}-m_1^{+}\big) - 2\big(l_1-m_1\big)l_1^{b}-4 l_1^{b+} b^{+l}b_{l}l_1^{b} \nonumber  \\
%%%%%%%%%%%%%%%%%%%%
  \hspace{-0.3em}&\hspace{-0.3em}\hspace{-0.3em} &\hspace{-0.3em} - 2 \imath\Xi  \big(l_1^{b+} - l_1^{b}\big) , \label{comml1m1+}\\
\hspace{-0.7em} \big[m_{1},\,l^+_{1}\big]\hspace{-0.3em} &\hspace{-0.3em}= &  \big[l_{1},\,m^+_{1}\big] \ \equiv \ \Big(\big[l_{1},\,m^+_{1}\big]\Big)^+  , \label{comml1+m1}\\
\hspace{-0.7em} \big[l_{1},\,m_{1}\big]\hspace{-0.3em} &\hspace{-0.3em}= &  2\big({b^+}^2 l_0 +4{b^+}^2 {l_1^{b+}}^2\big)b^{+l}b_{l}   -4 {b^+}^2 l_1^{b+}l_1^{b}, \label{comml1m1}
\end{eqnarray}
(with account of  $\big[l^+_{1},\,m^+_{1}\big] =  - \big(\big[l_{1},\,m_{1}\big]\big)^+$)
 implies an inclusion in itself of the operators of divergence and gradient with respect to the second group indices $(n)_l$ in the tensors ${\Phi}^{l}_{{(m)_k,(n)_l}}$,
\begin{equation}\label{defdivgradnl}
  \big(l_1^b,\, l_1^{b+}\big) \ \stackrel{\mathrm{def}}{=} - \imath\big(b_m,\, b^{+}_m\big)\partial^m \ \texttt{so that}\  \big[l_{1}^b,\,l^{b+}_{1}\big]=l_0.
\end{equation}
This fact requires a further careful study of the non-linear  HCS symmetry algebra, in question, in the representation with two pairs of dependent oscillators.

\section{On Problems with Lagrangian formulations  with\\  single  set of oscillators}\label{addalgebra2}
\renewcommand{\theequation}{\Alph{section}.\arabic{equation}}
\setcounter{equation}{0}

In this appendix, we  present the algebra $\mathcal{A}(\Xi;\mathbb{R}^{1,d-1})$   with use of the inverse degrees in $a^{+m}=-\imath \omega^m$ oscillators, which however leads to the problem with finiteness of the scalar product. To this end,  we endow $\mathcal{V}$ with a new scalar product, $\langle \ \big|\ \rangle^\prime$, by presenting $\mathcal{V}$ as $\mathcal{V}=\mathcal{V}^0+ \sum_{l>0}\mathcal{V}^l= \mathcal{V}^0+\mathcal{V}^-$ permitting the  representation for the vector $\big|\Phi\rangle$ as in (\ref{presentaPhi})  or (\ref{baspresentaPhi}) but for $b^{+n}= -  {\imath\omega^{n}}/{\omega^2} = - a^{+n}/{a^+}^2$:
\begin{eqnarray}
\label{sproductnew}
\hspace{-0.35em}&\hspace{-0.35em}&\hspace{-0.35em}  \langle{\Psi}\big|\Phi\rangle^\prime \hspace{-0.15em}  =  \hspace{-0.25em} \langle{\Psi}^{0}\big|\Phi^{0}\rangle^\prime + \langle{\Psi}^{-}\big|\Phi^{-}\rangle^\prime\ = \  \int
d^dx\Big\{\sum_{k_1,k_2=0}^{\infty}
         \hspace{-0.15em}\frac{\imath^{k_1}(-\imath)^{k_2}}{k_1!k_2!}
 \langle 0|\hspace{-0.15em}\prod_{j=1}^{k_2}\hspace{-0.15em} a^{m_{j}}\Psi^*_{(m)_{k_2}}\hspace{-0.15em}\times \\
 \hspace{-0.55em}&\hspace{-0.55em}&\hspace{-0.55em} \phantom{\big|\Phi\rangle \hspace{-0.15em}}\times
\Phi_{(n)_{k_1}}\hspace{-0.15em}
\prod_{i=1}^{k_1} a^{+n_i}|0\rangle + \sum_{k_1,k_2,l_1,l_2>0}^{\infty}
         \hspace{-0.15em}\frac{(-\imath)^{l_2+k_2}\imath^{l_1+k_1}}{k_1!l_1!k_2!l_2!}
 \langle 0|\hspace{-0.15em}\prod_{i_2,j_2=1}^{k_2,l_2} \frac{-a^{n'_{j_2}}}{a^2}a^{m'_{i_2}}{\Psi}^{l_2*}_{(m')_{k_2},(n')_{l_2}}\hspace{-0.15em}
 \nonumber \\
 \hspace{-0.55em}&\hspace{-0.55em}&\hspace{-0.55em} \phantom{\big|\Phi\rangle \hspace{-0.15em}} \times \hspace{-0.15em}{\Phi}^{l_1}_{(m)_{k_1},(n)_{l_1}}\prod_{i_1,j_1=1}^{k_1,l_1} {a^{+m_{i_1}}}  \frac{-a^{+n_{j_1}}}{a^{+2}}|0\rangle\Big\}\hspace{-0.15em} =  \sum_{k,l=0}^{\infty}\hspace{-0.15em}\frac{(-1)^{k+l}}{k!l!}\int d^dx {\Psi}{}^{l*}_{(m)_k,(n)_l}{\Phi}^{l{(m)_k,(n)_l}} , \nonumber
\\
 \hspace{-0.35em}&\hspace{-0.35em}&\hspace{-0.35em} =  \hspace{-0.15em}\sum_{k=0}^{\infty}\hspace{-0.15em}\frac{(-1)^k}{{k!}}\int \hspace{-0.15em} d^dx\hspace{-0.15em} \Big\{\hspace{-0.15em}
 \Psi^{0*}_{(n)_k}\Phi^{0(n)_k}\hspace{-0.15em} + \sum_{l>0}\frac{(-1)^l}{{l!}}\Big[{\Psi}^{l*}_{(n)_l} K_{l,l}{\Phi}^{l(n)_l}\hspace{-0.15em} + more \Big] \Big\}, \nonumber
\end{eqnarray}
with some real numbers  $K_{l,l}$. Here, the term "$more$" denotes the summands  proportional to a respective product $\Psi^{l*}_{(m)_k,(n)_l} {\Phi}^{l (m)_k,(n)_l}$ for $k>0$ and some possible others.

Indeed, the orthogonality properties among the vectors  $\langle0|a^m  (a^{+m}|0\rangle)$ and  $\frac{a^{+m}}{a^{+2}}|0\rangle \big(\langle 0|\frac{a^{m}}{a^{2}} \big)$ take the form:
\begin{align} \label{scalprod1}
   &  \langle0|\prod_{j=1}^{p}\hspace{-0.15em} a^{m_{j}} \, \prod_{i=1}^{q}\hspace{-0.15em} a^{+}_{n_{i}}|0\rangle = \delta_{pq}(-1)^p p!S^{(m)_p}_{(n)_p}, \qquad \langle0|\prod_{j=1}^{p}\hspace{-0.15em} a^{m'_{j}} \, \prod_{i_1,j_1=1}^{k_1,l_1} {a^{+m_{i_1}}}  \frac{a^{+n_{j_1}}}{a^{+2}}|0\rangle = 0,   \\
   &  \langle0|\prod_{j=1}^{p}\hspace{-0.15em} \frac{a_{m_{j}}}{a^2} \, \prod_{i=1}^{k} \frac{a^{+n_i}}{(a^+)^{2}}|0\rangle =
\hspace{-0.15em}   \left\{\hspace{-0.25em}\begin{array}{l}
  \delta_{p,k+2l}(-1)^p p!S_{(m)_p}^{(n)_p} K_{k+2l,k}\eta_{n_{k+1}n_{k+2}}...\eta_{n_{k+2l-1}n_{k+2l}},\ p>k \\
   \delta_{p+2l,k}(-1)^k k!S_{(m)_k}^{(n)_k} K_{p,p+2l}\eta^{m_{p+1}m_{p+2}}...\eta^{m_{p+2l-1}m_{p+2l}},\ p\leq k
                                                              \end{array}\right.
    \label{scalprod11}
     \end{align}
  (and with more complicated form for $\langle0|\prod_{j_1,i_1=1}^{p_1,k_1}\hspace{-0.15em} \frac{a_{m'_{i_1}}}{a^2} a^{n'_{j_1}} \, \prod_{i_2=1, j_2=1}^{p_2,k_2} a^{+n_{j_2}} \frac{a^{+n_{i_2}}}{(a^+)^{2}}|0\rangle$),  where  $K_{p,p+2l}$ = $K_{p+2l,p}$\footnote{Again, the non-diagonal terms in the scalar product (\ref{sproductnew}) are proportional to $$\Big(\sum_{p=1}Z_{p,l}{\Psi}^{2p+l *}_{(m)_{p}(m^\prime)_{k},}{}_{,(n^\prime)_l(n)_p}{}^{(n)_p}\Big){\Phi}^{p+l}{}^{(m^\prime)_{k},(m)_{p}(n^\prime)_l}+c.c.$$ as in (\ref{nondiag})
(for some rationals $Z_{p,l}\thicksim K_{p,p+2l}$) which may  vanish due to symmetry (\ref{difftrace})  of the new tensor fields in case choosing only  traceless components from it   (\ref{anyrrless}) following  to  (\ref{hatphi}) with projectors $\widetilde{P}_1{}_{m_hn_h}^{\sigma_h\rho_h}(a^+)$ = $P_1{}_{m_hn_h}^{\sigma_h\rho_h}\big(a^+, b^+=-a^+/{a^+}^2\big)$, so that the only diagonal terms with $K_{p,p}$ survive in (\ref{sproductnew}). The last argument provides the correct spectral property and non-degeneracy  for $\langle{\Psi}\big|\Phi\rangle^\prime$, in which  for any tensor ${\Phi}^{l}_{(m)_k,(n)_l}$  the only one  tensor  ${\Psi}^{l*}_{(m)_k,(n)_l}$ exists which gives input in the scalar product.} and explicitly for $l=0$:
  %%%%%%%%%%%%%%%%%%%%%%%%
 \begin{align}& K_{p,p} =  \int_0^{\infty} dt_1dt_2  \sum_{k=0} \frac{(t_1t_2)^k}{(k!)^2}  \prod_{j=1}^{pk}2j[d+2(j+p-1)],  \label{Kpp}
 \end{align}
$\forall p, k, q \in \mathbb{N}$ with the symmetrizer $S^{(m)_k}_{(n)_k}$.

   The first products in (\ref{scalprod1}) are standard, whereas to prove the validity of the second ones, we  apply the induction. For $p=1, \forall k \in \mathbb{N}$ we have
\begin{eqnarray}\nonumber
&&  \langle0| a^{m_{1}}  \prod_{i=1}^{k} \frac{a^{+n_i}}{(a^+)^{2}}|0\rangle =   \langle0| \Big([a^{m_{1}}, a^{+n_1}] \, \frac{1}{(a^+)^{2}}-a^{+n_1}[a^{m_{1}}, a^{+2}]\frac{1}{(a^+)^{4}} \Big)\prod_{i=2}^{k} \frac{a^{+n_i}}{(a^+)^{2}}|0\rangle \\
&& \quad = - \langle0| \Big(\eta^{m_{1}n_1}(a^+)^{2}-2a^{+n_1}a^{+m_{1}}\Big)\frac{1}{(a^+)^{4}} \prod_{i=2}^{k} \frac{a^{+n_i}}{(a^+)^{2}}|0\rangle =0,\label{profscalprod11}
\end{eqnarray}
 due to  $\langle0|a^{+n_1} = 0$. Let for $\forall p\leq p_0\in \mathbb{N}$ the same equations  as one (\ref{profscalprod11}) hold. Then, for $p=p_0+1, \forall k \in \mathbb{N}$ it follows, with account of the relation above:
\begin{eqnarray}\nonumber
\hspace{-0.35em}&\hspace{-0.35em}&\hspace{-0.35em}
 \langle 0| \Big(\prod_{j=1}^{p_0}\hspace{-0.15em} a^{m_{j}}\Big)  a^{m_{p_0+1}}\,\prod_{i=1}^{k} \frac{a^{+n_i}}{(a^+)^{2}}|0\rangle =  -  \langle0|\hspace{-0.15em} \Big\{\Big(\prod_{j=1}^{p_0}\hspace{-0.15em} a^{m_{j}}\Big)\Big(\eta^{m_{p_0+1}n_1}(a^+)^{2}-\hspace{-0.15em}2a^{+n_1}a^{+m_{p_0+1}}\Big)\hspace{-0.15em} \frac{1}{(a^+)^{4}} \\
\hspace{-0.35em}&\hspace{-0.35em}&\hspace{-0.35em} \ \  \hspace{-0.15em} - \frac{a^{+n_1}}{(a^+)^{2}}a^{m_{p_0+1}}\Big\}\prod_{i=2}^{k} \frac{a^{+n_i}}{(a^+)^{2}}|0\rangle
   = \langle0|\Big(\prod_{j=1}^{p_0}\hspace{-0.15em} a^{m_{j}}\Big) \frac{a^{+n_1}}{(a^+)^{2}}a^{m_{p_0+1}}\prod_{i=2}^{k} \frac{a^{+n_i}}{(a^+)^{2}}|0\rangle-   \nonumber \\
\hspace{-0.35em}&\hspace{-0.35em}&\hspace{-0.35em} \ \  \hspace{-0.15em} -  \langle0|\Big(\prod_{j=1}^{p_0}\hspace{-0.15em} a^{m_{j}}\Big)  \hspace{-0.15em}\Big(\hspace{-0.15em}\eta^{m_{p_0+1}n_1}\hspace{-0.15em}\eta_{m_{k + 1}m_{k + 2}}- \hspace{-0.15em} \delta^{m_{p_0+1}}_{\{m_{k + 1} }\delta^{n_1}_{m_{k + 2}\}}\hspace{-0.15em}\Big) \hspace{-0.15em}\prod_{i=2}^{k+2} \hspace{-0.15em}\frac{a^{+n_i}}{(a^+)^{2}}|0\rangle \nonumber \\
\hspace{-0.35em}&\hspace{-0.35em}&\hspace{-0.35em} \   \hspace{-0.15em} = -  \langle0|\hspace{-0.15em} \Big(\prod_{j=1}^{p_0}\hspace{-0.15em} a^{m_{j}}\Big)\Big\{ \Big(\hspace{-0.15em}\eta^{m_{p_0+1}n_2}\hspace{-0.15em}\eta_{m_{k + 3}m_{k + 4}}- \hspace{-0.15em} \delta^{m_{p_0+1}}_{\{m_{k + 3} }\delta^{n_2}_{m_{k + 4}\}}\hspace{-0.15em}\Big) \hspace{-0.15em}\frac{a^{+n_1}}{(a^+)^{2}}\prod_{i=3}^{k+4} \hspace{-0.15em}\frac{a^{+n_i}}{(a^+)^{2}}\nonumber \\
\hspace{-0.35em}&\hspace{-0.35em}&\hspace{-0.35em} \  \   \hspace{-0.15em} - \frac{a^{+n_1}}{(a^+)^{2}}\frac{a^{+n_2}}{(a^+)^{2}}a^{m_{p_0+1}}\prod_{i=3}^{k} \frac{a^{+n_i}}{(a^+)^{2}}\Big\}|0\rangle    = \ldots  =   \langle0|\hspace{-0.15em} \Big(\prod_{j=1}^{p_0}\hspace{-0.15em} a^{m_{j}}\Big)\prod_{i=1}^{k} \frac{a^{+n_i}}{(a^+)^{2}}a^{m_{p_0+1}}|0\rangle =   0,
   \label{profscalprod1lef}
\end{eqnarray}
due to the repeated applying of the induction hypothesis, e.g. for the first summand in the  relation before last (and for the second term in the previous relation), as well as with  commutating of $a^{m_{p_0+1}}$ with $\frac{a^{+n_i}}{(a^+)^{2}}$ for $i=3,...,k$. The Hermitian conjugated quantities for ones in (\ref{scalprod1}): $\langle0|\prod_{i=1}^{k} \frac{a^{n_i}}{a^{2}}\prod_{j=1}^{p}\hspace{-0.15em} a^{+m_{j}}  |0\rangle$,  vanish as well.

To establish validity of (\ref{scalprod11}) we should commute $\frac{1}{(a^{+})^{2p}}$ through $\frac{1}{a^{2k}}$, which may be done with help of the  integral representation for $\frac{1}{(a^{(+)})^{2k}} = \int_{0}^{\infty}dt\exp\{-t a^{(+)2k}\}$, starting from the case, $p=k$, by means of  the auxiliary relation:
\begin{eqnarray}
 \hspace{-0.75em} &\hspace{-0.75em}& \hspace{-0.75em}\langle0| \frac{1}{a^{2k}}\frac{1}{a^{+2k}}|0\rangle =    \langle0|\hspace{-0.15em} \int_0^{\infty} dt_1dt_2   \sum_{e\geq 0}\frac{(-t_1)^e (a^{2})^{ke}}{e!}\sum_{g\geq 0}\frac{(-t_2)^g (a^{+2})^{kg}}{g!}|0\rangle =   \label{scalprod123}\\
  \hspace{-0.75em}&\hspace{-0.75em}&\hspace{-0.75em}\ \  =  \hspace{-0.15em}\int_0^{\infty} dt_1 dt_2    \sum_{e,g\geq 0}\frac{(-t_1)^e(-t_2)^g \langle0|\left\{(a^{2})^{ke} (a^{+2})^{kg}\right\} |0\rangle }{e!g!}  =  \hspace{-0.15em}\int_0^{\infty} dt_1 dt_2   \times   \nonumber \\
  \hspace{-0.75em}&\hspace{-0.75em}&\hspace{-0.75em}\ \  \times \sum_{e\geq 0}\frac{(t_1t_2)^e}{(e!)^2} \langle0|\prod_{j=1}^{ke}4j(g_0+j-1) |0\rangle = \hspace{-0.15em}\int_0^{\infty} dt_1 dt_2    \sum_{e\geq 0}\frac{(t_1t_2)^e}{(e!)^2}  { \prod_{j=1}^{ke}2j[d+2(j-1)] }  \label{scalprod21}
 \end{eqnarray}
 with using of the expansion   above in Taylor series for $\exp\{-t a^{(+)2k}\}$,   spectral properties: $\langle0|(a^{2})^{ke} (a^{+2})^{kg} |0\rangle$ $\sim \delta_{ge}... $, and  that $\forall k \in \mathbb{N}_0$:
  \begin{equation}\label{calcaux}
  \hspace{-0.55em} \langle0|\hspace{-0.15em} (a^{2})^k(a^{+2})^k|0\rangle = \hspace{-0.15em}\prod_{j=1}^{k}\langle0|\hspace{-0.15em}4j(g_0+j-1)|0\rangle,\ \mathrm{and} \ \langle 0|\hspace{-0.15em}(g_0+j-1)|0\rangle  = (d/2+j-1).
 \end{equation}
Therefore, we have respectively for $p=k=1$ and $p=2$, $k=1$
\begin{eqnarray}
 \hspace{-0.75em} &\hspace{-0.75em}& \hspace{-0.75em}\langle0| \frac{a^{m}}{a^2} \,  \frac{a^{+}_n}{(a^+)^{2}}|0\rangle =    \langle0|\hspace{-0.15em} \int_0^{\infty} dt_1 dt_2    \sum_{e\geq 0}\frac{(t_1t_2)^e}{(e!)^2} a^{m}\big(a^{2e}a^{+2e}\big)a^{+}_n|0\rangle  =  - \delta^{m}_{n} \hspace{-0.15em}\int_0^{\infty} dt_1 dt_2     \times\nonumber \\
  \hspace{-0.75em}&\hspace{-0.75em}&\hspace{-0.75em}\ \ \times  \sum_{e\geq 0}\frac{(t_1t_2)^e}{(e!)^2} \prod_{j=1}^{e}\langle0|\hspace{-0.15em}4j(g_0+j)    |0\rangle  =  -   \delta^{m}_{n} \int_0^{\infty} dt_1 dt_2    \sum_{e\geq 0}\frac{(t_1t_2)^e}{(e!)^2} \prod_{j=1}^{e}2j[d+2j]  ;  \label{scalprod321} \\
  \hspace{-0.75em} &\hspace{-0.75em}& \hspace{-0.75em}\langle0| \frac{a^{m_1}a^{m_2}}{a^4} \,  \frac{a^{+}_n}{(a^+)^{2}}|0\rangle =    \hspace{-0.15em}\int_0^{\infty} dt_1 dt_2  \sum_{e\geq 0}(-1)^e\frac{(t_1t_2^2)^e}{e!(2e)!}\langle0| a^{m_1}a^{m_2}\big(a^{4e}a^{+4e}\big)a^{+}_n|0\rangle   =0; \label{scalprod3210}
 \end{eqnarray}
so that, for any $p= k+1$, $p,k \in \mathbb{N}$ the presentation (\ref{scalprod11}) is valid.  Whereas for   $p=k$, $\forall p \in \mathbb{N}$ the average values in (\ref{scalprod11}) calculated with account of (\ref{scalprod123}), (\ref{scalprod21}):
\begin{eqnarray}
 \hspace{-0.75em} &\hspace{-0.75em}& \hspace{-0.75em}\langle0| \prod_{j=1}^{p}\hspace{-0.15em} \frac{a^{m_{j}}}{a^2} \, \prod_{i=1}^{p} \frac{a^{+}_{n_i}}{(a^+)^{2}}|0\rangle = \hspace{-0.15em}\int_0^{\infty} dt_1 dt_2    \sum_{e,g\geq 0}\frac{(-t_1)^e(-t_2)^g}{e!g!} \langle0| \prod_{j=1}^{p}\hspace{-0.15em} {a^{m_{j}}}\left\{(a^{2})^{pe} (a^{+2})^{pg}\right\} \prod_{i=1}^{p} {a^{+}_{n_i}} |0\rangle \nonumber \\
  \hspace{-0.75em}&\hspace{-0.75em}&\hspace{-0.75em}\ \  =   \int_0^{\infty} dt_1 dt_2    \sum_{e\geq 0}\frac{(t_1t_2)^e}{(e!)^2}\left\{\langle0| \prod_{j=1}^{p}{a^{m_{j}}}\hspace{-0.15em} \prod_{j=1}^{pe} \hspace{-0.15em}[4j(g_0+j-1)]   \prod_{i=1}^{p}a^{+}_{n_i}  |0\rangle\right\}  \nonumber \\
  \hspace{-0.75em}&\hspace{-0.75em}&\hspace{-0.75em}\ \ =\int_0^{\infty} dt_1 dt_2    \sum_{e\geq 0}\frac{(t_1t_2)^e}{(e!)^2}\left\{\langle0| \hspace{-0.15em} \prod_{j=1}^{pe} \hspace{-0.15em}[4j(g_0+j+p-1)]  \prod_{j=1}^{p}{a^{m_{j}}} \prod_{i=1}^{p}a^{+}_{n_i}  |0\rangle\right\}  \nonumber \\
 \hspace{-0.75em}&\hspace{-0.75em}&\hspace{-0.75em}\ \ =  (-1)^p  p! S^{(m)_p}_{(n)_p} \int_0^{\infty} dt_1 dt_2    \sum_{e\geq 0}\frac{(t_1t_2)^e}{(e!)^2} \hspace{-0.15em} \prod_{j=1}^{pe} \hspace{-0.15em}2j(d+2(j+p-1)) ,  \label{scalprod32100f}
 \end{eqnarray}
 that proves the validity of (\ref{scalprod11}) with $ K_{p,p}$ in (\ref{Kpp}).
The case $p\ne k$ in (\ref{scalprod11}) due to the argument from the footnote~10 is not essential for the evaluation of the scalar product (\ref{sproductnew})

Let us evaluate the finiteness of the
quantities $K_{p,p}$  (\ref{Kpp}). It is enough to check it for  simplest case of $K_{1,1}$:
   \begin{align} \label{Kpppr}& K_{1,1} =  \int_0^{\infty} dt_1dt_2  \sum_{k=0} \frac{(t_1t_2)^k}{(k!)^2}  \prod_{j=1}^{k}2j[d+2j]\ >\  \int_0^{\infty} dt_1dt_2  \sum_{k=0} \frac{(t_1t_2)^k}{(k!)^2}  \prod_{j=1}^{k}4j^2  \nonumber \\
& \phantom{K_{p,p}} = \int_0^{\infty} dt_1dt_2  \sum_{k=0}{(4t_1t_2)^k} > \int_0^{\infty} dt_1 dt_2\exp\{4t_1t_2  \} = \infty.
    \end{align}
    Thus, the operation $ \langle\ \big|\ \rangle^\prime$ (\ref{sproductnew})  can not be consider as the scalar product with finite norm, therefore not endowing the vector space $\mathcal{V}$ with the Hilbert space structure. This point proves impossibility to use inverse degrees in powers of oscillators  for the purpose of BRST-BFV Lagrangian formulation of the form $S_\Xi \sim  \langle\Phi \big| Q \big|\Phi\rangle^\prime $.

\section{Towards tensionless limit in open bosonic string\\  with CSR}\label{addalgebra3}
\renewcommand{\theequation}{\Alph{section}.\arabic{equation}}
\setcounter{equation}{0}
In this appendix we will show the way to find the CSR fields in the spectrum of  open bosonic string within a special tensionless limit.

Let us recall some standard properties of  open bosonic string
oscillators that  satisfy the commutation relations
\begin{equation}
[\alpha^m_k,\, \alpha^n_l]  \ = \  - k \delta_{k,-l}\eta^{mn}, \ \ k,\,l \in \mathbb{Z} .  \label{commosc}
\end{equation}
The  Virasoro generators $L_k$, and the Virasoro
algebra for their commutators take the form
\begin{align}
&L_k   \ = \  - \frac{1}{2} \sum_{l=-\infty}^{\infty} \alpha^m_{k-l}\alpha_{m{}l} ,
&& [L_k,\, L_l]\ =\  (k-l)L_{k+l} + \frac{d}{12}k(k^2-1),\label{Virasoroalg}
\end{align}
with the zero mode rescalling as
\begin{equation}
\alpha^m_0  \ = \  - i \sqrt{2\alpha^\prime} \partial^{m} \ = \ \sqrt{2\alpha^\prime} p^{m} .  \label{commosc0}
\end{equation}
We define the reduced generators
\begin{align}
& l_0   \ = \  - p^2 =  -\frac{1}{2\alpha^\prime}\alpha^m_0 \alpha_{m{}0}, &&   l_{\pm 1} =   p^{m} \alpha_{m{}\pm 1} \mp  \imath \Xi , \quad  \Xi \equiv -(1/\imath\sqrt{2\alpha'}) \alpha^m_{-2} \alpha_{m{}3}, \label{redVirasoro0}\\
&
  l_{k} =   p^{m} \alpha_{m{}k}  =  \frac{1}{\sqrt{2\alpha^\prime}}\alpha^m_0\alpha_{m{}k},  &&  |k| >1  \label{redVirasoro},
\end{align}
 where the real-valued dimensional parameter $
 \Xi $, $\Xi = \Xi^+ $ should satisfy the property
\begin{equation}\label{hermitespin}
\Xi^+  = \big(1/ \imath\sqrt{2\alpha'}\big) \alpha^m_{-3} \alpha_{m{}2} \   \Longrightarrow \ \alpha^m_{-3} \alpha_{m{}2} = -\alpha^m_{-2} \alpha_{m{}3}
\end{equation}
with a strong  operator constraint  on the values of the oscillators $\alpha^m_{\pm3},  \alpha_{m{}\pm 2}$ (for comparison see,  e.g.
 \cite{tensionlessl}, where the reducible massless (half)-integer representation of $ISO(1,d-1)$ was deduced  for $\Xi=0$).

The algebra of the constraints  $l_0, l_{\pm 1}, l_l$ for $ |l| >1$ satisfies to the simpler algebra, related  to the algebra for continuous spin fields in $\mathbb{R}^{1,d-1}$ in the  Shuster-Toro-like  form \cite{ShusterToro}:
\begin{eqnarray}\label{CSalgebra}
[l_k,l_l] &  = & k \delta_{k+l,0}\Big(l_0 + \frac{1}{2\alpha'}  \delta_{|k|,1}\big\{2\alpha^\mu_{-2}\alpha_{\mu{}2} -3 \alpha^\mu_{-3}\alpha_{\mu{}3}\big\}\Big)\\
&& + \frac{1}{\sqrt{2\alpha'}} l \cdot l_{k+l} \Big(\delta_{k, 1}\{\delta_{l, 2} +\delta_{l, -3}\}+\delta_{k, -1}\{\delta_{l, -2} + \delta_{l, 3}\} \Big), \nonumber
\end{eqnarray}
The non-diagonal non-vanishing commutators above are
\begin{equation}\label{CSalgebranontriv}
[l_{\pm 2},l_{\pm 1}]\  = \   \mp (2/\sqrt{2\alpha'})  l_{\pm 3} , \qquad  [l_{\pm 3},l_{\mp 1}]\  = \ \mp  (3/\sqrt{2\alpha'})    l_{\pm 2} ,
\end{equation}
 which in the naive tensionless limit $\alpha^\prime \to \infty$ vanish as well as the terms, $(1/ \alpha')\big\{2\alpha^\mu_{-2}\alpha_{\mu{}2} -3 \alpha^\mu_{-3}\alpha_{\mu{}3}\big\}$, in the first commutator.

 The Grassmann-odd BRST charge $\mathcal{Q}$  subject to the ghost number $gh_H (\mathcal{Q})$=1 with the Grassmann-odd operators of  ghost coordinates $C_k$ and momenta $P_k$,    $gh_H(C_k) = - gh_H(P_k)=1$, satisfying  the anticommutator relations
  \begin{equation}
\{C_k,\, P_l\}  \ = \   \delta_{k,-l}  \label{commghost}
\end{equation}
are written in the known form \cite{Kato}
  \begin{equation}
\mathcal{Q}  \ = \   \sum_{-\infty}^{\infty} \Big(C_{-k}L_k -\frac{1}{2}(k-l) :C_{-k}C_{-l} P_{k+l}:\Big)  - C_0 . \label{Qstring}
\end{equation}
 Rescaling the ghost operators $(C, P) \to (c, p)$ without changing the commutation relations (\ref{commghost}):
  \begin{align}
& c_k = -\sqrt{2\alpha'}C_k,   \ \  p_k  = -(1/\sqrt{2\alpha'})P_k  ,\  k\ne 0 ,  && (c_0, p_0) =   (\alpha' C_0, (1/ \alpha')P_0) \label{ghosts }
\end{align}
 and make the tensionless limit in $\mathcal{Q}$:
 \begin{eqnarray}
\lim_{\alpha' \to \infty} \mathcal{Q} \hspace{-0.2em} & \hspace{-0.2em}= &  \hspace{-0.2em} \lim_{\alpha' \to \infty}  \bigg[(1/{\alpha'}) c_{0}\big\{{\alpha'}l_0 + \widetilde{L}_0\big\} - \sum_{k\ne 0}^{\infty}(1/\sqrt{2\alpha'})c_{-k}\big\{-\sqrt{2\alpha'}l_k + \widetilde{L}_k\big\} \label{Qstringtens}  \\
\hspace{-0.2em}&\hspace{-0.2em}& \hspace{-0.2em}+ \hspace{-0.2em}\sum_{k,l }\hspace{-0.2em}\Big(\frac{1}{2}(k-l)(1/{2\alpha'}) :c_{-k}c_{-l} \Big\{\sqrt{2\alpha'}p_{k+l}(1-\delta_{k,-l})-\alpha'\delta_{k,-l} p_0 \hspace{-0.2em}\Big\}:\hspace{-0.2em}\Big) \hspace{-0.1em} + (1/ \alpha') c_0\bigg]  \nonumber  \\
\hspace{-0.2em} & \hspace{-0.2em}= &  \hspace{-0.2em} c_{0}l_0 + \sum_{k\ne 0}^{\infty} \Big(c_{-k}l_k -\frac{k}{2}c_{-k}c_{k} p_{0}\Big) , \nonumber
\end{eqnarray}
 where the operators $\big(\widetilde{L}_0, \widetilde{L}_k\big)$ = $\big({L}_0-{\alpha'}l_0, \widetilde{L}_k+\sqrt{2\alpha'}l_k \big)$  do not contain the terms with ${\alpha'}$-dependence and the algebra of the operators $l_k $ (\ref{redVirasoro0}, \ref{redVirasoro}) has the form (\ref{CSalgebra}) for $\alpha' \to \infty$:
 \begin{equation}\label{CSalgebratens1}
[l_k,l_l]\  = \  k \delta_{k+l,0}l_0 ,
\end{equation}
which encoded by the nilpotent  BRST operator $Q$  for any $d$
\begin{equation}\label{BRSTCSR}
  Q \ = \ \sum_{-\infty}^{\infty} \Big(c_{-k}l_k -\frac{k}{2}c_{-k}c_{k} p_{0}\Big) = \lim_{\alpha' \to \infty} \mathcal{Q},
\end{equation}
which coincides with $Q_C$ (\ref{Qcchic}) for $|k|\leq 1$, for vanishing  $\eta_1$, $\eta_1^{+}$ and for
\begin{equation}\label{corrQCQ}
\Big(\hspace{-0.15em}\alpha^m_{-1}, \hspace{-0.1em}\alpha_{m{}1}; l_{-1},\hspace{-0.1em} l_1;  c_0, c_{-1}, c_{1}, p_{-1}, p_{1}, p_{0}\hspace{-0.15em}\Big)\equiv \Big(\hspace{-0.15em} -\imath\omega^m \hspace{-0.15em},\hspace{-0.1em}-\imath{\partial_\omega^m} ; m^+_{1}, m_1; \eta_0,\eta^{m+}_{1},\eta^{m}_{1}, \mathcal{P}^{m+}_{1}, \mathcal{P}^{m}_{1}, \imath \mathcal{P}_{0}\hspace{-0.15em}\Big).
\end{equation}
The operator $Q$ still contains  the  dependent oscillators $\alpha^m_{\pm 2},  \alpha^{n}_{\pm 3}$ due to identity in right-hand side of (\ref{hermitespin}).
In order to get the truncated BRST operator $\widetilde{Q}$ from $\mathcal{Q}$ without its presence, i.e. without the constraints
 $l_{\pm 2}, l_{\pm 3}$ as well as without the ghost variables $c_{\pm k}, p_{\pm k}$, $k=2,3$ we  modify the rescalling only for the latter ghosts as it was done for the zero mode ones:
 \begin{equation} (c_k, p_k) =   -(2\alpha' C_k, (1/ 2\alpha')P_k),\ \ \ k= \pm 2, \pm 3. \label{ghostsnew }
\end{equation}
 As a result, the operator  $\widetilde{Q} = \lim_{\alpha' \to \infty} \mathcal{Q}$ has the form
 \begin{eqnarray}
\lim_{\alpha' \to \infty} \mathcal{Q}  & = &  \widetilde{Q} \ = \ Q\vert_{\big( c_{\pm 2}, p_{\pm 2}; c_{\pm 3}, p_{\pm 3}\big)=0}     \label{Qstringtens2}
\end{eqnarray}
and is nilpotent for  any space-time dimension  $d$.  After rescalling for the oscillators $\alpha^m_{k_1} = \sqrt{|{k_1}|}\Big(a^m_{k_1}\theta_{{k_1},0}, a^{m+}_{-{k_1}}\theta_{0,{k_1}}\Big)$ for $|{k_1}|>0$ the relations (\ref{commosc}) are transformed to non-vanishing commutators
 \begin{equation}
[a^m_k,\, a^{n+}_l]  \ = \  -  \delta_{k,l}\eta^{mn}, \ k,l \in \mathbb{N}.  \label{commosc2}
\end{equation}
 The operators $Q$  and $\widetilde{Q}$ coincide when acting on the Hilbert subspace $\widetilde{\mathcal{H}}$ from the total Hilbert space ${\mathcal{H}}$ ($\widetilde{\mathcal{H}} \subset {\mathcal{H}}$), whose vectors do not depend on  $a^m_{- 2},  a^{n}_{- 3}$
 \begin{eqnarray}\label{BRSTCSR2}
  && Q\vert_{ \widetilde{\mathcal{H}}} \ = \ \widetilde{Q} \ = \
   c_{0}l_0 +   \sum_{k>0,k\ne 2,3}^{\infty}\Big(c_kl^+_k +c^+_kl_k - c_{k}c^+_{k} p_{0}\Big) ,
\end{eqnarray}
whereas the algebra takes the form
\begin{equation}\label{CSalgebratens}
[l_k,l^+_l]\  = \ \delta_{k,l}l_0, \ \ l_k = p^m a_{m {} k} - \imath  \Xi\delta_{1,k}, \ \ l^{+}_k = p^m a^{+}_{m {} k} + \imath  \Xi\delta_{1,k}.
\end{equation}
From the  nilpotency of  $\mathcal{Q}$ in $d=26$ and  the standard string BRST-complex with  free string equations and infinite chain of reducible gauge symmetries
\begin{equation}\label{qequations}
  \mathcal{Q} |\Phi\rangle =0 , \   \delta|\Phi\rangle =\mathcal{Q} |\Lambda\rangle ,  \ \delta|\Lambda\rangle =\mathcal{Q} |\Lambda^1\rangle, \ldots  ,
\delta|\Lambda^{p-1}\rangle =\mathcal{Q} |\Lambda^p\rangle, \ p\in \mathbb{N}.
\end{equation}
[for $gh_H(|\Phi\rangle, |\Lambda^p\rangle)$ = $(0,-p-1)$] it follows the same BRST-complex with nilpotent $\widetilde{Q}$ in  the tensionless limit for any $d$.

Recalling the representation in $\widetilde{\mathcal{H}}$ for the vacuum vector $|0\rangle$ has the form   $\big(a^m_k, c_k,$ $p_k, p_0\big)|0\rangle$ $ = 0$, $k>0$. Extracting the zero-mode ghosts in $\widetilde{Q}$  and in $|\Phi\rangle$, $|\Lambda^{p}\rangle$
 \begin{eqnarray}\label{BRSTCSR3}
  && \widetilde{Q} \ = \  c_{0}l_0 - M p_{0}+ \Delta Q, \qquad  \big(|\Phi\rangle, \, |\Lambda^{p}\rangle\big) = \big(|\phi\rangle, \, |\Lambda^{p}_0\rangle\big)+ c_0 \big(|\phi_1\rangle, \, |\Lambda^{p}_1\rangle\big)
\end{eqnarray}
for
\begin{equation}\label{MdQ}
M =  \sum_{k\ne 0, 2, 3}^{\infty}c_{k}c^+_{k}, \quad   \Delta Q = \sum_{k\ne 0, 2, 3}^{\infty}\Big(c_kl^+_k +c^+_kl_k\Big)
\end{equation}
we get the $c_0$-independent sequence
\begin{eqnarray}
\hspace{-0.9em}  &\hspace{-0.9em}&  \hspace{-0.9em} \left(\begin{array}{cc}
               l_0 & - \Delta Q\\
                \Delta Q & - M
             \end{array}\right)\left(\begin{array}{c}
                |\phi\rangle\\
               |\phi_1\rangle
             \end{array}\right)
   = 0 ,\ \  \delta\left(\begin{array}{c}
                |\phi\rangle\\
               |\phi_1\rangle
             \end{array}\right) = \left(\begin{array}{cc}
               \Delta Q  & -M \\
               l_0 & - \Delta Q
             \end{array}\right)\left(\begin{array}{c}
                 |\Lambda_0\rangle\\
                |\Lambda_1\rangle
             \end{array}\right),\ldots   \label{coind}
\end{eqnarray}
In case of scalar CSR ($k=1$)  the fields $|\phi\rangle $, $|\phi_1\rangle$, gauge parameter $|\Lambda_0\rangle$  (for $|\Lambda_1\rangle \equiv 0$ and $|\Lambda^{p}\rangle \equiv 0$, when $p>0$ due to $gh_H$ distribution)  can be presented in powers of  oscillators
\begin{eqnarray}\label{phi0D}
\hspace{-0.3em}&\hspace{-0.3em}& \hspace{-0.3em}  |\phi\rangle =  \Big(\sum_{l\geq 0}\frac{1}{l!}\varphi^{(m)_l}(x)+c^+_1p^+_1\sum_{l\geq 0}\frac{1}{l!}D^{(m)_l}(x)\Big)a^{+}_{m_1}...a^{+}_{m_l}|0\rangle\equiv  |\varphi\rangle +  c^+_1p^+_1  |D\rangle, \\
\hspace{-0.3em}&\hspace{-0.3em}& \hspace{-0.3em}  \Big(|\phi_1\rangle,\,|\Lambda_0\rangle \Big) =   p^+_1\sum_{l\geq 0}\frac{\imath}{l!}\Big( -C^{(m)_l}(x),\,\Lambda^{(m)_l}(x)\Big) a^{+}_{m_1}...a^{+}_{m_l} |0\rangle \equiv p^+_1  \Big(|C\rangle,\, |\lambda\rangle \Big). \label{phi1Lambda}
  \end{eqnarray}
 Because  the  number particle  operator, $g_0 = - \frac{1}{2}\{a^{+}_{m},\,a^{m}\}$,
   (usually associated  with the spin value of basic field in case of integer HS field)  no longer commute with enlarged divergence (gradient) $l_1^{(+)}$:
  \begin{equation}\label{g0diverg}
    [g_0,\, l_1] \ = \ - (l_1 + \imath\Xi ),\qquad
  [g_0,\, l_1^{+}] \ =  (l_1^+  - \imath\Xi ),  \end{equation}
  the ghost-independent equations and gauge transformations
  \begin{eqnarray}
  &&
               l_0|\varphi\rangle  - l_1^+ |C\rangle  =0 , \qquad   l_1|\varphi\rangle  - l_1^+ |D\rangle  =  - |C\rangle, \qquad   l_0|D\rangle  - l_1 |C\rangle
               =0, \label{tripletCS} \\
  && \delta\left( |\varphi\rangle, |C\rangle, |D\rangle\right) =  \left( l_1^+,  l_0,  l_1 \right) |\lambda\rangle  \label{gaugetrCS}
\end{eqnarray}
  contain all tensor fields $\varphi^{(m)_k}(x)$, $C^{(m)_k}(x)$, $D^{(m)_k}(x)$ starting from the scalar fields.

Equations (\ref{tripletCS}) are Lagrangian,  follow from the action, $\mathcal{S}(\varphi, C, D) = \int d c_0 \langle \Phi|\widetilde{Q}|\Phi\rangle$ and represent the triplet analogue of EoM for the fields from  scalar reducible CSR. The irreducible CSR should be selected by means of the specified  trace conditions imposed on $|\varphi\rangle, |C\rangle, |D\rangle, |\lambda\rangle$ which are realized by the operators
\begin{equation}\label{traceext}
\mathcal{L}_{11} = l_{11} + \mathcal{O}(c_1, p_1), \ \ \mathrm{for}    \ \ l_{11} = a^ma_m,
\end{equation}
  which should commute with BRST operator $\widetilde{Q}$: $[\widetilde{Q},\,\mathcal{L}_{11}] = 0$ to get consistent dynamics.

However, within the Fock space $\mathcal{H}$ generated by $a^+_m, c^+_1, p_1^+$  it seems impossible to realize such operator due to
\begin{equation}\label{l11l1+}
  [l_1^+,\, l_{11}] \ = \ 2(l_1+\imath\Xi)   \ \ \Big(\ \Longrightarrow \    [l_1,\, l^+_{11}] \ = \  -2(l^+_1-\imath\Xi) , \  \  l^+_{11}=a^+_ma^{+m} \Big).
\end{equation}
The problem  of non-closing for the commutators (\ref{l11l1+}) may be effectively resolved within a
special conversion procedure in a larger Hilbert space $\widetilde{\mathcal{H}}\otimes \mathcal{H}^{\prime}$ (we develop the respective study in \cite{PR2}).
  One can show, that the evaluation of the  Casimir operators (\ref{Casoperator})  (see, as well the footnotes 1,~5) on the field $|\phi\rangle$: $C_2  |\phi\rangle = P^2  |\phi\rangle = 0$ and $C_4  |\phi\rangle= (M^{mn}P_m)^2 |\phi\rangle = \Xi^2  |\phi\rangle + \delta_\lambda |\mathcal{F}\rangle $, with accuracy up to the gauge transformations of some vector $ |\mathcal{F}\rangle$  can be done following the recipe of \cite{ShusterToro} with allowance for the appropriate traceless conditions.

\end{document}